\documentclass[a4paper,12pt]{article}
\usepackage{amsmath}
\usepackage{latexsym}
\usepackage{amssymb}
\usepackage{graphicx}
\usepackage{color}
\usepackage{float}
\usepackage[caption = false]{subfig}
\usepackage{slashed}

\begin{document}

\begin{center}
{\LARGE \textbf{
$\tau^-\to\eta^{(\prime)}\pi^-\nu_\tau \gamma$ decays as backgrounds in the search for second class currents\\[2cm]}}

{\large \textbf{A.~Guevara$^{1}$, G.~L\'opez-Castro$^{1}$ and P.~Roig$^{1}$\\[1.2 cm]}
$^{1}$ Departamento de F\'isica, Centro de Investigaci\'on y de Estudios Avanzados del Instituto Polit\'ecnico Nacional,
Apartado Postal 14-740, 07000 Mexico City, Mexico}
\end{center}

\begin{abstract}
Observation of $\tau^-\to\eta^{(\prime)}\pi^-\nu_\tau $ decays at Belle-II would indicate either a manifestation of isospin symmetry breaking or genuine second class currents (SCC) effects. The corresponding radiative 
$\tau^-\to\eta^{(\prime)}\pi^-\nu_\tau \gamma$ decay channels are not suppressed by $G$-parity considerations and may represent a serious background in searches of SCC in the former. We compute the observables associated 
to these radiative decays using Resonance Chiral Lagrangians and conclude that vetoing photons with $E_\gamma>100$ MeV should get rid of this background in the Belle-II environment searching for the $\tau^-\to\eta \pi^-\nu_\tau $ channel. 
Similar considerations hold inconclusive for decays involving the $\eta^\prime$ given the theory uncertainties in the prediction of the $\tau^-\to\eta^\prime\pi^-\nu_\tau$ branching ratio. Still, additional kinematics-based cuts should be able 
to suppress this background in the $\eta^\prime$ case to a negligible level.
\end{abstract}

\vspace*{2.0cm}
PACS numbers: 11.30.Er 
, 12.39.Fe
, 12.40.Vv
, 13.35.Dx
\\
\hspace*{0.5cm}Keywords: Second class currents, semileptonic tau decays, chiral Lagrangians, vector meson dominance\\

\section{Introduction}\label{sec:Intro}
Searches for the tau lepton decays $\tau^- \to a^-_0(980)\nu_{\tau}$ and $\tau^-\to b^-_1(1235)\nu_{\tau}$ were suggested long ago  
in ref. \cite{Leroy:1977pq} as clean signatures of second-class currents (SCC) \cite{weinberg}~\footnote{The other two SCC have the 
quantum numbers of the $\eta/\eta^\prime$ and $\omega/\phi$ mesons, respectively. 
Thus, in their production via the charged weak current they need to come along with 
an associated $\pi^\pm$.}. SCC are defined as those having 
opposite $G$-parity to the weak currents in the standard model (SM). Since $G\equiv Ce^{i\pi I_2}$ (with $C$ the charge conjugation 
operator and $I_i$ the generators of isospin rotations), the above decay channels of $\tau$ leptons can be induced also by breaking 
of charge-conjugation and/or isospin symmetry. Breaking of isospin symmetry \cite{pi0eta} allows us to estimate that branching fractions of $G$-parity suppressed channels are four orders of magnitude 
smaller than similar decays that are allowed in the SM. The opposite G-parities 
of pions and $\eta$ mesons would yield a violation of this quantum number in $\eta^{(\prime)}\pi^-$ production through the $\bar{d}\gamma^\mu u$ current \textit{independently} of the intermediate 
(resonance) dynamics~\footnote{Although $\eta\pi^-$ is the predominant decay mode of the $a^-_0(980)$ \cite{PDG}, this final-state di-meson system need not be produced through an intermediate $a_0^-$ 
resonance.}. Therefore, the measurement of $\tau^- \to \pi^- \eta^{(\prime)}\nu_{\tau}$ would be an unambiguous signature of SCC: either \textit{induced} by isospin or $C$-parity 
breaking (within the SM) or \textit{genuine} (by beyond the SM currents). On the contrary, the detection of $\tau^-\to b^-_1(1235)\nu_{\tau}$ through the $b^-_1(1235)$ 
dominant decay products ($\omega(782)\pi$, where the $\omega$ decays in turn mostly to $\pi^+\pi^-\pi^0$) must be indirect, since the intermediate $\omega\pi$ system could
have been produced via a $\rho(770)$ resonance (which is an ordinary first class current process). Analyzing the angular distribution of the final-state pions allows to set an upper bound of $1.4\cdot10^{-4}$ on 
the SCC decay $\omega\pi$ at the $90\%$ confidence level \cite{Aubert:2009an}, to be compared with the measured rate of $\sim 2\%$ for this process \cite{PDG}. Theoretical expectations of SCC contributions 
to this decay mode within the SM have been explored in Ref.~\cite{Paver:2012tq}, estimating $BR\sim2.5\cdot10^{-5}$ based on spin-one meson dominance.\\

After unsuccessful searches of SCC in nuclear beta decays \cite{Grenacs:1985da}, there was a renewed interest on this topic  
after the claim by the HRS collaboration \cite{derrick1987} of having observed the decay channel $\tau^- \to \pi^-\eta\nu_{\tau}$ 
with a branching fraction of $(5.1\pm 1.5)$\% \cite{derrick1987}, an unexpectedly large rate. This result was followed by 
an effort of theorists to assess the size of this decay \cite{estimates-scc}, which led to $\mathcal{O}(10^{-6}-10^{-5})$ for the 
branching ratio into the $\eta\pi^{-}$ channel (and $<10^{-6}$ for the $\eta'\pi^{-}$ decay mode). Currently, the best upper limits available are based on searches by the BaBar
 collaboration \cite{babar2011} corresponding to $BR(\tau^-\to \pi^-\eta\nu_\tau)<9.9\times 10^{-5}$ and $BR(\tau^-\to \pi^-\eta^\prime\nu_\tau)<7.2\times 10^{-6}$ \cite{Aubert:2008nj}, which lie 
close to the estimates based on isospin symmetry breaking for the $BR(\tau^-\to \pi^-\eta^{(\prime)}\nu)$ decays \cite{estimates-scc, recent} 
~\footnote{Belle reported slightly smaller branching ratio upper limits 
\cite{Hayasaka:2009zz}, $BR<7.3\times 10^{-5}(<4.6\times 10^{-6})$ for the $\pi^-\eta(\eta^\prime)$ decay channels, at $90\%$ CL, in the 2009 Europhysics Conference on High Energy Physics.}. 
Future searches at superflavor factories (like Belle-II) will hopefully provide us with the discovery of these channels \cite{B2TIPReport}. In view of this experimental 
improvement and since the discovery of \textit{genuine} SCC would point to the existence of new physics, it becomes interesting to revisit 
these tau lepton decays. For this purpose it is very important to have a reliable theoretical estimate of the SM prediction on these 
channels as well as of all possible physical backgrounds in experimental searches.\\

 Along this line of research, two QCD based studies of the $\tau^- \to \pi^-\eta\nu_{\tau}$ decays have been published recently \cite{Descotes-Genon:2014tla, Escribano:2016ntp} (also discussing the 
 $\eta^\prime$ channel in the latter reference). 
 It is clear, however, that both the errors on the mixings in the $\pi^0-\eta-\eta^\prime$ system \cite{Kroll:2005sd} and the uncertainties of the parameters describing the dominant scalar form factor 
 (obtained from a fit to meson-meson scattering 
 data \cite{Guo:2012yt} in ref.~\cite{Escribano:2016ntp}) are currently limiting the accuracy of these predictions. Still, $\tau^- \to \pi^-\eta\nu_{\tau}$ decays are predicted with a branching fraction of $\sim1\times10^{-5}$ 
 (certainly at reach of even first-generation 
 B-factories), while $\tau^- \to \pi^-\eta^\prime\nu_{\tau}$ decays are expected with a branching ratio of $\left[10^{-7},10^{-6}\right]$ (which could even be challenging for Belle-II).\\
 
 If SCC were not discovered in $\tau^- \to \pi^-\eta\nu_{\tau}$ decays at first generation B-factories it was due to the tough background present \cite{Simon, Bevan:2014iga}. That happened even though Belle undertook a 
 thorough programme to measure the main of these 
 backgrounds to allow a data-driven rejection of them in the search for SCC: $\tau\to K^- \eta\nu_\tau$ \cite{Inami:2008ar} (with the $K$ misidentified as a $\pi$), $\tau\to\eta\pi^-\pi^0\nu_\tau$ 
 \cite{Inami:2008ar} (failing to reconstruct the 
 $\pi^0$ from its two photon decay products) and similarly $\tau\to\eta(K\pi)^-\nu_\tau$ \cite{Inami:2008ar}, $\tau^-\to(4\pi)^-\nu_\tau$ \cite{Hayashii} (if the $\eta$ meson in the SCC process is to be detected through 
 its 3-pion decays) and $\tau\to\pi^-\gamma\nu_\tau$ \cite{Denis} (due to an additional photon from elsewhere with a di-photon 
 invariant mass around $m_\eta$). Unfortunately $\tau\to\pi^-\nu_\tau$ (with continuum $\gamma\gamma$ contributions) was not measured at the B-factories, and among the most frequent tau decay modes: two 
 \cite{Fujikawa:2008ma} and three pion modes, 
 the latter was measured at BaBar \cite{Aubert:2008nj} but not at Belle, being these decay channels also a difficult background to reject. In parallel to this remarkable experimental effort, some of 
 these decays have also been studied recently 
 \cite{Escribano:2013bca, Escribano:2014joa, Dumm:2012vb, Guo, Dumm:2009va, Dumm:2013zh} aiming to reduce the associated uncertainties in the related Monte-Carlo simulation 
 \cite{Shekhovtsova:2012ra, Nugent:2013hxa, Was:2015laa}. A notable programme in this 
 direction was also pursued by the BaBar Collaboration \cite{babar2011, Aubert:2008nj, Lees:2012ks}.\\
 
  In this article we study the related $\tau^- \to \pi^-\eta^{(\prime)}\nu_{\tau}\gamma$ decays, which provide a physical background 
for undetected photons. Since the non-radiative decay is very suppressed in the SM owing to isospin breaking, photon 
radiation off external lines is further suppressed by at least two orders of magnitude ($\mathcal{O}(\alpha)$ suppression in the observables)~\footnote{We check in appendix A that this is indeed the case using a reasonable threshold for 
photon detection.}. Instead, the model-dependent contributions 
to this radiative decay (of order $k$ in photon four-momentum \cite{Low:1958sn}) are not suppressed by $G$ parity considerations and involve only 
the effective $\gamma W^*\pi^-\eta^{(\prime)}$ vertex. Corresponding to an isospin breaking analysis (where effects due to $m_u\neq m_d$ and $e\neq0$ have to be taken into account at the same order), we expect a 
similar rate for the G-parity violating $\tau^- \to \pi^-\eta\nu_{\tau}$ decays and for its radiative counterpart (with structure-dependent contributions suppressed only by 
$\mathcal{O}(\alpha)\sim\epsilon_{\pi\eta}^{(0)}=\frac{\sqrt{3}(m_d-m_u)}{4(m_s-(m_u+m_d)/2)}\sim 10^{-2}$). 
Another important aspect to note is the fact that while inner bremsstrahlung (IB) contributions peak at low photon energies, this is not the case for the model-dependent contributions we are interested in. 
In fact, we will see that this should enable us to get rid of the radiative background by cutting above certain reasonable photon energies \footnote{One cannot reject all photons since one of the preferred $\eta$ detection modes is 
its two photon decay. Also its decays involving $\pi^0$'s need them to be detected by means of two-photon decays.}.\\

 Noticeably, refs.~\cite{Descotes-Genon:2014tla, Escribano:2016ntp} disagree in the presence of a characteristic signature of the $\eta\pi$ decay mode as a peak corresponding to the $a_0(980)$ state. 
 While ref.~\cite{Descotes-Genon:2014tla} concludes that 
the strength of this particular signal depends on the energy dependence of the relevant phaseshift (and specifically on the energy at which it exhibits a dip), ref.~\cite{Escribano:2016ntp} -on the 
contrary- concludes that meson-meson scattering data 
require that any structure in the $a_0(980)$ resonance region be weak enough to appear as buried in the continuum. Nevertheless, this last reference concludes that a signature of scalar form factor 
contributions to the $\tau^- \to \pi^- \eta 
\nu_\tau$ decays should appear as a prominent sharp peak around the $a_0(1450)$ resonance, while basically no signal is expected above the GeV according to Ref.~\cite{Descotes-Genon:2014tla}. In view of 
these contradictory predictions it is therefore 
appropriate to discuss if the presence of $\tau^- \to \pi^- \eta \gamma \nu_\tau$ decays can be a relevant background, particularly concerning the scalar resonance signatures, a feature to which we will 
pay particular attention.\\

We carry out our computations in the framework of the Resonance Chiral Lagrangians and compare our results to a simplified calculation based on a meson dominance model. To the best of our knowledge, this decay has 
not been 
considered before in the literature. Our results confirm the isospin breaking counting as radiative and non-radiative processes are predicted with $BR\sim10^{-5}$ for the $\pi\eta$ case. However, a feasible cut (even at 
Belle-II) for $E_\gamma>100$ 
MeV allows to suppress this particular background enough to allow the possible detection of SCC in $\tau^- \to \pi^-\eta\nu_{\tau}$ decays. We will see, however, that this is not clear for the $\eta^\prime$ case, 
where the theory uncertainties on $BR(\tau^- \to \pi^-\eta\nu_{\tau})$ are large enough to cast doubts on the need of a cut around $50$ MeV to reject the radiative background. This cut does not seem realistic for Belle-II 
because a lot of activity will appear in the electromagnetic calorimeter at such low energies.\\

The paper is distributed as follows. We start deriving the expression for the matrix element of the $\tau^- \to \pi^-\eta\gamma\nu_{\tau}$ decays and splitting the model-(in)dependent contributions in 
section \ref{sec:General}. In the structure-dependent part we then deduce 
the basis of hadronic form factors that will be used throughout the paper. In section \ref{sec:MDM} we consider a meson dominance model to get a first prediction of these form factors and recall the 
phenomenological determination of the 
relevant couplings. In section \ref{sec:RChT} we begin discussing how the Chiral Lagrangians are extended to include resonances so that they can be applied at $\sim1$ GeV energies, corresponding 
to semileptonic tau decays, and give all 
relevant pieces of the Lagrangians that will be used to obtain the hadronic matrix element of $\tau^- \to \pi^-\eta^{(\prime)}\gamma\nu_{\tau}$ decays. In this case a much larger number of couplings 
emerges than in the meson dominance model. We will 
recap how some of them can be fixed demanding that the Green functions and related form factors obtained in the meson theory match their QCD counterparts obtained by doing the operator product expansion. 
Still some of them need to be determined phenomenologically 
which does not appear possible to us for a number of them, on which we could only made an estimation based on the scaling of the low-energy constants of the Chiral Lagrangian. In section \ref{sec:radbkg} 
we use the results in the two previous sections to examine the backgrounds that $\tau^- \to \pi^-\eta^{(\prime)}
\gamma\nu_{\tau}$ decays constitute in the search for SCC in the $\tau^- \to \pi^-\eta^{(\prime)} \nu_{\tau}$ decays. Finally, in section \ref{sec:Concl} we state our conclusions and discuss the prospects 
for discovering SCC in the considered decays 
at Belle-II. The analytical expressions for the one- and two-resonance mediated contributions to the form factors in the Resonance Chiral Lagrangian formalism and the corresponding energy-dependent resonance widths are 
included in the appendices.\\

\section{Matrix element and form factors}\label{sec:General}
The $\tau^- \to \pi^- \eta^{(\prime)} \gamma \nu_\tau$ decays have a richer dynamics than their non-radiative counterpart (see e. g. Refs.~\cite{Descotes-Genon:2014tla, Escribano:2016ntp}). In the SM both weak 
currents, vector and axial-vector, can contribute to 
their decay amplitude. Since the final state is not a $G$-parity eigenstate, these decays are not suppressed by isospin breaking/$C$-parity violation. As anticipated before, this decay channel is only 
suppressed by the fine structure constant, and may be of similar magnitude than the SCC channel. The radiative decay can pollute the non-radiative one if photons are undetected (either low-energy photons 
which come mostly from IB and should not be an issue since they are doubly 
suppressed by $G$-parity and $\alpha$; or present in inclusive measurements, where they can indeed become problematic if not properly cut above certain energy at which they can already be isolated 
properly from continuum contributions).\\

We chose the following convention of four-momenta: $\tau^-(P)\to \pi^-(p)\eta^{(\prime)}(p_0)\nu(p')\gamma(k,\epsilon)$. Thus, the general form of the decay amplitude for this radiative decay is

\begin{eqnarray}\label{m.e.}
 \mathcal{M}&=&\frac{e G_F V_{ud}^*}{\sqrt{2}}\epsilon^{*\mu}\left\lbrace \frac{H_\nu(p_0, p)}{-2P\cdot k}\bar{u}(p')\gamma^\nu(1-\gamma_5)(M_\tau+\slashed P-\slashed k)\gamma_\mu u(P)+
 \right.\nonumber\\
 && \left. +(V_{\mu\nu}-A_{\mu\nu})\bar{u}(p')\gamma^\nu(1-\gamma_5)u(P)\frac{}{}\right\rbrace\,.
\end{eqnarray}

The first term refers to the photon emission off the tau lepton. We recall that $H_\mu=H_\mu(p_0, p)$ is the hadronic current whose general form is
\begin{equation}\label{H_mu}
 H_{\mu}\equiv \langle \eta^{(\prime)} \pi^-| \bar{d}\gamma_{\mu}u|0\rangle = f_+(t)\left((p_0-p)_{\mu}-\frac{\Delta^2}{t}q_{\mu} \right) + f_0(t)\frac{\Delta^2}{t}q_{\mu}\ ,
\end{equation}
where $\Delta^2=m_{\eta^{(\prime)}}^2-m_\pi^2$, $q=p_0+p$ is the momentum transfer and $t=q^2$. With the above parametrization one can identify $f_+(t)$ and $f_0(t)$ with the form factors associated to the 
$L=1$ and 
$L=0$ waves of the $\eta^{(\prime)} \pi^-$ system, respectively \cite{Descotes-Genon:2014tla, Escribano:2016ntp}. Within the standard theory, this decay can be \textit{induced} by isospin breaking giving 
contributions to both $L=0,\ 1$ waves. \textit{Genuine} SCC (due, for 
example, to charged Higgs exchange) will contribute only to $L=0$ wave.\\

\begin{figure}[ht!]
 \centering
 \includegraphics[scale=0.75]{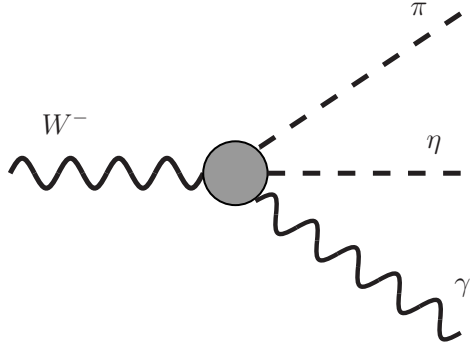}\caption{Effective hadronic vertex (grey blob) that defines the $V_{\mu\nu}$ and $A_{\mu\nu}$ tensors.}\label{fig:hadvertex}
\end{figure}

The hadronic $V_{\mu\nu}$ and $A_{\mu\nu}$ tensors in eq.~(\ref{m.e.}) are associated to the effective vector and axial-vector hadronic vertices with 
photon emission, shown in figure \ref{fig:hadvertex}. The vector tensor can be decomposed into a structure-independent (SI, which depends only upon the non-radiative decay amplitude) and a structure dependent (SD) piece: 
$V_{\mu\nu}=V_{\mu\nu}^{SI}+V_{\mu\nu}^{SD}$. On the contrary, 
$A_{\mu\nu}$ receives only SD contributions.\\

The model-independent contribution to the effective hadronic vector vertex is given by

\begin{eqnarray} \label{decomposition_Vmu}
 V_{\mu\nu}^{SI}&=&\frac{p_\mu}{p\cdot k} H_\nu(p_0,p+k)+\left[f_+(t')-\frac{\Delta^2}{t'}\left(f_0(t')-f_+(t')\right)\right]g_{\mu\nu} \nonumber\\
 &+& \frac{f_+(t)-f_+(t')}{(p_0+p)\cdot k} \left[ (p_0-p)_\nu-\frac{\Delta^2}{t}q_\nu\right] (p_0+p)_\mu \nonumber\\
 &+&\frac{\Delta^2}{t t'} \left[2(f_0(t')-f_+(t'))+\frac{t'}{(p_0+p)\cdot k}(f_0(t)-f_0(t'))\right] (p_0+p)_\mu q_\nu\,,
\end{eqnarray}
where $t'=(p_0+p+k)^2=t+2(p_0+p)\cdot k$
. It is easy to check that:
\begin{itemize}
 \item The Ward identity $k^\mu V_{\mu\nu}^{SI}=H_\nu(p_0,p)$ is satisfied. This ensures the current conservation for the corresponding SI part of eq.~(\ref{m.e.}).
 \item In the limit of equal hadron masses ($\Delta=0$), eq.~(\ref{decomposition_Vmu}) coincides with the SI part of eq.~(2.4) in ref.~\cite{Cirigliano:2002pv}.
 \item Note that in $\tau^-\to\pi^-\pi^0\gamma\nu_\tau$, it is justified to neglect $\Delta^2/t$ terms \cite{Cirigliano:2001er}. This is not the case for the $\tau^-\to\pi^-\eta^{(\prime)}\gamma\nu_\tau$ decays under study, 
 because $\Delta^2/t$ is not small in this case.
\end{itemize}

The first term in eq.~(\ref{m.e.}) and the SI piece in eq.~(\ref{decomposition_Vmu}) furnish the Low's amplitude with terms 
up to $\mathcal{O}(k^0)$. The SD terms, of $\mathcal{O}(k)$ in the decay 
amplitude, can be parametrized as follows 
\cite{BEG, Cirigliano:2002pv}:

\begin{eqnarray} \label{decomp}
V_{\mu\nu}&=&v_1(p.kg_{\mu\nu}-p_\mu k_{\nu})+v_2\left(g_{\mu\nu}p_0.k-p_{0\mu}k_{\nu}\right) \nonumber \\
&&+ v_3(p_\mu p_0.k-p_{0\mu}p.k)p_\nu+v_4(p_\mu p_0.k-p_{0\mu}p.k)p_{0\nu} \nonumber \\
A_{\mu\nu}&=&i\varepsilon_{\mu\nu\rho\sigma}\left(a_1p_0^{\rho}k^{\sigma}+
a_2k^{\rho}W^{\sigma}\right)+
i\varepsilon_{\mu\rho\sigma\tau}k^{\rho}p^{\sigma}p_0^{\tau}\left(a_3W_{\nu}+
a_4(p_0+k)_{\nu}\right)\,,
\end{eqnarray}
where $W=P-p'=p+p_0+k$. These tensors depend upon four vector ($v_i$) and four axial-vector ($a_i$) form factors, respectively; each one corresponding to coefficients of gauge-invariant structures. 
This decomposition is not unique and the 
non-vanishing form factors are determined by the specific theory input used to describe the $\pi^-\eta^{(\prime)}\gamma$ weak vertex. The Lorentz invariant form factors $v_i$, $a_i$ depend upon three Lorentz 
scalars. We can choose them as $W^2$, 
$(W-p_0)^2=(p+k)^2$ and $(W-p)^2=(p_0+k)^2$ (or any other convenient set). In writing the axial-vector part of the amplitude, the Schouten's identity has been used.\\

Since the form factors describing the non-radiative decay $\tau^-\to\pi^-\eta\nu_\tau$ are suppressed by isospin breaking giving $BR\lesssim 10^{-5}$, see above, we expect that 
Low's amplitude contribution to the rate of the radiative decay will be suppressed as 
$\epsilon_{\pi\eta}^2 \alpha$ ($BR's\lesssim 10^{-7}$)~\footnote{This type of contributions can also be neglected in $\tau^-\to\pi^-\eta^\prime\nu_\tau\gamma$ decays.}. Thus, in the following we will focus only in 
the SD contributions contained in eq.~(\ref{decomp}). In order to ascribe some systematic error to our predictions we will start using a simple 
meson dominance model whose results will be compared later on to those obtained from the more elaborated Resonance Chiral Lagrangian approach.\\

\section{Meson dominance model prediction}\label{sec:MDM}

\subsection{Framework}\label{sec:TheoMDM}
\begin{figure}[ht!]
 \includegraphics[scale=0.6]{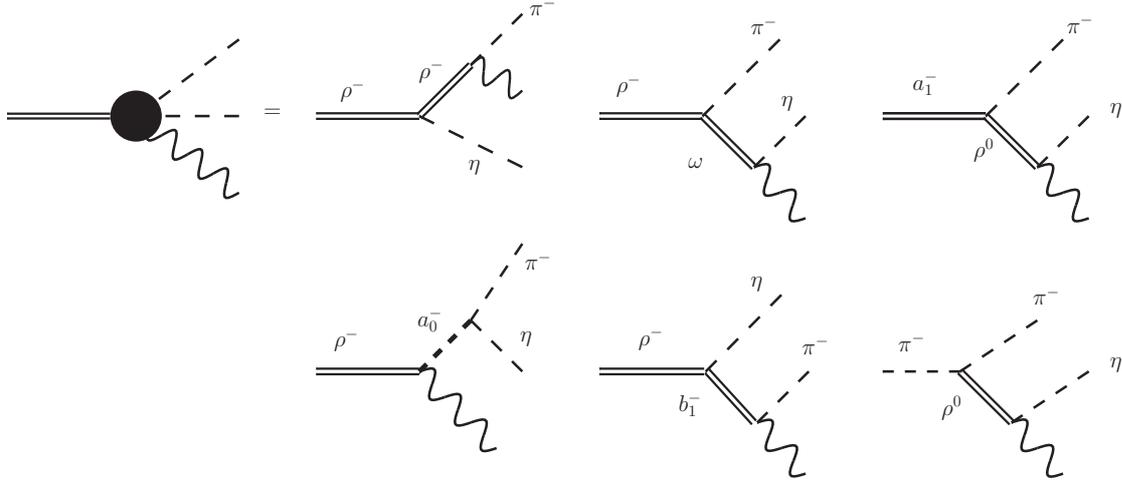}\caption{Contributions to the effective weak vertex in the MDM model. The wavy line denotes the photon.}\label{fig:vertex}
\end{figure}

The vertex of our interest, shown in Figure 1, involves the interactions of mesons with the weak and electromagnetic currents. 
In the Meson Dominance Model (MDM) one assumes that the weak and electromagnetic couplings are dominated by the exchange of a few light mesons (and their excitations). This approach is useful provided one 
is able to determine the relevant couplings 
from other independent sources (data fitting or model assumptions). The different contributions to the effective hadronic vertex in the MDM are depicted in figure \ref{fig:vertex}. The contribution of the $a_0(980)$ meson 
(and its excitation, with mass around $1450$ MeV), 
may produce a peak in the $\pi^-\eta$ invariant mass distribution which can mimic the effect of SCCs (albeit we recall there is some disagreement in the predicted scalar resonance effects according to 
different studies).\\

In MDM the structure of the vertices is more simple than the one obtained using Chiral Lagrangians. The Feynman rules required for the calculations are (momenta are indicated within parentheses; $V$, 
$A$, $P$, $S$ stand for the vector, 
axial-vector, pseudoscalar and scalar mesons, respectively):
\begin{eqnarray}\label{FeynmanrulesMDM}
 V'^\mu(r) \to V^\alpha(s) P(t) &:& i g_{V'VP}\epsilon^{\mu\alpha\rho\sigma}s_\rho t_\sigma\,,\label{FRVVP}\\ 
 V^\mu(r) \to \gamma^\alpha(s) P(t) &:& i g_{VP\gamma}\epsilon^{\mu\alpha\rho\sigma}s_\rho t_\sigma\,,\label{FRVAgamma}\\ 
 A^\mu(r) \to V^\alpha(s) P(t) &:& i g_{VAP} (r\cdot s g_{\mu\alpha}-r_\alpha s_\mu)\,,\label{FRVAP}\\
 V^\mu(r) \to \gamma^\alpha(s) S(t) &:& i g_{VS\gamma} (r\cdot s g_{\mu\alpha}-r_\alpha s_\mu)\,,\label{FRVSgamma}\\ 
 S(r) \to P(s) P'(t) &:& i g_{SPP'}\,. \label{FRSPP}
\end{eqnarray}

To simplify calculations, let us assume that:
\begin{itemize}
 \item The contribution from the intermediate $b_1(1235)$ meson can be neglected given that the $b_1$ couplings to both possible contributing vertices are suppressed: $BR(b_1\to\pi\gamma)=(1.6\pm0.4)\cdot10^{-3}$ and, 
 conservatively,  $BR(b_1\to\rho\eta)<10\%$ \cite{PDG}. We will also follow this hypothesis along the Chiral Lagrangian analysis in the next section.
 \item The contribution with the pion pole (last diagram in figure \ref{fig:vertex}) is very suppressed because the pion is far off its mass-shell. This approximation, on the contrary, can not be taken 
 using (Resonance) Chiral Lagrangians. We note that, as a result of this approach, all MDM contributions are in fact mediated by two-resonance exchanges.
\end{itemize}

\subsection{Form factors in the meson dominance model}\label{sec:FFsMDM}
The following contributions to the effective weak vertex are found (the superscripts denote the ordering of diagrams in the right-hand side of figure \ref{fig:vertex}), from left to right and from top to 
bottom; we have used the following definition ${\cal H}_\nu=(V_{\mu\nu}^{SD}-A_{\mu\nu})\epsilon^{*\mu}$):

\begin{eqnarray}
{\cal H}_{\nu}^a&=&\frac{i\sqrt{2}m^2_{\rho}}{g_{\rho}} g_{\rho^-\rho^-\eta}g_{\rho^-\pi^-\gamma}\frac{1}{D_{\rho}(W^2)}\frac{1}{D_{\rho}((p+k)^2)} 
\varepsilon_{\nu\alpha\rho\sigma}(p+k)^{\rho}p_0^{\sigma} \varepsilon^{\alpha\mu\gamma\delta}k_{\gamma}p_{\delta}\epsilon^*_{\mu} \,, \label{hme-mdmfirst}\\
{\cal H}_{\nu}^b&=&\frac{i\sqrt{2}m^2_{\rho}}{g_{\rho}} g_{\rho^-\omega\pi^-}g_{\omega\eta\gamma}\frac{1}{D_{\rho}(W^2)}\frac{1}{D_{\omega}((p_0+k)^2)} 
\varepsilon_{\nu\alpha\rho\sigma}(p_0+k)^{\rho}p^{\sigma} \varepsilon^{\alpha\mu\gamma\delta}k_{\gamma}p_{0\delta}\epsilon^*_{\mu} \,, \\
{\cal H}_{\nu}^c &=& \frac{i\sqrt{2}m_{a_1}^2}{g_{a_1}}g_{\rho^0 a_1^-\pi^-}g_{\rho^0\eta\gamma} \left((p_0+k).Wg_{\nu\alpha}-W_{\alpha}(p_0+k)_{\nu} \right) \nonumber \\ 
&&\ \ \ \ \ \times \frac{1}{D_{a_1}(W^2)D_{\rho}((p_0+k)^2)} \varepsilon^{\alpha\mu\gamma\delta}k_{\gamma}p_{0\delta}\epsilon^*_{\mu} \label{AconinMDM} \,, \\
{\cal H}_{\nu}^d &=& \frac{i\sqrt{2}m_{\rho}^2}{g_{\rho}}g_{\rho^-a_0^-\gamma}g_{a_0^-\pi^-\eta} \left(W.kg_{\mu\nu}-k_{\nu}W_{\mu} \right) \epsilon^{*\mu} \frac{1}{D_{\rho}(W^2)D_{a_0}((p+p_0)^2)}\, .\label{hme-mdmlast}
\end{eqnarray}
In the above expressions, we have defined $D_{X}(Q^2)$ as the denominator of the meson propagator, which may 
(or not) have an energy-dependent width; $g_X$ represents the weak couplings of spin-one mesons, 
defined here as $\langle X| J_{\mu} |0\rangle =i\sqrt{2}m^2_X/g_X\eta_{\mu}$ ($\eta_{\mu}$ is the polarization 
four-vector of meson $X$) and $g_{XYZ}$ denotes the trilinear coupling among mesons $XYZ$. The effects of the $\rho$ meson excitations can be taken into account through the following replacement
\begin{equation}
\frac{\sqrt{2}m_\rho^2}{g_\rho}\frac{1}{D_\rho(W^2)}\to\frac{\sqrt{2}}{g_{\rho\pi\pi}}\frac{1}{1+\beta_\rho}\left[BW_\rho(W^2)+\beta_\rho BW_{\rho'}(W^2)\right]\,,
\end{equation}
where 
\begin{equation}
BW_\rho(W^2)\,=\,\frac{m_\rho^2}{m_\rho^2-W^2-im_\rho\Gamma_\rho(W^2)} \,,
\end{equation}
with $BW_\rho(0)=1$ and $\beta_\rho$ encodes the strength of the $\rho'=\rho(1450)$ meson contribution. The $\rho\to\pi\pi$ coupling is denoted $g_{\rho\pi\pi}$ and $BW_{a_0}(X^2)$, $BW_{a_1}(X^2)$ and $BW_\omega(X^2)$ are defined in analogy to 
$BW_\rho(W^2)$.\\

Note that all the amplitudes in eqs.~(\ref{hme-mdmfirst}) to (\ref{hme-mdmlast}) are of $\mathcal{O}(k)$ in agreement with Low's theorem. All of them correspond to contributions to the vector current, except 
eq.~(\ref{AconinMDM}), which is due to the axial-vector current.\\

The MDM leads to the following form factors:
\begin{eqnarray}\label{FFsMDM}
v_1^{MDM}&=&iC_{\rho}\left[-\frac{g_{\rho^-\rho^-\eta}g_{\rho^-\pi^-\gamma}}{D_{\rho}[(p+k)^2]}p.p_0 +\frac{g_{\rho^-\omega\pi^-}g_{\omega\eta\gamma}}{D_{\omega}[(p_0+k)^2]}p_0.(p_0+k)
+\frac{g_{\rho^-a_0^-\gamma}g_{a_0^-\pi^-\eta}}{D_{a_0}[(p+p_0)^2]}\right]\,, \\
v_2^{MDM}&=& iC_{\rho}\left[\frac{g_{\rho^-\rho^-\eta}g_{\rho^-\pi^-\gamma}}{D_{\rho}[(p+k)^2]}p.(p+k) -\frac{g_{\rho^-\omega\pi^-}g_{\omega\eta\gamma}}{D_{\omega}[(p_0+k)^2]}p.p_0
+\frac{g_{\rho^-a_0^-\gamma}g_{a_0^-\pi^-\eta}}{D_{a_0}[(p+p_0)^2]}\right] \,,\\
v_3^{MDM}&=& iC_{\rho}\left[-\frac{g_{\rho^-\rho^-\eta}g_{\rho^-\pi^-\gamma}}{D_{\rho}[(p+k)^2]} \right]\,, \\
v_4^{MDM}&=& iC_{\rho}\left[ \frac{g_{\rho^-\omega\pi^-}g_{\omega\eta\gamma}}{D_{\omega}[(p_0+k)^2]}\right] \,, \\
a_1^{MDM}&=& C_A \left[\frac{g_{\rho^0a_1^-\pi^-}g_{\rho^0\eta\gamma}}{D_{\rho}[(p_0+k)^2]}\right] (p_0+k).W\,, \\ 
a_2^{MDM} &=& 0\,, \\
a_3^{MDM} &=& 0\,, \\
a_{4}^{MDM} &=& - \frac{a_1^{MDM}}{(p_0+k).W}\, .
\end{eqnarray}
In the above equations the shorthand notation $C_X(W^2)=\sqrt{2}m_X^2/[g_XD_X(W^2)]$ has been used.\\

\subsection{Determination of the relevant couplings}\label{sec:CoupsMDM}
The coupling constants required in MDM are defined in equations (\ref{FRVVP})-(\ref{FRSPP}). Comparisons of the calculated and measured rates 
allows to determine the relevant coupling constants assuming they are real and positive as indicated in the following.\\

 \begin{itemize}
  \item We can use the $\tau^-\to(\rho,a_1)^-\nu_\tau$ decays to extract the (axial-)vector weak coupling constants defined as indicated before. 
%
We use the decay width for $\tau^- \to X^-\nu_{\tau}$
 \begin{equation}
  \Gamma(\tau^-\to\nu_\tau X^-)\,=\,\frac{G_F^2|V_{ud}|^2}{8\pi M_\tau^3}\frac{M_X^2}{g_X^2} \left(M_\tau^2-M_X^2\right)^2 (M_\tau^2+2M_X^2)\,.
 \end{equation}
 
%
For the $a_1(1260)$ we assume 
$  BR(\tau^-\to a_1^-\nu_\tau)=0.1861\pm0.0013$ \cite{PDG}.
  Similarly, we can extract $g_\rho$ from $\tau^-\to\rho^-\nu_\tau$ decays; instead, we compare the measured value of the $\rho^0\to\ell^+\ell^-$ decay width with
   \begin{equation}
   \Gamma(\rho^0\to\ell^+\ell^-)\,=\,\frac{4\pi}{3}\left(\frac{\alpha}{g_\rho}\right)^2\left(1+\frac{2m_\ell^2}{M_V^2}\right)\sqrt{M_V^2-4m_\ell^2}\,.
  \end{equation}
  
  \item We extract the coupling constants $g_{VP\gamma}$ from the $V^\mu\to\gamma^\alpha(s) P(t)$ decays, 
using the decay width
 \begin{equation}
  \Gamma(V\to P\gamma)\,=\,\frac{|g_{VP\gamma}|^2}{96\pi M_V^3}(M_V^2-M_P^2)^3\,.
 \end{equation}
 This expression, together with $\Gamma(\rho/\omega \to \pi/\eta\, \gamma)$ \cite{PDG} allows to determine four of the required coupling constants. 
 
 \item In order to fix the $\rho a_1\pi$ coupling we consider the decay amplitude 
$ \mathcal{M}\,=\,i g_{\rho a_1\pi}(r\cdot s g_{\mu\alpha}-r_\alpha s_\mu)\eta_{a_1}^\mu \eta_\rho^{*\alpha}\,,$ for $a_1^\mu (r,\eta_{a_1}) \to\rho^\alpha(s,\eta_\rho)\pi(t)$ decays. This gives the
decay rate ($\lambda(a,b,c)$ is the ordinary K\"allen's function)
\begin{equation}
 \Gamma(a_1\to\rho\pi)\,=\,\frac{|g_{\rho a_1\pi}|^2}{96\pi M_{a_1}^3}\left[\lambda(M_{a_1}^2,M_\rho^2,m_\pi^2)+6M_\rho^2M_{a_1}^2\right] \lambda^{1/2}(M_{a_1}^2,M_\rho^2,m_\pi^2)\,.
\end{equation}
According to the PDG \cite{PDG} $a_1\to\rho\pi$ decays make up $61.5\%$ \cite{Asner:1999kj} of the total decay width of $a_1(1260)$, which we take as  $\Gamma_{a_1}=(475\pm175)$ MeV \cite{PDG}. Using isospin symmetry to relate 
the two decay modes of charged $a_1$ mesons lead us to the result in  Table \ref{ParsMDM}. 

\item
The following partial widths of $a_0(980)$ meson
\begin{eqnarray}
 \Gamma(a_0\to\gamma\gamma)\,&=&\,\frac{|g_{a_0\gamma\gamma}|^2}{32\pi}M_{a_0}^3\,, \\
 \Gamma(a_0\to\pi\eta)&=&\frac{|g_{a_0\pi\eta}|^2}{16\pi M_{a_0}^3}\lambda^{1/2}(M_{a_0}^2,m_\eta^2,m_\pi^2)\,,
\end{eqnarray}
can be used to extract the required coupling constants involving the $a_0$ meson.
%
Neither of these individual $a_0$ decay rates have been measured separately. Instead, measurements of their product have been reported by several groups with good agreement among them. The average value reported in PDG~\cite{PDG} is
\begin{equation}
 \Gamma(a_0\to\gamma\gamma)\times \frac{\Gamma(a_0\to\pi\eta)}{\Gamma_{a_0}}\,=\,\left(0.21^{+0.08}_{-0.04}\right) \mathrm{keV}\,.
\end{equation}
We can extract the product of coupling constants of the $a_0$ by comparing the previous equations and using  $\Gamma_{a_0}\,=\,\left(75.6\pm1.6^{+17.4}_{-10.0}\right)$ MeV~\cite{Uehara:2009cf} for the total decay width.

\item The coupling $g_{\rho\omega\pi}$ was fixed using the relation

\begin{equation}
 g_{\rho\omega\pi}\,=\,\frac{G^8}{\sqrt{3}}\left[\mathrm{sin}\theta_V+\sqrt{2}r\mathrm{cos}\theta_V\right]\,,
\end{equation}
where $G^8\ (G^0)$ is the $SU(3)$ invariant coupling of one pseudoscalar meson with two octets (one octet and one singlet) of vector mesons, and $r\equiv G^0/G^8$. Using the rates of $V \to P\gamma$ decays and assuming ideal 
$\omega-\phi$ mixing, $\theta_V=$tan$^{-1}\left(\frac{1}{\sqrt{2}}\right)$, one gets $G^8\,=\,(1.052\pm0.032)\cdot 10^{-2}$ MeV$^{-1}$ and $r=1.088\pm0.018$ \cite{FloresTlalpa:2008zz}.
 
\item 
The following MDM relations between strong and electromagnetic couplings
\begin{equation}
 g_{\rho\rho\eta}=\frac{g_\rho}{e}g_{\rho\eta\gamma}\,,
 \quad g_{a_0\rho\gamma}=\frac{g_\rho}{e}g_{a_0\gamma\gamma}\,.\label{MDMrelations}
\end{equation}
can be used to extract other relevant coupling constants.

\item 
Finally for the decays involving the $\eta^\prime$ meson, the couplings $g_{a_0\pi\eta'}$, $g_{\rho\rho\eta'}$, $g_{\omega\eta'\gamma}$ and $g_{\rho\eta'\gamma}$ need to be determined. Employing 
the above formulae it is straightforward to obtain the last two from the measured $\Gamma(\eta'\to\omega\gamma)$ and $\Gamma(\eta'\to\rho\gamma)$ decays \cite{PDG}. $g_{\rho\rho\eta'}$ is fixed in terms 
of $g_{\rho\eta'\gamma}$ in analogy to eq.~(\ref{MDMrelations}). It is not possible to determine $g_{a_0\pi\eta'}$ easily, because the involved masses forbid all possible one-to-two body decays. However, 
according to \cite{Guo:2012yt}, $g_{a_0\pi\eta'}<<g_{a_0\pi\eta}$. We will take $g_{a_0\pi\eta'}/g_{a_0\pi\eta}\leq 0.1$ as a conservative estimate.\\
\end{itemize}

 In table \ref{ParsMDM} we show the values of the coupling constants obtained using the above procedure. The errors are propagated from the experimental ones adding them in quadrature. In section \ref{sec:radbkgMDM} we 
 will present the MDM predictions for the $\tau^-\to\pi^-\eta^{(\prime)}\gamma\nu_\tau$ decays using these inputs.\\

  \begin{table}
   \centering
   \begin{tabular}{|c|c|} \hline
   Coupling constant&Fitted value\\ \hline
   $g_{\rho}$&$5.0\pm0.1$ \\ \hline
   $g_{a_1}$&$7.43\pm0.03$\\ \hline
   $eg_{\rho\eta\gamma}$&$(4.80\pm0.16)\times10^{-1}$ GeV$^{-1}$\\ \hline
   $g_{\rho\rho\eta}$&$(7.9\pm0.3)$ GeV$^{-1}$\\ \hline
   $eg_{\omega\eta\gamma}$&$(1.36\pm0.06)\times10^{-1}$ GeV$^{-1}$ \\ \hline
   $eg_{\rho\pi\gamma}$&$(2.19\pm0.12)\times10^{-1}$ GeV$^{-1}$\\ \hline
   $g_{\rho\omega\pi}$&$(11.1\pm0.5)$ GeV$^{-1}$\\ \hline
   $g_{a_1\rho\pi}$&$(3.9\pm1.0)$ GeV$^{-1}$\\ \hline
   $eg_{\rho a_0\gamma}$&$(9.2\pm1.6)\times10^{-2}$ GeV$^{-1}$\\ \hline
   $g_{a_0\pi\eta}$&$(2.2\pm0.9)$ GeV\\ \hline \hline
   $eg_{\rho\eta'\gamma}$& $(4.01\pm0.13)\times10^{-1}$ GeV$^{-1}$\\ \hline
   $eg_{\omega\eta'\gamma}$& $(1.30\pm0.08)\times10^{-1}$ GeV$^{-1}$\\ \hline
   $g_{\rho\rho\eta'}$& $(6.6\pm0.2)$ GeV$^{-1}$\\ \hline
   $g_{a_0\pi\eta'}/g_{a_0\pi\eta}$& $\leq 0.1$ \\ \hline
   \end{tabular}\caption{Our fitted values of the coupling parameters. Those involving a photon are given multiplied by the unit of electric charge.}\label{ParsMDM}

  \end{table}

\section{Resonance Chiral Lagrangian prediction}\label{sec:RChT}

\subsection{Theoretical framework}\label{sec:theoRChT}
Chiral Perturbation Theory ($\chi PT$) \cite{ChPT} is the quantum effective field theory dual to QCD at very low energies ($E\lesssim M_\rho$, being $M_\rho$ the mass of the $\rho(770)$ 
state). Therefore it provides an adequate description of semileptonic tau decays, albeit for low invariant masses of the meson system in the low multiplicity modes only \cite{Colangelo:1996hs}. 
Even in this situation, however, it only covers a small window of the available phase space in tau decays. A phenomenological approach to tackle this problem is to restore to the use of Chiral 
Lagrangians extended by including the lightest resonances as active fields, the so-called Resonance Chiral Lagrangians ($R\chi L$) \cite{RChL}. An advantage of this setting is that it 
reduces to the $\chi PT$ results in the chiral limit extending the applicability of the theory to GeV energies. This is done without assuming any symmetry related to the resonance 
dynamics (like for instance, hidden local symmetry, see e. g. \cite{HLS}) and ensuring that the Green functions and related form factors of $R\chi L$ comply with their known asymptotic 
suppression in QCD \cite{BL}. Then, the $R\chi L$ bridge between these two known limits of QCD on both energy ends: the chiral and perturbative regimes of the strong interaction. 
Extending the energy range of $\chi PT$ to larger energies implies that its perturbative expansion (in powers of the ratio of momenta and masses of the pseudoGoldstone bosons over the 
chiral symmetry breaking scale, $\Lambda_\chi\sim$ GeV) breaks down in the resonance region. Subsequently, $R\chi L$ face the problem of finding a suitable expansion parameter to build 
a perturbative expansion upon. A successful candidate is the inverse of the number of colors of the QCD gauge group in the limit where this is taken to be large \cite{LargeNc}. Remarkably, 
when this setting is applied to meson physics it agrees well both at the qualitative and quantitative levels with the related phenomenology~\cite{Pich:2002xy} (see also Refs.~\cite{NLO_1/N_C} 
where the extension of $R\chi L$ beyond the leading order in $1/N_C$ has been studied). In the following we recall the building blocks of the $R\chi L$ and present the operators relevant for 
our computation.\\

The light-quark ($q\,=\,u,\,d,\,s$) sector of QCD exhibits -in the approximate limit of massless quarks- a global $SU(3)_L\otimes SU(3)_R$ symmetry: the chiral symmetry of low-energy QCD 
in which the left- and right-handed quark components are transformed separately in (three-)flavor space. This symmetry is, nevertheless, not seen in the spectrum, where states belonging 
to flavor multiplets of opposite parity differ noticeably in mass (for instance $a_1(1260)$ vs. $\rho(770)$). Consequently, the chiral symmetry of the QCD Lagrangian must be realized in the 
Nambu-Goldstone boson way and only the vector subgroup $SU(3)_V$ of the chiral group is a symmetry of the QCD vacuum so that the meson multiplets fill irreps of $SU(3)_V$. The pattern of spontaneous 
symmetry breakdown is $SU(3)_L\otimes SU(3)_R\to SU(3)_V$ and the breaking of the $SU(3)_A$ generators should result in eight Goldstone bosons. These are in fact pseudo-Goldstone bosons (as a 
consequence of the explicit breaking of the chiral symmetry by the small $m_q$ values) to be identified with the lightest multiplet of pseudoscalar mesons. We discuss the parametrization of the 
corresponding fields in the following.\\

The coset space $SU(3)_L\otimes SU(3)_R\to SU(3)_V$ is conveniently parametrized by \cite{CCWZ}
\begin{equation}\label{eq_u(phi)}
 u(\phi)\,=\,\mathrm{exp}\left\lbrace \frac{i}{\sqrt{2}F}\Phi\right\rbrace\,,
\end{equation}
where (we include the generator of U(1) as the zeroth Gell-Mann matrix)
\begin{equation}\label{eq_Phi}
 \Phi\,=\,\frac{1}{\sqrt{2}}\sum_{i=0}^8 \lambda^i\phi_i\,=\,
 \left(
 \begin{array}{ccc}
  \frac{\pi^0+C_q\eta+C_{q'}\eta'}{\sqrt{2}} & \pi^+ & K^+\\ \vspace*{0.2cm}
  \pi^-& \frac{-\pi^0+C_q\eta+C_{q'}\eta'}{\sqrt{2}} & K^0\\ \vspace*{0.2cm}
  K^- & \overline{K}^0 & -C_s\eta+C_{s'}\eta'\\
 \end{array}
 \right)
\end{equation}
and $F\sim F_\pi\sim 92.2$ MeV is the pion decay constant in the chiral limit. In eq.~(\ref{eq_Phi}) we have considered the $\eta-\eta'$ mixing in the two-angle mixing scheme (which is 
consistent with the large-$N_C$ limit of QCD \cite{KL} in which $SU(3)_V\otimes U(1)_V$ becomes $U(3)_V$) and worked in the quark-flavor basis \cite{qfbasis}. Within this setting, the mixing parameters are
\begin{eqnarray}\label{definitions_Cq_Cs}
 C_q & \equiv & \frac{F}{\sqrt{3}\mathrm{cos}(\theta_8-\theta_0)}\left(\frac{\mathrm{cos}\theta_0}{f_8}-\frac{\sqrt{2}\mathrm{sin}\theta_8}{f_0}\right)\,,\quad C_{q^\prime}\equiv\frac{F}{\sqrt{3}\mathrm{cos}(\theta_8-\theta_0)}\left(\frac{\sqrt{2}\mathrm{cos}\theta_8}{f_0}+\frac{\mathrm{sin}\theta_0}{f_8}\right)\,,\nonumber\\
 C_s & \equiv & \frac{F}{\sqrt{3}\mathrm{cos}(\theta_8-\theta_0)}\left(\frac{\sqrt{2}\mathrm{cos}\theta_0}{f_8}+\frac{\mathrm{sin}\theta_8}{f_0}\right)\,,\quad C_{s^\prime}\equiv\frac{F}{\sqrt{3}\mathrm{cos}(\theta_8-\theta_0)}\left(\frac{\mathrm{cos}\theta_8}{f_0}-\frac{\sqrt{2}\mathrm{sin}\theta_0}{f_8}\right)\,,\nonumber\\
\end{eqnarray}
with~\cite{qfbasis} 
\begin{equation}\label{values_mixingangless&decayscts}
 \theta_8\,=\,\left(-21.2\pm1.6\right)^\circ,\quad  \theta_0\,=\,\left(-9.2\pm1.7\right)^\circ,\quad f_8\,=\, \left(1.26\pm0.04\right)F,\quad f_0\,=\, \left(1.17\pm0.03\right)F\,.
\end{equation}

As stated above, resonances are included without assuming any gauge symmetry related to their dynamics, and only $U(3)$ flavor symmetry is used to write
\begin{equation}\label{eq_Vmunu}
 V_{\mu\nu}\,=\,
 \left(
 \begin{array}{ccc}
 \frac{\rho^0}{\sqrt{2}}+\frac{\omega_8}{\sqrt{6}}+\frac{\omega_0}{\sqrt{3}} & \rho^+ & K^{*+}\\ \vspace*{0.2cm}
  \rho^-& \frac{-\rho^0}{\sqrt{2}}+\frac{\omega_8}{\sqrt{6}}+\frac{\omega_0}{\sqrt{3}} & K^{*0} \\ \vspace*{0.2cm}
  K^{*-}& \overline{K}^{*0}& \frac{-2\omega_8}{\sqrt{6}}+\frac{\omega_0}{\sqrt{3}}\\
 \end{array}
 \right)_{\mu\nu}\;.
\end{equation}
The antisymmetric tensor formalism for spin-one fields has been employed in eq.~(\ref{eq_Vmunu}). It turns out to be more convenient than the Proca formalism in this context, since upon 
integration of the resonance fields, the $\mathcal{O}(p^4)$ couplings of the even-intrinsic parity $\chi PT$ Lagrangian are saturated by these resonance contributions \footnote{Particularly, they turn out to be 
saturated by the spin-one resonance contributions. In this sense vector meson dominance \cite{Sakurai:1960ju} emerges as a result of the analysis and not as an \textit{a priori} assumption.}. Consequently, such 
next-to-leading order chiral Lagrangian in the normal parity sector is not included in our computations to avoid double counting \cite{RChL}.\\

The flavor states $\omega_0$ and $\omega_8$ are related to the physical $\omega(782)$ and $\phi(1020)$ particles by a rotation given by their mixing angle $\theta_V$:
\begin{equation}\label{eq_omegaphimixing}
 \left(
 \begin{matrix}
  \omega_8 \\
  \omega_0\\
 \end{matrix}
 \right)_{\mu\nu}
  \,=\,\left(\begin{matrix}
 \mathrm{cos}\theta_V & \mathrm{sin}\theta_V\\
 -\mathrm{sin}\theta_V & \mathrm{cos}\theta_V\\
 \end{matrix}\right)
 \left(
 \begin{matrix}
  \phi \\
  \omega\\
 \end{matrix}
 \right)_{\mu\nu}
 \,,
\end{equation}
with $\theta_V=$ tan$^{-1}\left(\frac{1}{\sqrt{2}}\right)$ in the ideal mixing scheme that we will follow. As a consequence of this precise value for the $\theta_V$, all possible contributions with intermediate exchanges of $\phi(1020)$ resonance 
to the (vector) form factors vanish.\\

The introduction of axial-vector resonances ($A_{\mu\nu}$) is performed analogously. Spin-zero 
resonances ($S$ and $P$) share the same flavor content as $V_{\mu\nu}$ and $A_{\mu\nu}$ as well (i.e., concerning the $SU(2)$ triplets, we will have the correspondences $a_0(980)\leftrightarrow\pi(1300)
\leftrightarrow \rho(770) \leftrightarrow a_1(1260)$ for the $S$, $P$, $V$ and $A$ states, respectively).\\

In addition to the fields corresponding to pseudoGoldstone bosons and resonances, it is convenient to add external hermitian matrix fields $s$, $p$, $v_\mu$ and $a_\mu$ transforming 
locally under the chiral group (as scalar, pseudoscalar, vector and axial-vector, respectively). These are coupled to the quark currents in order to provide a way of computing the 
corresponding Green functions of quark currents.\\

With these fields and external sources, the $R\chi L$ is built including resonance fields and the following basic covariant tensors \cite{CCWZ, ChPT}
\begin{eqnarray}\label{eq_chiraltensors}
&& u_\mu\,=\,u_\mu^\dagger\,=\,i\left\lbrace u^\dagger(\partial_\mu-ir_\mu)u-u(\partial_\mu-i\ell_\mu)u^\dagger\right\rbrace\,,\nonumber\\
&& \chi_{\pm}\,=\,u^\dagger \chi u^\dagger \pm u \chi^\dagger u\,,\nonumber\\
&& f_{\pm}^{\mu\nu}\,=\,uF_L^{\mu\nu}u^\dagger \pm u^\dagger F_R^{\mu\nu} u\,,\nonumber\\
&& h_{\mu\nu}\,=\,\nabla_\mu u_\nu + \nabla_\nu u_\mu\,.
\end{eqnarray}
In eq.~(\ref{eq_chiraltensors}), $\chi=2B_0(s+ip)$ includes the scalar and pseudoscalar external sources. The low-energy constant $B_0$ is related to the quark condensate in the chiral 
limit by means of $\left\langle0|\overline{q}^iq^j|0\right\rangle=-B_0 F^2\delta^{ij}$. (Axial)-vector and left/right sources are related by $v^\mu=\frac{1}{2}(r^\mu+\ell^\mu)$ and 
$a^\mu=\frac{1}{2}(r^\mu-\ell^\mu)$, respectively. $F_{L,R}^{\mu\nu}$ correspond to the usual field-strength tensors:
\begin{equation}
 F_y^{\mu\nu}\,=\,\partial^\mu y^\nu-\partial^\nu y^\mu-i\left[y^\mu,y^\nu\right],\quad y=\ell,r.
\end{equation}
The covariant derivative entering the last of eqs.~(\ref{eq_chiraltensors}) is given by
\begin{equation}
 \nabla_\mu X = \partial_\mu X + \left[\Gamma_\mu, X\right]\,,
\end{equation}
with the chiral connection
\begin{equation}
 \Gamma_\mu\,=\,\frac{1}{2}\left\lbrace u^\dagger(\partial_\mu-ir_\mu)u+u(\partial_\mu-i\ell_\mu)u^\dagger\right\rbrace\,.
\end{equation}

With these building blocks, the $R\chi L$ Lagrangian is built including the most general set of chiral-invariant operators which also respect Lorentz, $P$ and $C$ invariance, together with hermiticity. 
Schematically, the operators are
\begin{equation}\label{operators}
 \mathcal{O}_i^{R_iR'_j...}\sim\left\langle \prod_{i,j,...} R_iR'_j...\chi^{n}(\Phi)\right\rangle\,,
\end{equation}
where $\left\langle ...\right\rangle$ stands for a trace in flavour space and $\chi^{n}(\Phi)$ is a chiral tensor of $\mathcal{O}(p^n)$ in the chiral counting made up combining the chiral tensors 
appearing in eqs.~(\ref{eq_chiraltensors}). $\prod_{i,j,...} R_iR'_j...$ includes $i(j,...)$ copies of resonance multiplets of type $R_{i(j,...)}$ ($S$, $P$, $V$ and $A$, since here we are restricting 
ourselves to the lowest-lying states for given quantum numbers).\\

The construction of our Lagrangian will be driven by the $N_C\to\infty$ limit of large-$N_C$ QCD. In general, terms with a single trace are leading order in $N_C$, while every additional trace brings in 
a $1/N_C$ suppression factor (see, however, Appendix A of Refs. \cite{Cirigliano:2006hb, Kampf:2011ty}). We will start with the pseudoGoldstone boson Lagrangian, which is
\begin{equation}
 \mathcal{L}^{pGb}\equiv \mathcal{L}_{\chi PT}^{\mathcal{O}(p^2)}+\mathcal{L}_{\chi PT,WZW}^{\mathcal{O}(p^4)}\,,
\end{equation}
where the first(second) term belongs to the even-(odd-)intrinsic parity sector and $\mathcal{O}(p^n)$ indicates the order in the chiral expansion. We note that $\mathcal{L}_{\chi PT}^{\mathcal{O}(p^4)}$ 
in the even-intrinsic parity sector must not be included in $\mathcal{L}^{pGb}$ to avoid double counting, as explained before. The lowest-order Lagrangian in the chiral expansion is
\begin{equation}\label{LO_Lagrangian}
 \mathcal{L}_{\chi PT}^{\mathcal{O}(p^2)}\,=\,\frac{F^2}{4}\left\langle u^\mu u_\mu + \chi_+ \right\rangle\,.
\end{equation}
$\mathcal{L}_{\chi PT,WZW}^{\mathcal{O}(p^4)}$ corresponds to the Wess-Zumino-Witten chiral anomaly functional \cite{ChiralAnomaly} $Z[U,v,a]$, which can be read from Ref.~\cite{Kampf:2011ty} 
(using $U=u^2$).\\

For the terms with resonances, we start with those derived in Refs.~\cite{RChL}. 'Kinetic' terms (they also include interactions, via the covariant derivative) for resonances $R=Z,O$, of order 
$\mathcal{O}(N_C)$, are
\begin{equation}\label{kinetic}
 \mathcal{L}_{kin}^R\,=\,-\frac{1}{2}\left\langle \nabla^\mu O_{\mu\nu}\nabla_\alpha O^{\alpha\nu}\right\rangle+\frac{1}{4}M_O^2\left\langle O_{\mu\nu} O^{\mu\nu}\right\rangle+
 \frac{1}{2}\left\langle \nabla^\beta Z\nabla_\beta Z\right\rangle-\frac{1}{2}M_Z^2\left\langle Z Z\right\rangle\,,
\end{equation}
where $Z$ and $O$ are resonances of spin zero ($Z=S,P$) and one ($O=V,A$), respectively~\footnote{We will neglect the interaction with tensor resonances since they are rather weak \cite{Ecker:2007us}. See, however ref.~\cite{Shekhovtsova:2016luj}.}. 
The interaction terms linear in resonance fields which -upon their integration out- contribute to the low-energy constants of the $\chi PT$ 
Lagrangian at $\mathcal{O}(p^4)$ were also derived in Refs.~\cite{RChL}. These are
\begin{eqnarray}\label{linear_interactions}
{\cal L}
^R & = & c_d \left\langle S u^\mu u_\mu\right\rangle + c_m \left\langle S\chi_+\right\rangle + i d_m \left\langle P \chi_-\right\rangle+i \frac{d_{m0}}{N_F}\left\langle P\right\rangle\left\langle \chi_-\right\rangle\nonumber\\
& & + \frac{F_V}{2\sqrt{2}} \langle V_{\mu\nu} f_+^{\mu\nu}\rangle + i\,\frac{G_V}{\sqrt{2}} \langle V_{\mu\nu} u^\mu u^\nu\rangle+\frac{F_A}{2\sqrt{2}}\left\langle A_{\mu\nu}f_-^{\mu\nu}\right\rangle\,\,.
\end{eqnarray}
The last two operators on the first line involving pseudoscalar resonances do not play any role in our study \footnote{Although it may seem that the operator with coefficient $d_{m0}$ is suppressed 
with respect to the others in 
eq.~(\ref{linear_interactions}) because of its additional trace, this is not the case since it is enhanced due to $\eta^\prime$ exchange \cite{Kampf:2011ty}.} because they couple the pseudoscalar 
resonances to spin-zero sources 
instead of to the weak $V-A$ current.\\

Resonant operators contributing at $\mathcal{O}(p^6)$ in the chiral expansion (in the low-energy limit) were studied systematically in Refs.~\cite{Cirigliano:2006hb} and \cite{Kampf:2011ty} for the even- and odd-intrinsic parity sectors, 
respectively. We will be discussing those entering our study of $\tau^-\to\pi^-\eta^{(\prime)}\gamma\nu_\tau$ decays in the following.\\

We will consider first the even-intrinsic parity sector and start with the operators containing one resonance field. There, only one of the operators involving a scalar resonance matters to our analysis: 
$O_{15}^S=\langle Sf^{\mu\nu}_+{f_{+}}_{\mu\nu}\rangle$ \cite{Cirigliano:2006hb}, while again no operators including pseudoscalar resonances contribute (in either intrinsic parity sector).\\

The corresponding Lagrangian with one vector resonance field was derived in Ref.~\cite{Cirigliano:2006hb}:
\begin{equation}\label{Lagrangian_V_Towards}
 {\cal L}_{(4)}^{V} = \sum_{i=1}^{22} \, \lambda_i^{V} \, {\cal O}^V_i\,,
\end{equation}
with the operators
\begin{eqnarray}\label{Operators_V_Towards}
 {\cal O}^V_1\,=\, i \, \langle \, V_{\mu \nu} \,  u^{\mu} u_{\alpha} u^{\alpha} u^{\nu} \, \rangle & , & {\cal O}^V_2\,=\, i \, \langle \,  V_{\mu \nu} \,  u^{\alpha} u^{\mu} u^{\nu} u_{\alpha} \, \rangle\,,\nonumber\\
 {\cal O}^V_3\,=\, i \, \langle \, V_{\mu \nu} \, \{ \,  u^{\alpha} , u^{\mu} u_{\alpha} u^{\nu} \, \} \,\rangle & , & {\cal O}^V_4\,=\, i \, \langle \, V_{\mu \nu} \, \{ \, u^{\mu} u^{\nu},  u^{\alpha} u_{\alpha} \, \} \, \rangle\,,\nonumber\\
 {\cal O}^V_5\,=\, i \, \langle \, V_{\mu \nu} \, f_{-}^{\mu \alpha} \, f_{-}^{\nu \beta} \, \rangle \, g_{\alpha \beta} & , & {\cal O}^V_6\,=\, \langle \, V_{\mu \nu} \, \{ \, f_{+}^{\mu \nu} \, , \, \chi_{+} \, \} \, \rangle\,,\nonumber\\
 {\cal O}^V_7\,=\, i \, \langle \, V_{\mu \nu} \, f_{+}^{\mu \alpha} \, f_{+}^{\nu \beta}  \, \rangle \, g_{\alpha \beta} & , & {\cal O}^V_8\,=\, i \, \langle \, V_{\mu \nu} \, \{ \, \chi_{+} \, , \, u^{\mu} u^{\nu} \, \} \, \rangle\,,\nonumber\\
 {\cal O}^V_9\,=\, i \, \langle \, V_{\mu \nu} \, u^{\mu} \, \chi_{+} \, u^{\nu} \, \rangle & , & {\cal O}^V_{10}\,=\, \langle \, V_{\mu \nu} \, [ \, u^{\mu} \, , \, \nabla^{\nu} \chi_{-} \, ] \, \rangle\,,\nonumber\\
 {\cal O}^V_{11}\,=\, \langle \, V_{\mu \nu} \, \{ \, f_{+}^{\mu \nu} \, , \, u^{\alpha} u_{\alpha} \, \} \, \rangle & , & {\cal O}^V_{12}\,=\, \langle \, V_{\mu \nu} \, u_{\alpha} \,  f_{+}^{\mu \nu} \, u^{\alpha} \, \rangle\,,\nonumber\\
 {\cal O}^V_{13}\,=\, \langle \, V_{\mu \nu} \, ( \, u^{\mu} \, f_{+}^{\nu \alpha} \, u_{\alpha} \, + \, u_{\alpha} \, f_{+}^{\nu \alpha} \, u^{\mu} \, ) \, \rangle & ,
 & {\cal O}^V_{14}\,=\, \langle \, V_{\mu \nu} \, ( \, u^{\mu} u_{\alpha} \, f_{+}^{\alpha \nu} \, + \, f_{+}^{\alpha \nu} \, u_{\alpha} u^{\mu} \, ) \,\rangle\,,\nonumber\\
 {\cal O}^V_{15}\,=\, \langle \, V_{\mu \nu} \, ( \, u_{\alpha} u^{\mu} \, f_{+}^{\alpha \nu} \, + \, f_{+}^{\alpha \nu} \, u^{\mu} u_{\alpha} \, ) \, \rangle & ,
 & {\cal O}^V_{16}\,=\,i \, \langle \, V_{\mu \nu} \, [ \, \nabla^{\mu} f_{-}^{\nu \alpha} \, , \, u_{\alpha} \, ] \, \rangle \,,\nonumber\\
 {\cal O}^V_{17}\,=\, i \, \langle \, V_{\mu \nu} \, [ \, \nabla_{\alpha} f_{-}^{\mu \nu } \, , \, u^{\alpha} \, ] \, \rangle & , & {\cal O}^V_{18}\,=\, i \, \langle \, V_{\mu \nu} \, [ \, \nabla_{\alpha} f_{-}^{\alpha \mu} \, , \, u^{\nu} \, ] \, \rangle\,,\nonumber\\
 {\cal O}^V_{19}\,=\, i \, \langle \, V_{\mu \nu} \, [ \, f_{-}^{\mu \alpha} \, , \, h^{\nu}_{\alpha} \, ] \, \rangle & , & {\cal O}^V_{20}\,=\, \langle \, V_{\mu \nu} \, [ \, f_{-}^{\mu \nu} \, , \, \chi_{-} \, ] \, \rangle\,,\nonumber\\
 {\cal O}^V_{21}\,=\, i \, \left\langle \, V_{\mu\nu} \, \nabla_{\alpha} \nabla^{\alpha} \, \left( u^{\mu} \, u^{\nu} \right) \, \right\rangle & , & 
 {\cal O}^V_{22}\,=\, \langle \, V_{\mu \nu} \, \nabla_{\alpha} \nabla^{\alpha} \, f_{+}^{\mu \nu} \, \rangle\,.
\end{eqnarray}

Two-resonance operators which conserve intrinsic parity are discussed in the following. We begin with the basis of operators for vertices with one $V$ and one $A$ resonances and a pseudoscalar meson \cite{GomezDumm:2003ku} (here denoted $P$ in the 
operators indexes, like in the quoted reference) in the normal parity sector. This is
 \begin{equation}\label{VAP}
  \mathcal{L}^{VAP}
  =\sum_{i=1}^5\lambda^i\mathcal{O}^i_{VAP},
 \end{equation}
 where the operators are
\begin{eqnarray}\label{operatorsVAP}
  \mathcal{O}_{VAP}^1&=&\langle[V^{\mu\nu},A_{\mu\nu}]\chi_-\rangle,\nonumber\\
  \mathcal{O}_{VAP}^2&=&i\langle[V^{\mu\nu},A_{\nu\alpha}]h_\mu^{\hspace*{0.7ex}\alpha}\rangle,\nonumber\\
  \mathcal{O}_{VAP}^3&=&i\langle[\nabla^\mu V_{\mu\nu},A^{\nu\alpha}]u_\alpha\rangle,\nonumber\\
  \mathcal{O}_{VAP}^4&=&i\langle[\nabla^\alpha V_{\mu\nu},A_\alpha^{\hspace*{0.7ex}\nu}]u^\mu\rangle,\nonumber\\
  \mathcal{O}_{VAP}^5&=&i\langle[\nabla^\alpha V_{\mu\nu},A^{\mu\nu}]u_\alpha\rangle.
\end{eqnarray}

There is only one relevant operator with both a $V$ and a $S$ field, $O_3^{SV}=\left\langle \, \{ \, S \, , \, V_{\mu \nu} \, \} \, f_{+}^{\mu \nu} \, \right\rangle $, with coupling $\lambda_3^{SV}$ \cite{Cirigliano:2006hb}.\\

Finally, we include the relevant operators with two $V$ resonances in this even-intrinsic parity sector \cite{Cirigliano:2006hb}
\begin{equation}
 \mathcal{L}^{VV}\,=\,\sum_{i=1}^{\;\;\;\;18} \lambda_i^{VV} \mathcal{O}_i^{VV}\,,
\end{equation}
where
\begin{eqnarray}
 O_1^{VV}&=&\langle V_{\mu\nu}V^{\mu\nu}u^\alpha u_\alpha\rangle\,,\nonumber\\
 O_2^{VV}&=&\langle V_{\mu\nu} u^\alpha V^{\mu\nu} u_\alpha\rangle\,,\nonumber\\
 O_3^{VV}&=&\langle V_{\mu\alpha}V^{\nu\alpha}u^\mu u_\nu\rangle\,,\nonumber\\
 O_4^{VV}&=&\langle V_{\mu\alpha}V^{\nu\alpha}u_\nu u^\mu\rangle\,,\nonumber\\
 O_5^{VV}&=&\langle V_{\mu\alpha}(u^\alpha V^{\mu\beta}u_\beta+u_\beta V^{\mu\beta}u^\alpha)\rangle\,,\nonumber\\
 O_6^{VV}&=&\langle V_{\mu\nu}V^{\mu\nu}\chi_+\rangle\,,\nonumber\\
 O_7^{VV}&=&i\langle V_{\mu\alpha}V^{\alpha\nu}f_{+\beta\nu}\rangle g^{\beta\mu}\,.
\end{eqnarray}

Next we turn to the odd-intrinsic parity sector, where the two terms involving a scalar and an axial-vector resonance \cite{Kampf:2011ty} are
\begin{equation}
 O_1^{SA}\,=\,i\epsilon_{\mu\nu\alpha\beta}\left\langle \left[A^{\mu\nu}, S\right]f_+^{\alpha\beta}\right\rangle\,,\quad O_2^{SA}\,=\,\epsilon_{\mu\nu\alpha\beta}\left\langle A^{\mu\nu} \left[S,u^\alpha u^\beta\right]\right\rangle\,. 
\end{equation}

In this intrinsic parity sector, operators with only vector resonances and sources and at most one pseudoscalar (again denoted $P$ in the naming of the operators) were derived in reference 
\cite{RuizFemenia:2003hm}
 \begin{equation}\label{VJP&VVP}
 \mathcal{L}^{V,odd}
 =\sum_{a=1}^7\frac{c_a}{M_V}\mathcal{O}^a_{VJP}+\sum_{a=1}^4d_a\mathcal{O}^a_{VVP},
 \end{equation}
 where the operators are 
 \begin{eqnarray}\label{operatorsVJP&VVP}
  \mathcal{O}_{VJP}^1&=&\varepsilon_{\mu\nu\rho\sigma}\langle\{V^{\mu\nu},f^{\rho\alpha}_+\}\nabla_\alpha u^\sigma\rangle\,,\nonumber\\
  \mathcal{O}_{VJP}^2&=&\varepsilon_{\mu\nu\rho\sigma}\langle\{V^{\mu\alpha},f^{\rho\sigma}_+\}\nabla_\alpha u^\nu\rangle\,,\nonumber\\
  \mathcal{O}_{VJP}^3&=&i\varepsilon_{\mu\nu\rho\sigma}\langle\{V^{\mu\nu},f^{\rho\sigma}_+\}\chi_-\rangle\,,\nonumber\\
  \mathcal{O}_{VJP}^4&=&i\varepsilon_{\mu\nu\rho\sigma}\langle V^{\mu\nu}[f^{\rho\sigma}_-,\chi_+]\rangle\,,\nonumber\\
  \mathcal{O}_{VJP}^5&=&\varepsilon_{\mu\nu\rho\sigma}\langle\{\nabla_\alpha V^{\mu\nu},f^{\rho\alpha}_+\}u^\sigma\rangle\,,\nonumber\\
  \mathcal{O}_{VJP}^6&=&\varepsilon_{\mu\nu\rho\sigma}\langle\{\nabla_\alpha V^{\mu\alpha},f^{\rho\sigma}_+\}u^\nu\rangle\,,\nonumber\\
  \mathcal{O}_{VJP}^7&=&\varepsilon_{\mu\nu\rho\sigma}\langle\{\nabla^\sigma V^{\mu\nu},f^{\rho\alpha}_+\}u_\alpha\rangle\,;\\ \nonumber\\
  \mathcal{O}_{VVP}^1&=&\varepsilon_{\mu\nu\rho\sigma}\langle\{V^{\mu\nu},V^{\rho\alpha}\}\nabla_\alpha u^\sigma\rangle\,,\nonumber\\
  \mathcal{O}_{VVP}^2&=&i\varepsilon_{\mu\nu\rho\sigma}\langle\{V^{\mu\nu},V^{\rho\sigma}\}\chi_-\rangle\,,\nonumber\\
  \mathcal{O}_{VVP}^3&=&\varepsilon_{\mu\nu\rho\sigma}\langle\{\nabla_\alpha V^{\mu\nu},V^{\rho\alpha}\}u^\sigma\rangle\,,\nonumber\\
  \mathcal{O}_{VVP}^4&=&\varepsilon_{\mu\nu\rho\sigma}\langle\{\nabla^\sigma V^{\mu\nu},V^{\rho\alpha}\}u_\alpha\rangle\,.
\end{eqnarray}

In our case, however, we will not only need odd-intrinsic parity couplings of a $V$ resonance, a $J$ source and a pseudoGoldstone; but also such vertices with two pseudoscalars \footnote{Obviously, in 
this case $J$ has opposite parity than in the 
case with one pseudoGoldstone since both vertices are of odd-intrinsic parity.}. In this case, as warned in Ref.~\cite{RuizFemenia:2003hm}, the set $\left\lbrace \mathcal{O}^a_{VJP}\right\rbrace_{a=1}^{\;\;\;\;7}$ is no longer a basis 
\footnote{Analogous comment applies to eq.~(\ref{VAP}), as pointed out in Ref~\cite{GomezDumm:2003ku}.} and one needs to use the operator basis with a $V$ resonance derived in Ref.~\cite{Kampf:2011ty}; 
i. e.
\begin{equation}\label{V_KN}
\widetilde{\mathcal{L}^{V,odd}}
    =\varepsilon^{\mu\nu\alpha\beta}\sum_i\kappa_i^V{\mathcal{O}_i^V}_{\mu\nu\alpha\beta},
  \end{equation}
with the operators
 \begin{eqnarray}\label{operatorsV_KN}
&& ({\mathcal{O}^{V}_1})^{\mu\nu\alpha\beta}=i\langle V^{\mu\nu}(h^{\alpha\sigma}u_\sigma u^\beta-u^\beta u_\sigma h^{\alpha\sigma}) \rangle\,,\nonumber\\
&& ({\mathcal{O}^{V}_2})^{\mu\nu\alpha\beta}=i\langle V^{\mu\nu}(u_\sigma h^{\alpha\sigma} u^\beta-u^\beta h^{\alpha\sigma}u_\sigma)\rangle\,,\nonumber\\
&& ({\mathcal{O}^{V}_3})^{\mu\nu\alpha\beta}=i\langle V^{\mu\nu}(u_\sigma u^\beta h^{\alpha\sigma}-h^{\alpha\sigma}u^\beta u_\sigma)\rangle\,,\nonumber\\
&& ({\mathcal{O}^{V}_4})^{\mu\nu\alpha\beta}=i\langle [V^{\mu\nu},\nabla^\alpha\chi_+]u^\beta\rangle\,,\nonumber\\
&& ({\mathcal{O}^{V}_5)}^{\mu\nu\alpha\beta}=i\langle V^{\mu\nu}[f_-^{\alpha\beta},u_\sigma u^\sigma]\rangle\,,\nonumber\\
&& ({\mathcal{O}^{V}_6})^{\mu\nu\alpha\beta}=i\langle V^{\mu\nu}(f_-^{\alpha\sigma}u^\beta u_\sigma-u_\sigma u^\beta f_-^{\alpha\sigma})\rangle\,,\nonumber\\
&& ({\mathcal{O}^{V}_7})^{\mu\nu\alpha\beta}=i\langle V^{\mu\nu}(u_\sigma f_-^{\alpha\sigma}u^\beta-u^\beta f_-^{\alpha\sigma}u_\sigma)\rangle\,,\nonumber\\
&& ({\mathcal{O}^{V}_8})^{\mu\nu\alpha\beta}=i\langle V^{\mu\nu}(f_-^{\alpha\sigma}u_\sigma u^\beta-u^\beta u_\sigma f_-^{\alpha\sigma})\rangle\,,\nonumber\\
&& ({\mathcal{O}^{V}_9})^{\mu\nu\alpha\beta}=\langle V^{\mu\nu}\lbrace\chi_-,u^\alpha u^\beta\rbrace\rangle\,,\nonumber\\
&& ({\mathcal{O}^{V}_{10}})^{\mu\nu\alpha\beta}=\langle V^{\mu\nu} u^\alpha \chi_- u^\beta\rangle\,,\nonumber\\
&& ({\mathcal{O}^{V}_{11}})^{\mu\nu\alpha\beta}=\langle V^{\mu\nu} \lbrace f_+^{\alpha\rho},f_-^{\beta\sigma}\rbrace\rangle g_{\rho\sigma}\,,\nonumber\\
&& ({\mathcal{O}^{V}_{12}})^{\mu\nu\alpha\beta}=\langle V^{\mu\nu} \lbrace f_+^{\alpha\rho},h^{\beta\sigma}\rbrace\rangle g_{\rho\sigma}\,,\nonumber\\
&& ({\mathcal{O}^{V}_{13}})^{\mu\nu\alpha\beta}=i\langle V^{\mu\nu} f_+^{\alpha\beta}\rangle\langle\chi_-\rangle\,,\nonumber\\
&& ({\mathcal{O}^{V}_{14}})^{\mu\nu\alpha\beta}=i\langle V^{\mu\nu} \lbrace f_+^{\alpha\beta},\chi_-\rbrace\rangle\,,\nonumber\\
&& ({\mathcal{O}^{V}_{15}})^{\mu\nu\alpha\beta}=i\langle V^{\mu\nu} [f_-^{\alpha\beta},\chi_+]\rangle\,,\nonumber\\
&& ({\mathcal{O}^{V}_{16}})^{\mu\nu\alpha\beta}=\langle V^{\mu\nu} \lbrace \nabla^\alpha f_+^{\beta\sigma},u_\sigma\rbrace\rangle\,,\nonumber\\
&& ({\mathcal{O}^{V}_{17}})^{\mu\nu\alpha\beta}=\langle V^{\mu\nu} \lbrace \nabla_\sigma f_+^{\alpha\sigma},u^\beta\rbrace\rangle\,,\nonumber\\
&& ({\mathcal{O}^{V}_{18}})^{\mu\nu\alpha\beta}=\langle V^{\mu\nu} u^\alpha u^\beta\rangle\langle\chi_-\rangle\,.
  \end{eqnarray}

The operators in eq.~(\ref{operatorsVJP&VVP}) can be written in terms of those in eq.~(\ref{operatorsV_KN}). This yields the following identities among the corresponding couplings \cite{Roig:2013baa}

\begin{eqnarray}\label{eq: relation different basis}
& & \kappa_1^{VV}= \frac{-d_1}{8 n_f}\,,\qquad \kappa_2^{VV}
= \frac{d_1}{8} + d_2\,,\qquad \kappa_3^{VV}=d_3\,,\qquad
\kappa_4^{VV}=d_4\,,
\nonumber\\
& &
-2 M_V \kappa_5^V \, = \,  M_V \kappa_6^V \, =\, M_V \kappa_7^V = \frac{c_6}{2}\,,
\qquad \qquad
M_V \kappa_{11}^V = \frac{c_1 - c_2 - c_5 + c_6 + c_7}{2}\,,
\nonumber\\
& & M_V \kappa_{12}^V = \frac{c_1 - c_2 - c_5 + c_6 - c_7}{2}\,,
\qquad n_f M_V \kappa_{13}^V = \frac{- c_2 + c_6}{4}\,,
\qquad M_V \kappa_{14}^V = \frac{c_2 + 4 c_3 - c_6}{4}\,,\nonumber\\
& &
 M_V \kappa_{15}^V = c_4\,,
\qquad
M_V \kappa_{16}^V = c_6 + c_7\,,
\qquad
M_V \kappa_{17}^V = -c_5 + c_6\,.
\end{eqnarray}

The analogous Lagrangian to eq.~(\ref{V_KN}) involving an $A$ resonance \cite{Kampf:2011ty} is the last missing piece needed for our computations. This is
\begin{equation}\label{A_KN}
\mathcal{L}^{A,odd}
    =\varepsilon^{\mu\nu\alpha\beta}\sum_i\kappa_i^A{\mathcal{O}_i^A}_{\mu\nu\alpha\beta},
  \end{equation}
with the operators
 \begin{eqnarray}\label{operatorsA_KN}
&& ({\mathcal{O}^{A}_1})^{\mu\nu\alpha\beta}=\langle A^{\mu\nu}[u^\alpha u^\beta,u_\sigma u^\sigma]\rangle,\nonumber\\
&& ({\mathcal{O}^{A}_2})^{\mu\nu\alpha\beta}=\langle A^{\mu\nu}[u^\alpha u^\sigma u^\beta, u_\sigma]\rangle,\nonumber\\
&& ({\mathcal{O}^{A}_3})^{\mu\nu\alpha\beta}=\langle A^{\mu\nu}\{\nabla^\alpha h^{\beta\sigma},u_\sigma\}\rangle,\nonumber\\
&& ({\mathcal{O}^{A}_4})^{\mu\nu\alpha\beta}=i\langle A^{\mu\nu}[f_+^{\alpha\beta},u^\sigma u_\sigma]\rangle,\nonumber\\
&& ({\mathcal{O}^{A}_5)}^{\mu\nu\alpha\beta}=i\langle A^{\mu\nu}(f_+^{\alpha\sigma}u_\sigma u^\beta-u^\beta u_\sigma f_+^{\alpha\sigma})\rangle,\nonumber\\
&& ({\mathcal{O}^{A}_6})^{\mu\nu\alpha\beta}=i\langle A^{\mu\nu}(f_+^{\alpha\sigma} u^\beta u_\sigma-u_\sigma u^\beta f_+^{\alpha\sigma})\rangle,\nonumber\\
&& ({\mathcal{O}^{A}_7})^{\mu\nu\alpha\beta}=i\langle A^{\mu\nu}(u_\sigma f_+^{\alpha\sigma} u^\beta -u^\beta f_+^{\alpha\sigma}u_\sigma)\rangle,\nonumber\\
&& ({\mathcal{O}^{A}_8})^{\mu\nu\alpha\beta}=\langle A^{\mu\nu}\{f_-^{\alpha\sigma},h^{\beta}_{\sigma}\}\rangle,\nonumber\\
&& ({\mathcal{O}^{A}_9})^{\mu\nu\alpha\beta}=i\langle A^{\mu\nu}f_-^{\alpha\beta}\rangle\langle \chi_-\rangle,\nonumber\\
&& ({\mathcal{O}^{A}_{10}})^{\mu\nu\alpha\beta}=i\langle A^{\mu\nu} u^{\alpha}\rangle \langle\nabla^\beta\chi_-\rangle ,\nonumber\\
&& ({\mathcal{O}^{A}_{11}})^{\mu\nu\alpha\beta}=i\langle A^{\mu\nu}\{f_-^{\alpha\beta},\chi_-\}\rangle,\nonumber\\
&& ({\mathcal{O}^{A}_{12}})^{\mu\nu\alpha\beta}=i\langle A^{\mu\nu}\{\nabla^{\alpha}\chi_-,u^{\beta}\}\rangle,\nonumber\\
&& ({\mathcal{O}^{A}_{13}})^{\mu\nu\alpha\beta}=\langle A^{\mu\nu}[\chi_+,u^{\alpha}u^{\beta}]\rangle,\nonumber\\
&& ({\mathcal{O}^{A}_{14}})^{\mu\nu\alpha\beta}=i\langle A^{\mu\nu}\{f_+^{\alpha\beta},\chi_+\}\rangle,\nonumber\\
&& ({\mathcal{O}^{A}_{15}})^{\mu\nu\alpha\beta}=\langle A^{\mu\nu}\{\nabla^{\alpha}f_-^{\beta\sigma},u_{\sigma}\}\rangle,\nonumber\\
&& ({\mathcal{O}^{A}_{16}})^{\mu\nu\alpha\beta}=\langle A^{\mu\nu}\{\nabla_{\sigma}f_-^{\alpha\sigma},u^{\beta}\}\rangle.
  \end{eqnarray}
  
 We recall that the basis for odd-intrinsic parity operators with two vector resonances and a pseudoscalar meson was given in eq.~(\ref{VJP&VVP}).\\
 
  \subsection{Short-distance QCD constraints on the $R\chi L$ couplings}\label{sec:QCDconstraints}
  We have discussed in the previous section how symmetry determines the structure of the operators in the $R\chi L$ though it leaves, however, the corresponding couplings undetermined (as in $\chi PT$ or any other effective field 
  theory with a corresponding fundamental theory in the strongly coupled regime). It was soon 
  observed \cite{Weinberg:1967kj, RChL} that demanding that the Green functions (and related form factors) computed in the meson theory to match their known asymptotic behaviour according to the operator product expansion \cite{Wilson:1969zs} 
  of QCD relates some of the $R\chi L$ couplings and thus increases the predictive power of the theory. We will quote in the following the results of this programme interesting to our study.\\
  
  In the odd-intrinsic parity sector, the analysis of three-point $VVP$ Green function and associated form factors yields \cite{RuizFemenia:2003hm, Kampf:2011ty, Roig:2013baa}
  \begin{eqnarray}
M_V( 2 \kappa_{12}^V + 4 \kappa_{14}^V +\kappa_{16}^V -\kappa_{17}^V) \quad =&
4 \,c_3 \,+\, c_1 & = \quad  0\,,
\nonumber
\\[3mm]
M_V (2\kappa_{12}^V + \kappa_{16}^V -2\kappa_{17}^V) \quad =&
c_1 \, - \, c_2 \, + \, c_5 \, & = \quad  0 \, ,
\nonumber
\\[3mm]
-\, M_V \kappa_{17}^V \quad = &  c_5 \, - \, c_6 \, & =
\quad  \frac{N_C \,M_V}{64 \,\sqrt{2}\, \pi^2\,F_V} \, ,
\nonumber
\\[3mm]
8\kappa_2^{VV} \quad =&  d_1 \, + \, 8 \,d_2 & = \quad
\frac{F^2}{8\,F_V^2} - \frac{N_C \,M_V^2}{64\, \pi^2\, F_V^2}\,,
\nonumber
\\[3mm]
\kappa_3^{VV}\quad = & d_3& =\quad  -\, \frac{N_C}{64\pi^2} \frac{M_V^2}{F_V^2} \,,
\nonumber
\\[3mm]
& 1 \, + \, \frac{32 \,\sqrt{2} \, F_V \,d_m\, \kappa_3^{PV}}{F^2} & = \quad  0\,,
\nonumber
\\[3mm]
& F_V^2 & = \quad  3\, F^2\,.
\label{eq: Consistent set of relations}
\end{eqnarray}
It is remarkable that the last of eqs.~(\ref{eq: Consistent set of relations}) involves couplings belonging to the even-intrinsic parity $R\chi L$, despite it was obtained demanding consistency to the high-energy constraints derived in the 
odd-intrinsic parity sector \cite{RuizFemenia:2003hm, Kampf:2011ty, Roig:2013baa, Guo:2008sh, Dumm:2009kj, Guo:2010dv, Guevara:2013wwa}. Let us also mention that the short-distance QCD constraint $\kappa_2^S\,=\,0$ \cite{Kampf:2011ty} forbids a diagram similar to the third one in fig.~\ref{fig:1R-FV} where this time the coupling to the 
current would conserve intrinsic parity (it would be thus a contribution to the axial-vector form factors, since $a_0^-\to\pi^-\eta$ belongs to the unnatural intrinsic parity sector)~\footnote{For completeness we quote the corresponding 
operator, $O_2^S\,=\,\epsilon_{\mu\nu\alpha\beta}\langle i S\left[f_+^{\mu\nu},f_-^{\alpha\beta}\right]\rangle$.}. Another relevant short-distance constraint in the odd-intrinsic parity sector which is derived from the study of the 
$VAS$ Green function \cite{Kampf:2011ty} is $\kappa^{14}_A=0$. Interestingly, this same analysis also yields the relation $\kappa^V_4=2\kappa_{15}^V$, where $\kappa^V_4$ does not enter the relations (\ref{eq: relation different 
basis}). Other high-energy constraints derived in the quoted study are not relevant to our computation.\\

 In the even-intrinsic parity sector, the study of $VAP$ and $SPP$ Green functions~\footnote{Four-point functions have been studied in Ref.~\cite{Ananthanarayan:2004qk}.} and their form factors allowed to derive the 
 following restrictions \cite{Cirigliano:2004ue, Cirigliano:2005xn, Cirigliano:2006hb}
 \begin{eqnarray}\label{Relations_even}
  \lambda'&\equiv&\frac{1}{\sqrt{2}}\left(\lambda_2-\lambda_3+\frac{\lambda_4}{2}+\lambda_5\right)=\frac{F^2}{2\sqrt{2}F_AG_V}\,,\nonumber\\
  \lambda''&\equiv&\frac{1}{\sqrt{2}}\left(\lambda_2-\frac{\lambda_4}{2}-\lambda_5\right)=\frac{2G_V-F_V}{2\sqrt{2}F_A}\,,\nonumber\\
  \lambda_0&\equiv&-\frac{1}{\sqrt{2}}\left(4\lambda_1+\lambda_2+\frac{\lambda_4}{2}+\lambda_5\right)=\frac{\lambda'+\lambda''}{4}\,,\nonumber\\
  \kappa_{1}^{SA}&\equiv&\frac{F^2}{32\sqrt{2}c_mF_A},\quad
 \end{eqnarray}
supplemented by $ F_V G_V=F^2\,,\; F_A=\sqrt{2}F$ and $F_V=\sqrt{3}F$ (this one in accord with the result found in the odd-intrinsic parity sector)~\cite{Weinberg:1967kj, RChL, Peris}. Since $\lambda^V_{21}=0=\lambda^V_{22}$ 
\cite{Cirigliano:2006hb}, we will not consider the contribution of the corresponding operators. The well-known relation $c_d c_m\,=\,F^2/4$ \cite{JOP} arising in the study of strangeness-changing scalar form factors will 
also be employed.\\
  
  Although not all the operators appearing in section \ref{sec:theoRChT} do actually contribute to the considered decays, the number of asymptotic relations looks too small compared to the number of free couplings to allow a meaningful general 
  phenomenological study of the $\tau^-\to\pi^-\eta^{(\prime)}\gamma \nu_\tau$ decays within $R\chi L$. Also there is not enough phenomenological information on the couplings of eqs.~(\ref{operatorsV_KN}) and (\ref{operatorsA_KN}), for instance. 
  Due to that we will first consider only the diagrams with at most one resonance and then comment on the possible extension to include two-resonance diagrams in section \ref{sec:radbkgRChT}.\\
  
  \subsection{Form factors according to Resonance Chiral Lagrangians}\label{sec:FFsRChT}
\begin{figure}[ht!]\centering
  \includegraphics[scale=0.8]{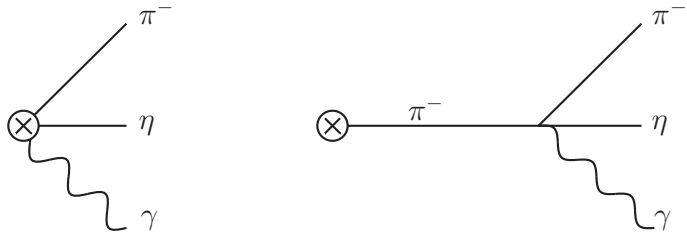}\caption{Contributions from the Wess-Zumino-Witten functional \cite{ChiralAnomaly} to $\tau^-\to\pi^-\eta\gamma\nu_\tau$ decays. The cross circle indicates the 
  insertion of the charged weak current.}\label{fig:ChPT}
 \end{figure}

\begin{figure}[ht!]
  \centering
  \includegraphics[scale=0.5]{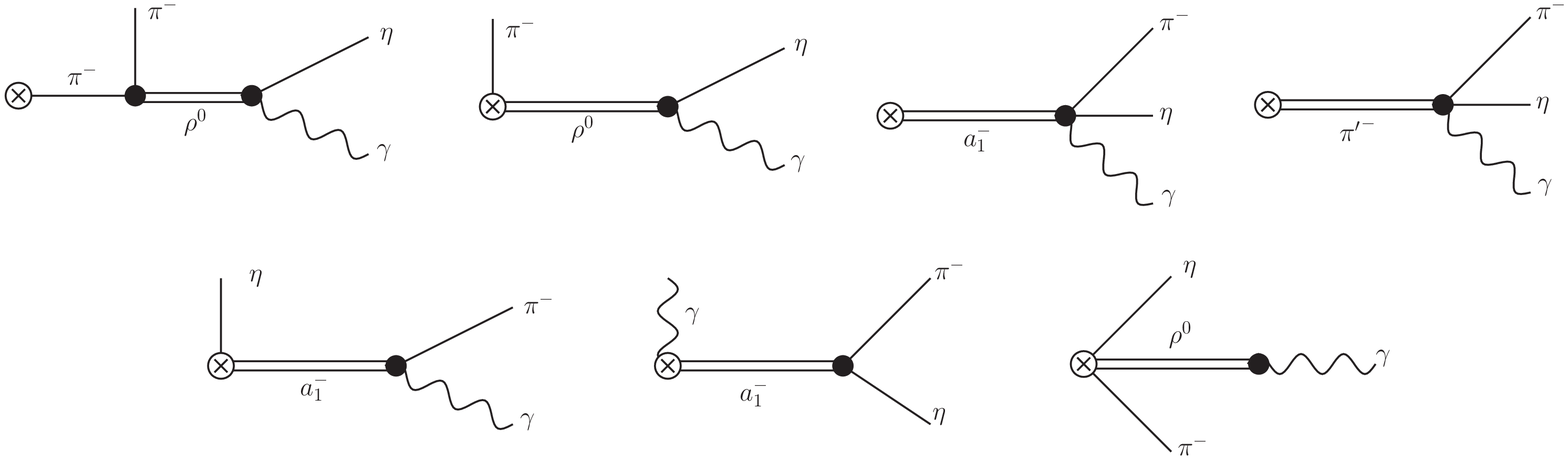}\caption{One-resonance exchange contributions from the $R\chi L$ to the axial-vector form factors of the $\tau^-\to\pi^-\eta\gamma\nu_\tau$ decays. Vertices involving 
  resonances are highlighted with a thick dot.}\label{fig:1R-FA}
 \end{figure}

\begin{figure}[ht!]
  \centering
  \includegraphics[scale=0.5]{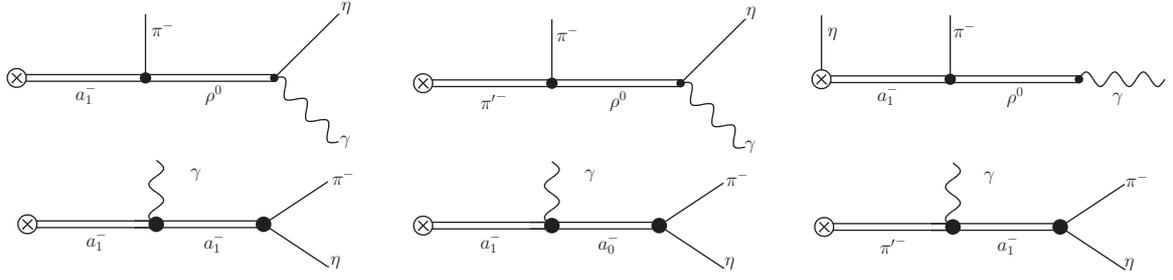}\caption{Two-resonance exchange contributions from the $R\chi L$ to the axial-vector form factors of the $\tau^-\to\pi^-\eta\gamma\nu_\tau$ decays. Vertices involving 
  resonances are highlighted with a thick dot.}\label{fig:2R-FA}
 \end{figure}

\begin{figure}[ht!]
  \centering
  \includegraphics[scale=0.45]{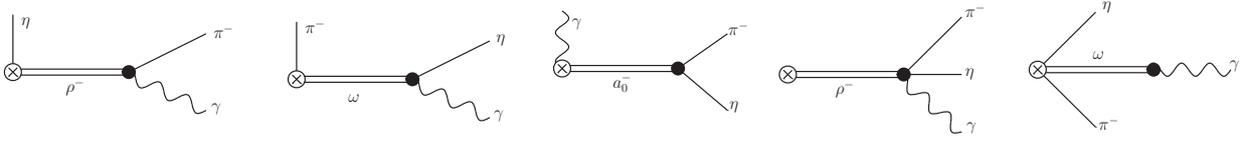}\caption{One-resonance exchange contributions from the $R\chi L$ to the vector form factors of the $\tau^-\to\pi^-\eta\gamma\nu_\tau$ decays. Vertices involving 
  resonances are highlighted with a thick dot.}\label{fig:1R-FV}
 \end{figure}

\begin{figure}[ht!]
  \centering
  \includegraphics[scale=0.35]{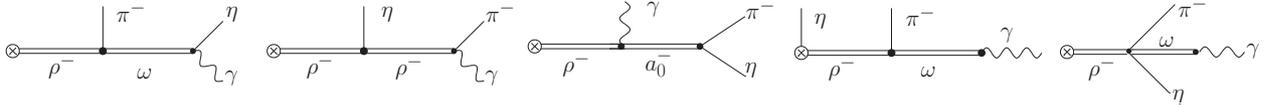}\caption{Two-resonance exchange contributions from the $R\chi L$ to the vector form factors of the $\tau^-\to\pi^-\eta\gamma\nu_\tau$ decays. Vertices involving 
  resonances are highlighted with a thick dot.}\label{fig:2R-FV}
 \end{figure}

  The relevant Feynman diagrams are shown in figures \ref{fig:ChPT} to \ref{fig:2R-FA}~\footnote{We remind that only diagrams which do not violate G-parity are considered.}. Fig.~\ref{fig:ChPT} corresponds to the model-independent 
  contribution given by the chiral $U(1)$ anomaly, fixed by QCD~\footnote{We note that this contribution is absent in the MDM approach.}. The left-hand side diagram is the purely local contribution while, in the one on the right, 
  the Wess-Zumino-Witten functional provides the $\pi\pi\eta\gamma$ vertex (and all hadronic information corresponding to the coupling of the pion to the axial-vector current is encoded in the pion decay constant). The anomalous 
  vertices violate intrinsic parity, as these two diagrams do. Figs.~\ref{fig:1R-FA} to \ref{fig:2R-FV} are, on the contrary, model-dependent. Figs.~\ref{fig:1R-FA} and \ref{fig:2R-FA} (\ref{fig:1R-FV} and \ref{fig:2R-FV}) correspond 
  to the one- and two-resonance mediated contributions to the axial-vector (vector) form factors in eqs.~(\ref{m.e.}) to (\ref{decomp}), respectively.\\

\vspace*{1.5cm}
  
   As a general fact, the axial-vector form factors in radiative tau decays to two pseudoscalars violate intrinsic parity as it can be checked for all contributing diagrams in figs.~\ref{fig:1R-FA} and \ref{fig:2R-FA}. The last vertex 
   in all diagrams in the first line of fig.~\ref{fig:1R-FA} is of odd-intrinsic parity (as well as it happens with the second diagram in the second line of this figure). In the first and third diagrams of the second line of 
   fig.~\ref{fig:1R-FA} intrinsic parity is violated in the coupling to the weak (thus axial-vector) current. The odd-intrinsic parity violating vertices appearing in the diagrams in fig.~\ref{fig:2R-FA} are $\rho^0\to\eta\gamma$, 
   $a^{-\mu}\to a_1^-\eta$ ($a^\mu$ stands for the axial-vector current), $a_1^-\to\pi^-\eta$ and $a_1^-\to a_0^-\gamma$.\\

\vspace*{1.5cm}

 \vspace*{1.5cm}
     
  We note that the first two diagrams of figs.~\ref{fig:1R-FV} contain only odd-intrinsic parity violating vertices while the last three diagrams in this figure contain only even-intrinsic 
  parity vertices in such a way that intrinsic-parity is not violated in neither of them (as it corresponds to the vector form factors). Similarly, in fig.~\ref{fig:2R-FV}, the first, second and fourth diagram contain two 
  intrinsic parity violating vertices and the third and fifth diagram contain only even-intrinsic parity vertices. Thus, again intrinsic parity is conserved in these diagrams as well.\\
     
\vspace*{1.5cm}

 \vspace*{1.5cm}

%
%

 Using the $R\chi L$ introduced in section \ref{sec:theoRChT}, it is straightforward to verify that all three diagrams involving the $\pi'$ resonance vanish (in figs.~\ref{fig:1R-FA} and \ref{fig:2R-FA}). Also the last 
 diagram of fig.~\ref{fig:1R-FV} is null but all other diagrams in figures \ref{fig:ChPT} to \ref{fig:2R-FV} contribute nontrivially to the considered $\tau^-\to\pi^-\eta^{(\prime)}\gamma\nu_\tau$ decays. Since the left-handed 
 weak current has both vector and axial-vector components, one could expect to have two different contributions per given topology, with intrinsic parity conserving and violating coupling to the weak charged current, 
 respectively. However, we point out that using the Lagrangian introduced in section \ref{sec:theoRChT} this only happens for the last diagrams in figs.~\ref{fig:1R-FA} and \ref{fig:1R-FV}. In our computation we have 
 neglected subleading contributions in the chiral counting, namely the coupling to the weak current in the second diagram of fig.~\ref{fig:1R-FA} receives contributions from the piece of the  Lagrangian in eq.~(\ref{linear_interactions}). 
 Correspondingly, we are not considering the contributions given by the Lagrangian in eq.~(\ref{Lagrangian_V_Towards}), which are suppressed by one chiral order.\\
 
 Comparing the $R\chi L$ diagrams in figs.~\ref{fig:ChPT} to \ref{fig:2R-FV} with the MDM diagrams in fig.~\ref{fig:hadvertex}, we see first that the model-independent contribution of both diagrams in fig.~\ref{fig:ChPT} (axial-form factors at 
 lowest order in the chiral expansion) is not included in the MDM approach. Among the 13 contributions in figs.~\ref{fig:1R-FA} and ~\ref{fig:2R-FA} (which are subleading in the chiral regime) only one is considered in MDM 
 \footnote{The diagram with the pion pole also appears in fig.~\ref{fig:vertex}, but it is neglected.} (the first diagram in figure \ref{fig:2R-FA}). Finally, 10 diagrams appear in figs.~\ref{fig:1R-FV} and \ref{fig:2R-FV} but only 
 three of them (those including the vertices $\rho-\omega-\pi$, $\rho-a_0-\gamma$ and $\rho-\rho-\eta$) enter the MDM description.\\
 
 We would like to make a final comment regarding gauge invariance before quoting our form factor results using $R\chi L$. It can be checked that the contribution of $\mathcal{O}_{10}^A$ to the third diagram in fig.~\ref{fig:1R-FA} is not gauge 
invariant by itself. However, for this particular operator, the cancellation of gauge-dependent pieces involves the diagrams with radiation off the $a_1$ and off the weak vertex in figs.~\ref{fig:1R-FA} and \ref{fig:2R-FA}. As a result of this 
mechanism, we note the presence of $D_{a_1}(W^2)$ and $D_{a_1}[(p+k)^2]$ factors and the absence of $D_{a_1}[(p+p_0)^2]$ terms in the corresponding contributions to the axial-vector form factors~\footnote{We note that, among the $\mathcal{O}_i^A$ 
operators, only $\mathcal{O}_{10}^A$ couples to $\pi^-\eta^{(\prime)}$. This vertex does not contribute to the corresponding non-radiative decays because at least an additional independent momentum is needed for a non-vanishing contraction with 
the Levi-Civita symbol.}.\\

For convenience, we will quote the individual contributions to each form factor figure by figure (following the order of the diagrams in a given figure). We will start with the axial-vector form factors. The diagrams in fig.~\ref{fig:ChPT} give 
\begin{equation}
 a_1^{\chi PT}\,=\,\frac{N_C C_q}{6 \sqrt{2} \pi^2 F^2}\,,\quad a_3^{\chi PT}\,=\,\frac{a_1^{\chi PT}}{D_\pi\left[W^2\right]}\,,\label{FFsRChT_ChPT}
\end{equation}
which is a model-independent result coming from the QCD anomaly.\\

The contribution of the remaining diagrams (figures \ref{fig:1R-FA} and \ref{fig:2R-FA} for the axial-vector form factors and \ref{fig:1R-FV} and \ref{fig:2R-FV} for the vector form factors) is collected in appendix B. The corresponding off-shell 
width of meson resonances used in our numerical analysis can be found in appendix C. We will discuss in the next section if further insight can be gained on the $R\chi L$ couplings values restoring to phenomenology and using the expected scaling 
of the low-energy constants of the $\chi PT$ Lagrangian.\\
%

\subsection{Phenomenological estimation of $R\chi L$ couplings}
Although the relations in section \ref{sec:QCDconstraints} only reduce the number of unknowns in eqs.~(\ref{FFsRChT_ChPT}) and (\ref{a1-1R-RChL}) to (\ref{v4-2R-RChL}), some of the remaining free couplings can still be estimated phenomenologically. The 
high-energy constraint $c_dc_m=F^2/4$ leaves either $c_d$ or $c_m$ as independent. We will use $c_d = \left(19.8^{+2.0}_{-5.2}\right)$ MeV \cite{Guo:2012yt}. In this way all relevant couplings in eq.(\ref{linear_interactions}) have been determined.\\
$\lambda_{15}^S$ is the only leading operator contributing to $a_0\to\gamma\gamma$. From $\Gamma(a_0\to\gamma\gamma)=(0.30\pm0.10)$ keV$=\frac{64\pi\alpha^2}{9}M_{a_0}^3|\lambda_{15}^S|^2$ we can estimate $|\lambda_{15}^S|=(1.6\pm0.3)\cdot 10^{-2}$ 
GeV$^{-1}$. We note that the coupling relevant for the $a_1-a_0-\gamma$ vertex, $\kappa_1^{SA}$ is fixed by a short-distance constraint in eqs.~(\ref{Relations_even}).\\

We turn now to the $\lambda_i$ couplings in eq.~(\ref{VAP}). Short-distance constraints leave two such couplings undetermined. The three combinations of them that are predicted by high-energy conditions have the following numerical values:
\begin{equation}\label{values_lambdasVAP}
 \lambda^\prime\,\sim\,0.4\,,\quad \lambda^{\prime\prime}\,\sim\,0.04\,,\quad \lambda_0\,\sim\,0.12\,.
\end{equation}
The same linear combination of $\lambda_4$ and $\lambda_5$ enters all couplings in eq.~(\ref{values_lambdasVAP}). Therefore we can take one them as independent ($\lambda_4$ for us). We will choose as the other independent coupling $\lambda_2$, which 
enters all couplings in eq.~(\ref{values_lambdasVAP}). A conservative estimate would be $|\lambda_2|\sim|\lambda_4|\leq0.4$, to which we will stick in our numerical analysis.\\

According to ref.~\cite{Cirigliano:2006hb} the $\lambda_i^V$ couplings can be estimated from the expected scaling of the NNLO low-energy constants of the $\chi PT$ Lagrangian (we also employ short-distance QCD constraints on the $R\chi L$ couplings 
to write the following expression conveniently) as
\begin{equation}\label{estimate_lambda_i^V}
 \lambda_i^V\,\sim\,3C_i^R\frac{M_V^2}{F}\sim 0.05\; \mathrm{GeV}^{-1}\,,
\end{equation}
that can be considered an upper bound on $|\lambda_i^V|$ because the employed relation $C_i^R\sim\frac{1}{F^2(4\pi)^4}$ is linked to $L_i^R\sim\frac{1}{(4\pi)^2}\sim5\cdot10^{-3}$, which is basically the size of $L_9^R$ and $|L_{10}^R|$ but clearly 
larger than the remaining eight $L_i^R$ \cite{RChL, Bijnens:2011tb}. There is not that much information on the values of the $C_i^R$ (see, however Ref.~\cite{Jiang:2009uf}). We will take $|\lambda_i^V|\leq 0.04$ GeV$^{-1}$ for the variation of these 
couplings ($i=6,11,12,13,14,15$ are relevant to our analysis), although it may be expected that only one or two of them (if any) are close to that (upper) limit. Proceeding similarly we can estimate $\lambda_i^{VV}\sim \frac{M_V^4}{2F^2}C_i^R$ and 
$\lambda_i^{SV}\sim\sqrt{2}\frac{M_S^2 M_V^2}{c_m F}C_i^R$. This sets a reasonable upper bound $|\lambda_i^{SV}|\sim|\lambda_i^{VV}|\lesssim 0.1$ that we will assume in the numerics.\\

We discuss next the values of the $c_i$ ($\kappa_i^V$) couplings in eqs.~(\ref{VJP&VVP}) and (\ref{V_KN}). Eqs.~(\ref{eq: Consistent set of relations}) predict the vanishing of two linear combinations of $c_i$'s. The numerical value for the predicted 
$c_6-c_5$ is $-0.017$. There are some determinations of $c_3$. It was estimated (although with a sign ambiguity) studying $\tau^-\to\eta\pi^-\pi ^0\nu_\tau$ decays \cite{Dumm:2012vb}. Taking into account the determinations by Y.~H.~Chen \textit{et. al.} 
\cite{Chen:2012vw,Chen:2013nna,Chen:2014yta} as well, we will use $c_3=0.007^{+0.020}_{-0.012}$. $c_4$ was first determined studying $\sigma(e^+e^-\to KK\pi)$ in Ref.~\cite{Dumm:2009va}, even though with a value yielding inconsistent results for the 
$\tau^-\to K^-\gamma\nu_\tau$ branching ratio \cite{Guo:2010dv}. We will take the determination $c_4=-0.0024\pm0.0006$ \cite{Chen:2013nna} as the most reliable one. Two other independent $c_i$ combinations appear in our form factors. We will take them as $c_5$ and 
$c_7$ whose modulus we will vary in the range $[0,0.03]$. Using eqs.~(\ref{eq: relation different basis}) to relate the $c_i$ and $\kappa_i^V$ couplings we can find reasonable guesses on the latter from $|c_i|\lesssim 0.03$. Thus, we will take 
$|\kappa_i^V|\leq 0.04$ GeV$^{-1}$ for their variation.\\

There is very little information on the $\kappa_i^A$ couplings. As a reasonable estimate we will make them vary in the same interval as the $\lambda_i^V$ and $\kappa_i^V$ couplings.\\

The numerical values of the two $d_i$ couplings (VVP operators) which were determined in eq.~(\ref{eq: Consistent set of relations}) are $d_1+8d_2\sim0.15$ and $d_3\sim-0.11$. $d_2$ has been determined jointly with $c_3$ (discussed above). According 
to the quoted references we will employ $d_2=0.08\pm0.08$. Then only $d_4$ would remain free. Given the previous values for the other $d_i$'s we will assume $|d_4|<0.15$.\\

  We will discuss in the next section the phenomenology of $\tau^-\to\pi^-\eta^{(\prime)}\gamma\nu_\tau$ decays, focusing on the background they constitute to the searches 
 of SCC in their corresponding non-radiative decays. We will start discussing the simplified case of $MDM$, according to eqs.(\ref{FFsMDM}), to turn next to the $R\chi L$ prediction corresponding to eqs.~(\ref{FFsRChT_ChPT}) and (\ref{a1-1R-RChL}) to 
 (\ref{v4-2R-RChL}).\\

\section{$\tau^-\to\eta^{(\prime)}\pi^-\nu_\tau \gamma$ as background in the searches for $\tau^-\to\eta^{(\prime)}\pi^-\nu_\tau$}\label{sec:radbkg}

\subsection{Meson dominance predictions}\label{sec:radbkgMDM}
\begin{figure}[ht!]
  \subfloat[$100$ normally sampled points in the MDM parameter space are plotted.]{\includegraphics[scale=0.3,angle=-90]{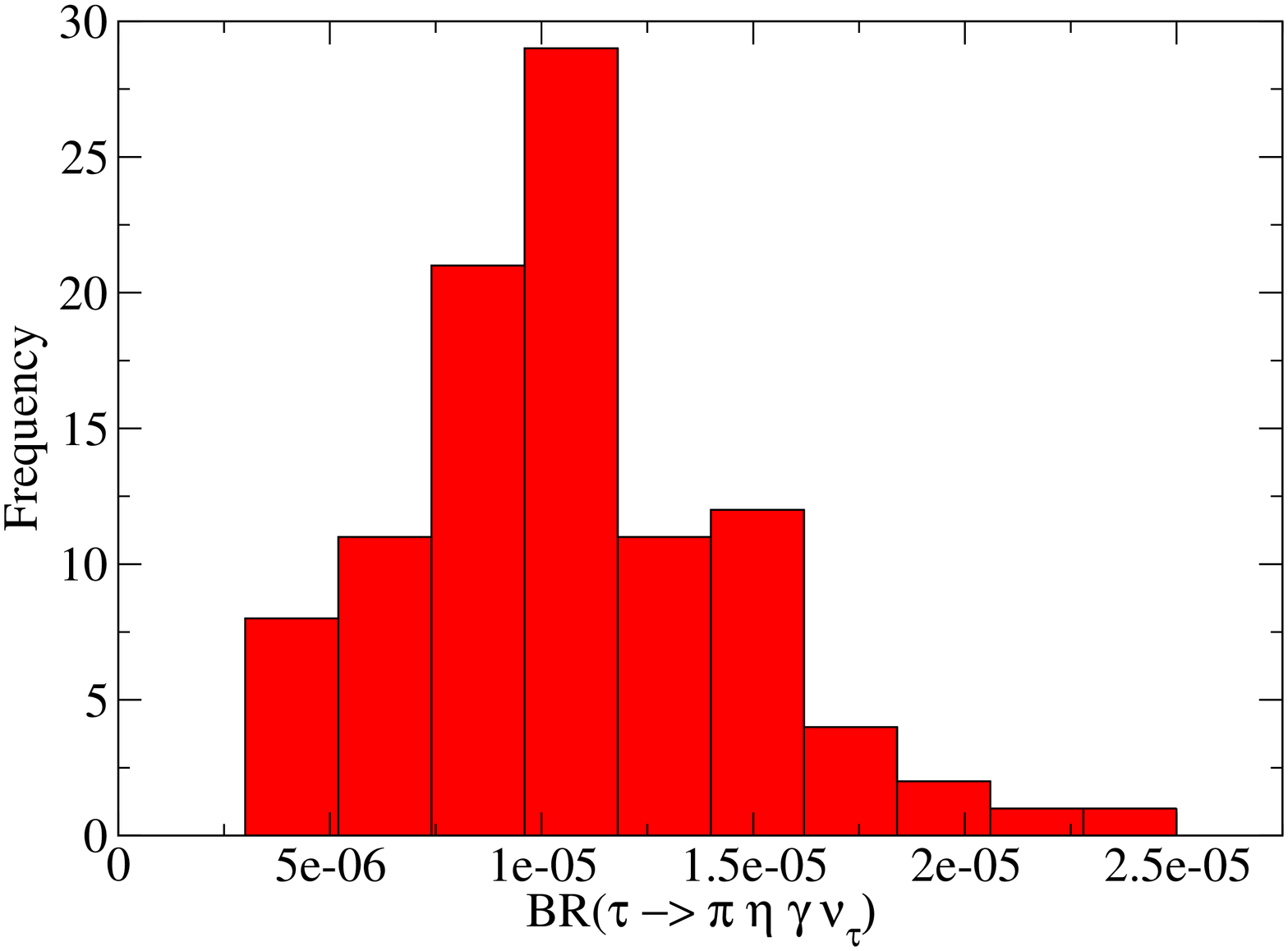}}
  \hspace*{0.2cm}
  \subfloat[Same as in (a) with $1000$ points.]{\includegraphics[scale=0.3,angle=-90]{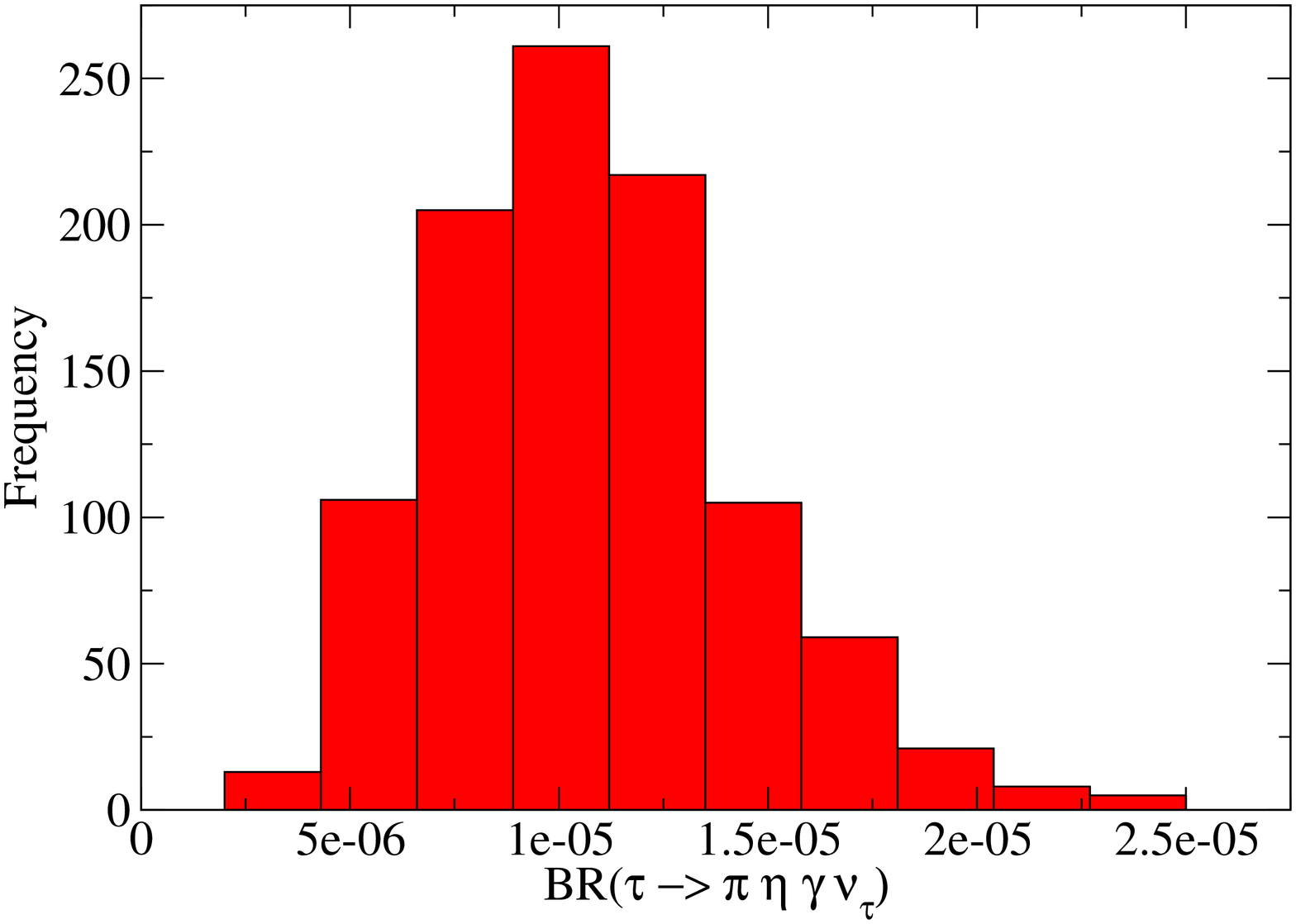}}
  \caption{Predictions for the $BR(\tau^-\to\eta\pi^-\nu_\tau \gamma)$ depending on the simulation sample size.}\label{fig:VMD100/1000}
  \end{figure}

We will get our MDM predictions on the $\tau^-\to\eta^{(\prime)}\pi^-\nu_\tau \gamma$ decays varying the couplings appearing in table \ref{ParsMDM} within a one-sigma range assuming a Gaussian distribution for them. Lacking any information on their correlations, 
we will take them as independent, which would (conservatively) increase the statistical error of our predictions. All other inputs are set to PDG values \cite{PDG} taking into account the corresponding 
errors. For the decay mode with an $\eta^\prime$ meson we need to change the couplings $g_{\rho\eta\gamma}$, $g_{\omega\eta\gamma}$, $g_{\rho\rho\eta}$ and $g_{a_0\pi\eta}$ by those in the last four rows of table \ref{ParsMDM}. For our 
phenomenological analysis we will be mainly concerned in examining the backgrounds that the $\tau^-\to\eta^{(\prime)}\pi^-\nu_\tau \gamma$ constitute in the search for SCC in the corresponding non-radiative processes, with branching fractions of 
the order of $1.7\cdot10^{-5}$ ($\eta$ mode) and $\left[10^{-7},10^{-6}\right]$ ($\eta^\prime$ channel) \cite{Escribano:2016ntp}. We will first plot the predicted branching ratios when sampling these 10 parameters within one-sigma uncertainties (using normal 
distributions). This 
information is collected in figures \ref{fig:VMD100/1000}. In the left panel we can see the result of taking 100 points in the parameters space scan, while 1000 points were used to obtain the figure on the right hand side. The corresponding mean 
and standard deviation of both data samples branching ratios are $(1\pm1)\cdot 10^{-5}$ (100 points) and $(1.1\pm0.3)\cdot 10^{-5}$ (1000 points). We do not assign a theory error to these values, since our only purpose is to have a simple estimate 
to compare with the $R\chi L$ predictions in section \ref{sec:radbkgRChT} (whose systematic uncertainty will be discussed).\\ \\

\vspace*{1.5cm}
\begin{figure}[ht!]
  \subfloat[Normalized spectrum (corresponding to the data in figure \ref{fig:VMD100/1000} (a)) in the invariant mass of the $\eta\pi^-$ system is plotted.]{\includegraphics[scale=0.3,angle=-90]{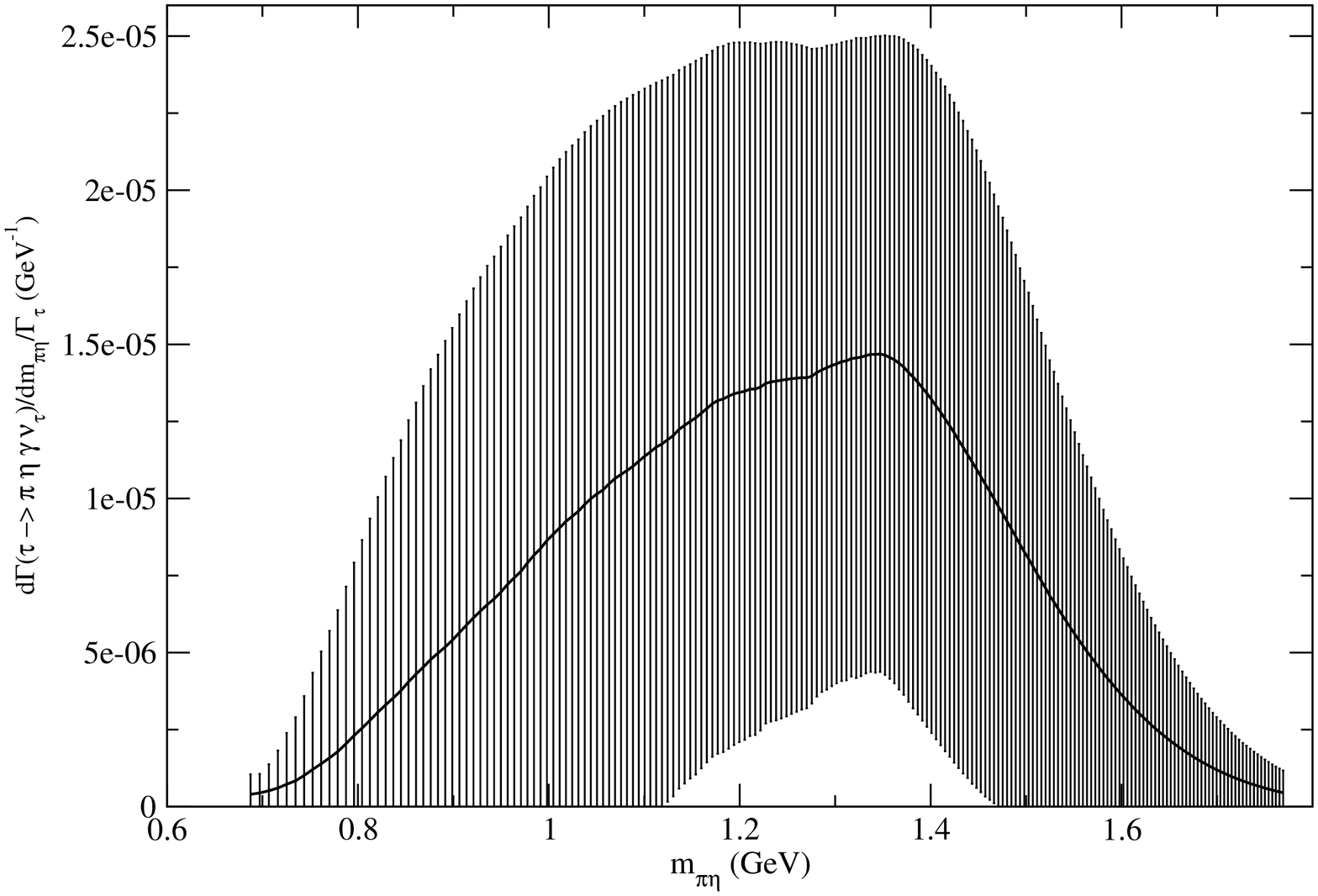}}\hspace*{0.2cm}
  \subfloat[For the same points as in (a), the normalized spectrum in $E_\gamma$ is drawn.]{\includegraphics[scale=0.3,angle=-90]{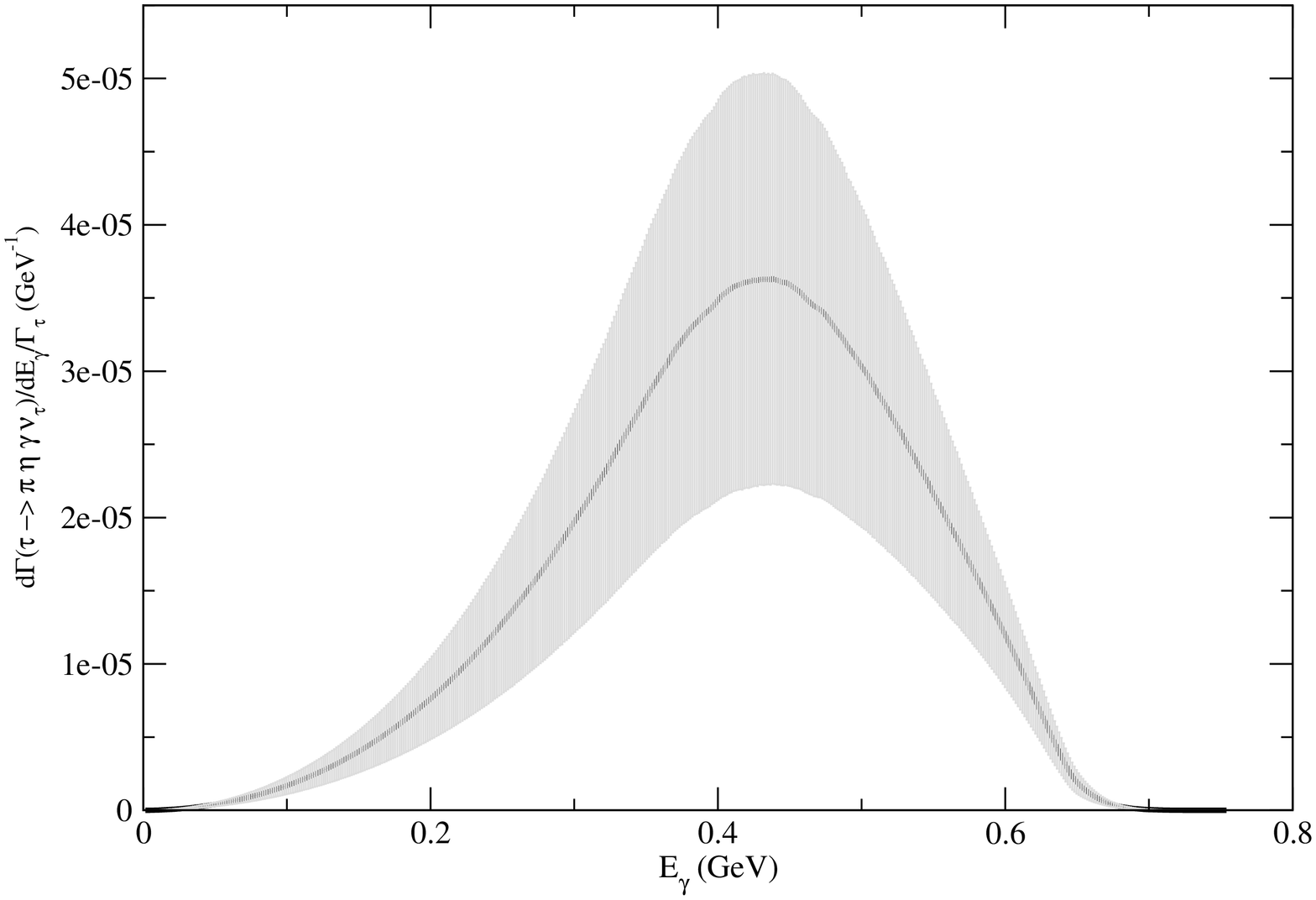}}
  \caption{Normalized spectra of the $\tau^-\to\eta\pi^-\nu_\tau \gamma$ decays according to MDM.}\label{fig:VMDSpectra_eta}
 \end{figure}

\vspace*{1.5cm}
For the previously simulated $100$ data points we have obtained their spectra in both the $\pi^-\eta$ invariant mass (left) and on the photon energy (right). In the first (second) case 200 (500) points constitute the spectra for every simulated point in the 
sampled parameters space. The corresponding normalized spectra (the differential decay distributions are divided by the tau full width) are plotted in figs.~\ref{fig:VMDSpectra_eta}. 
Although the (normalized) spectra in $m_{\pi\eta}$ tend to peak around $1.15-1.35$ GeV, there is not any marked dynamics responsible for that. The dependence on $E_\gamma$ shown in 
the right-hand plot turns out to be essential for getting rid of these backgrounds in related SCC searches. Indeed, while photon spectra are peaked at low energies in IB contributions, this is not the case for the SD ones. In our 
case IB has a negligible impact on the considered decays rate because it is doubly suppressed by $G$-parity violation and by a factor $\alpha$  (See also our appendix A, giving more details on these features). Thus, the relevant photon emission 
in $\tau^-\to\eta^{(\prime)}\pi^-\nu_\tau \gamma$ decays exhibits a 
soft dependence on $E_\gamma$ which vanishes smoothly at both energy ends. Consequently, one can envisage that cutting out photons above a certain energy value will allow to reduce drastically this background in SCC searches.\\ \\

\begin{figure}[ht!]
  \subfloat[Photons with $E_\gamma>100$ MeV are rejected.]{\includegraphics[scale=0.3,angle=-90]{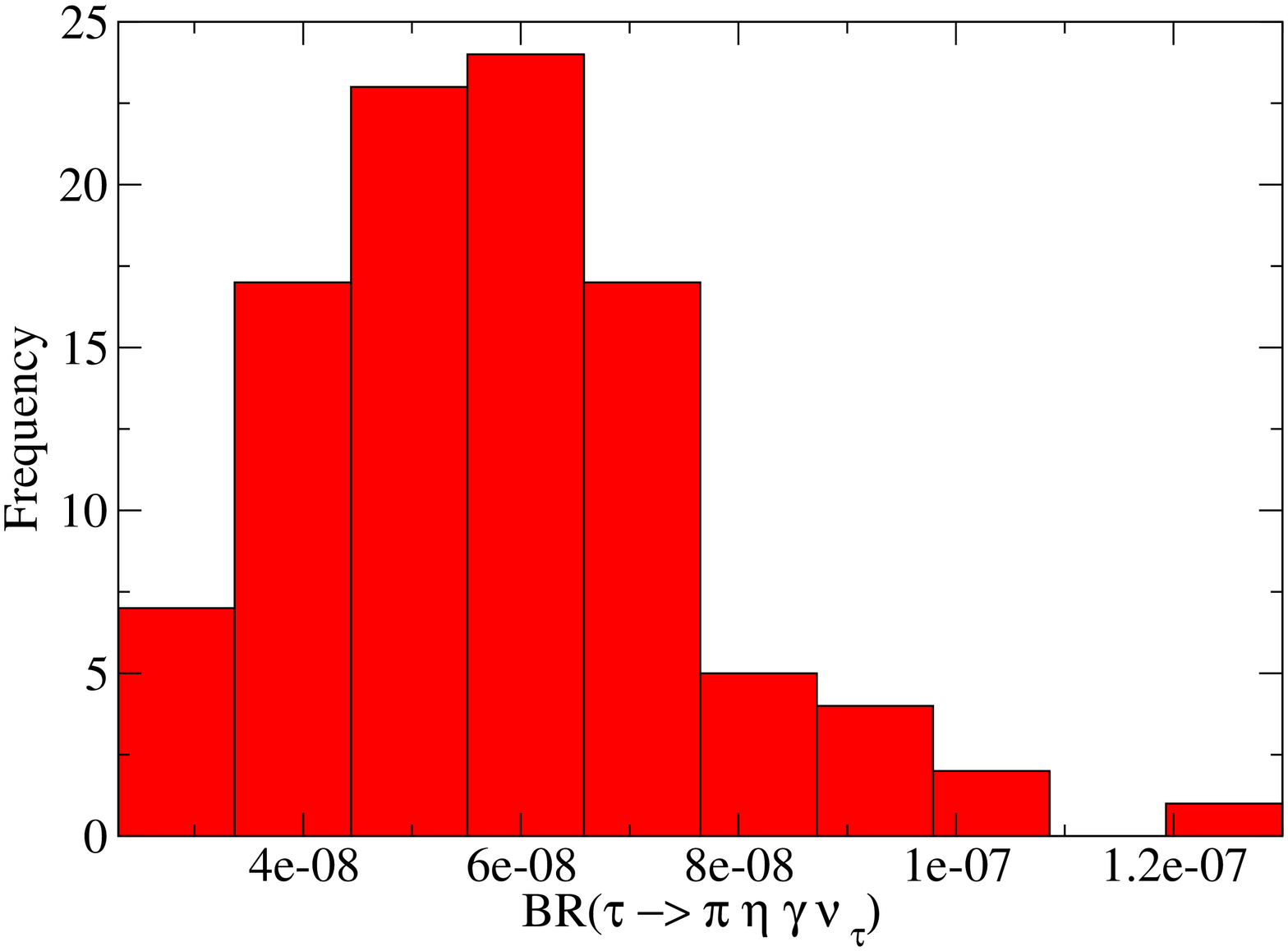}}
  \hspace*{0.2cm}
  \subfloat[Same as in (a) with $50$ MeV as the photon energy cut.]{\includegraphics[scale=0.3,angle=-90]{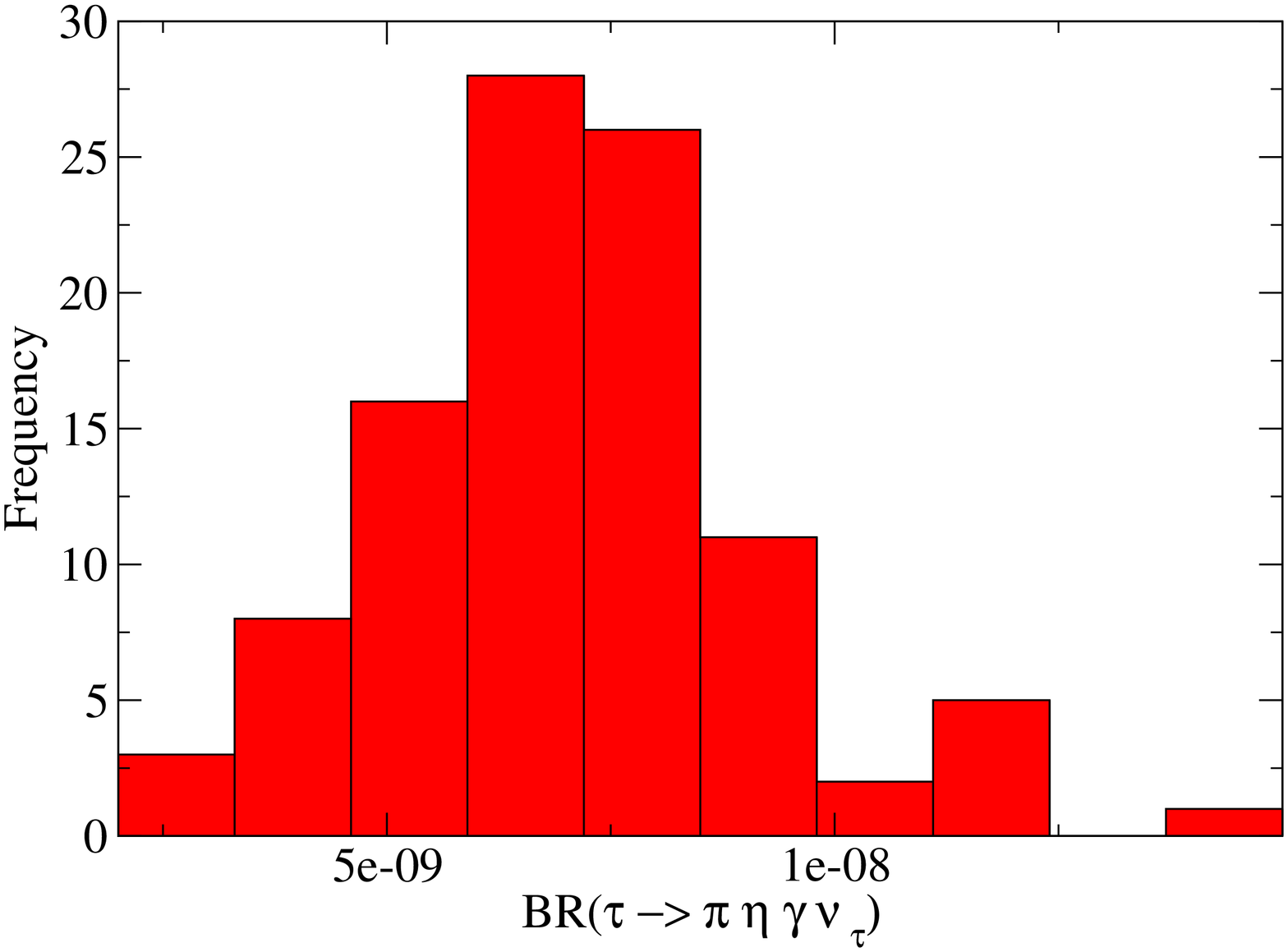}}
  \caption{Predictions for the $BR(\tau^-\to\eta\pi^-\nu_\tau \gamma)$ as a function of the photon energy cut.}\label{fig:VMDcuts}
  \end{figure}

In figs.~\ref{fig:VMDcuts} we show the effect of cutting photons above $100$ MeV (left) and $50$ MeV (right).
 It is fair to acknowledge that $50$ MeV can be a too aggressive cut for Belle-II, because the typical calorimeter activity 
will be considerably larger than at BaBar/Belle. As far as we know, $100$ MeV represents a perfectly feasible cut. It is seen that even for this cut, $\tau^-\to\eta\pi^-\nu_\tau \gamma$ decays are suppressed to a level where they do not affect the 
search for the corresponding non-radiative decay channel. The corresponding branching fractions upper bounds (obtained with a larger simulation sample, not shown in the figure) are $\leq0.6\cdot10^{-7}$ (cut for $E_\gamma>100$ MeV) and 
$\leq0.7\cdot 10^{-8}$ (for $E_\gamma>50$ MeV). In any case this would be at least two orders of 
magnitude smaller than the associated non-radiative decay. We will see in section \ref{sec:radbkgRChT} if these expectations, based on na\"ive MDM, hold in a more elaborated treatment of strong interactions in the chiral and resonance regions.
It is noteworthy that no peak associated to the $a_0(980)$ resonance exchange is appreciated in our spectra.\\ \\

\vspace*{1.5cm}
\vspace*{1.5cm}

\begin{figure}[ht!]
  \subfloat[$100$ normally sampled points in the MDM parameter space are plotted.]{\includegraphics[scale=0.3,angle=-90]{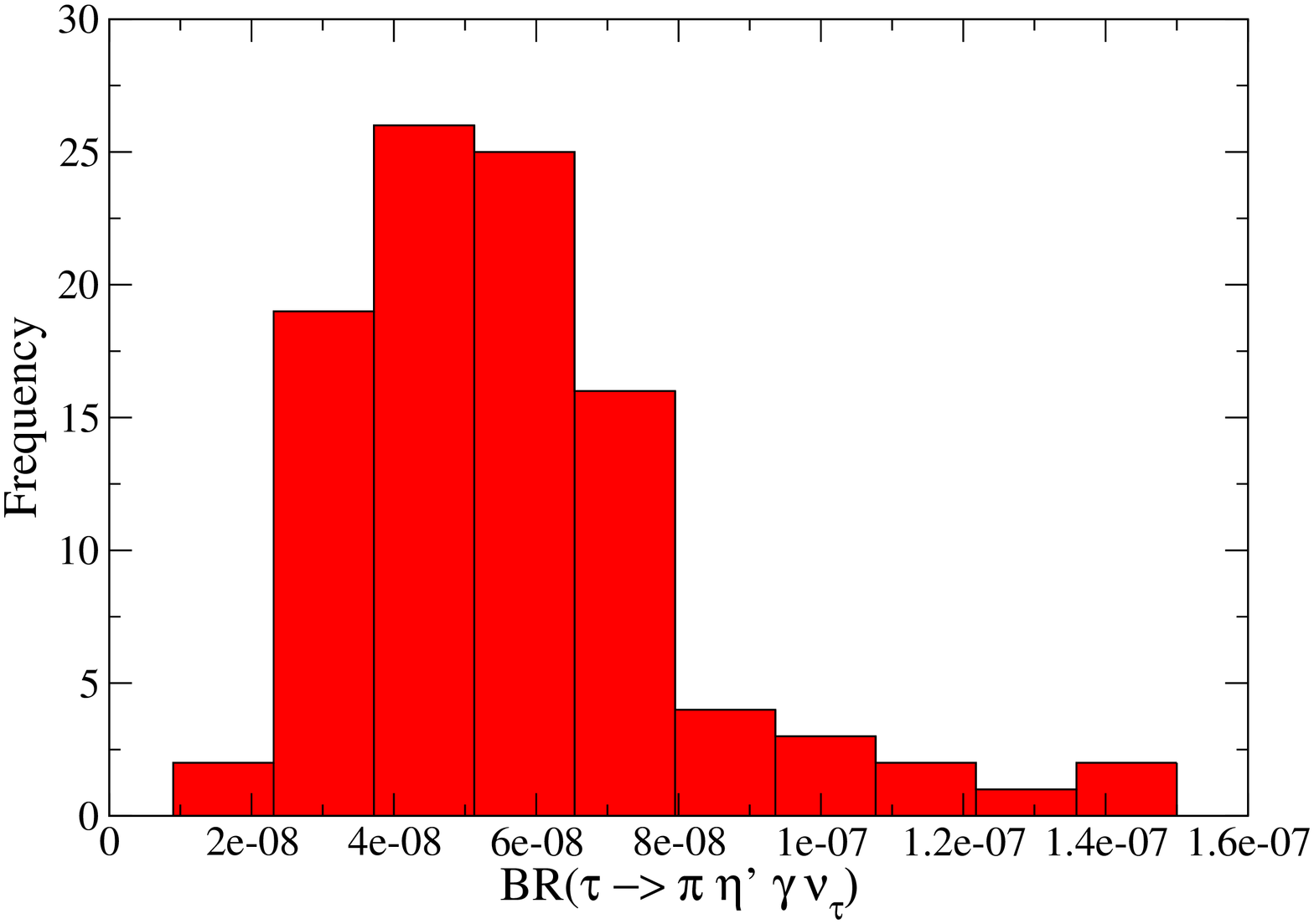}}
  \hspace*{0.2cm}
  \subfloat[Same as in (a) with $1000$ points.]{\includegraphics[scale=0.3,angle=-90]{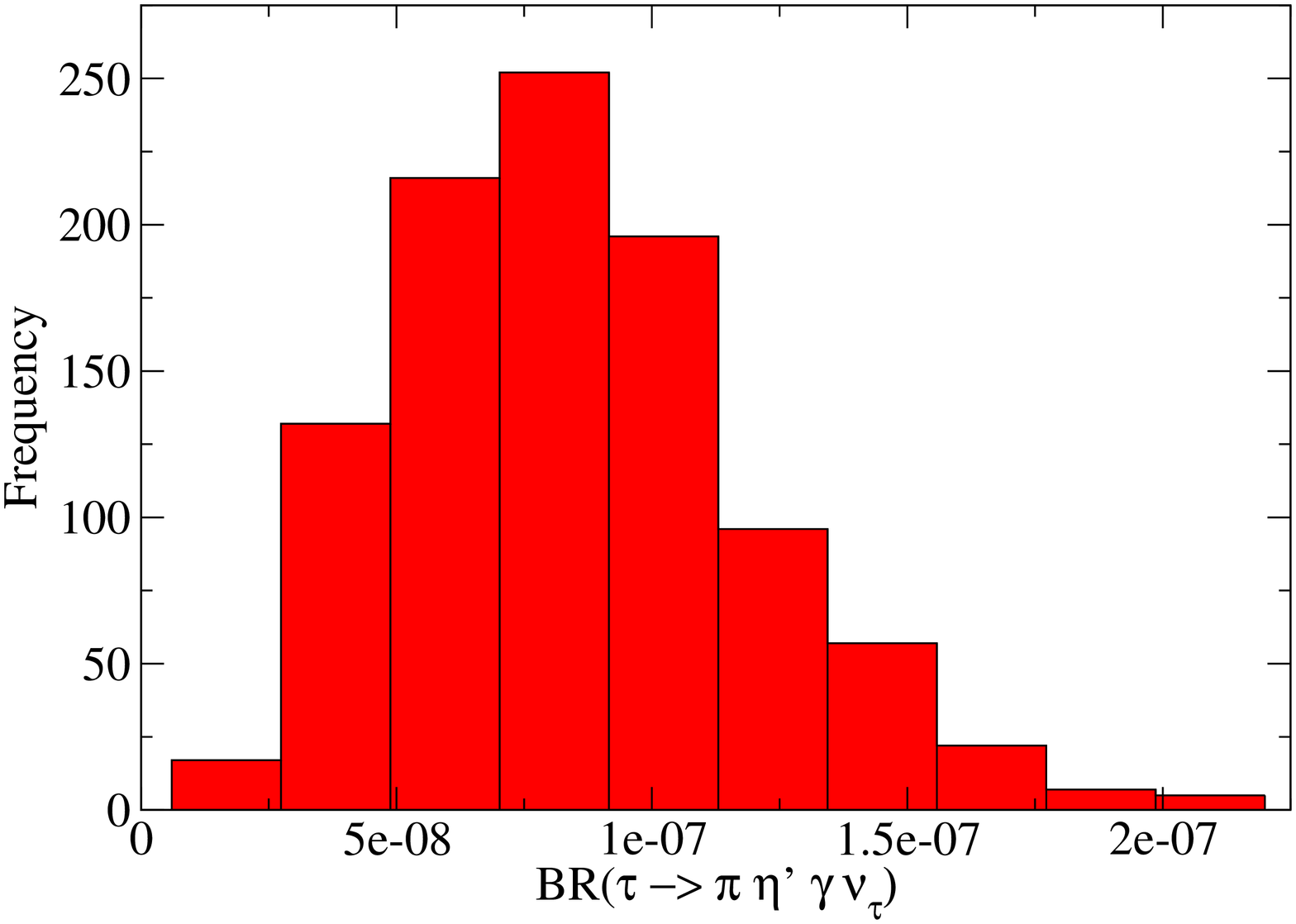}}
  \caption{Predictions for the $BR(\tau^-\to\eta^\prime\pi^-\nu_\tau \gamma)$ depending on the simulation sample size.}\label{fig:VMD100/1000p}
  \end{figure}

Now we turn to the predictions of MDM for the partner $\tau^-\to\eta^{\prime}\pi^-\nu_\tau \gamma$ decays. We will proceed analogously as for the 
$\eta$ meson channel. We first plot the branching ratio for $100$ ($1000$) normally sampled points in the 
parameter space in fig.~\ref{fig:VMD100/1000p}. The corresponding mean branching fractions are $\sim6\cdot10^{-8}$ ($(0.8\pm0.8)\cdot10^{-7}$), where the error is again only statistical and reducible.\\ \\

 \vspace*{1.5cm}
\vspace*{1.5cm}

\begin{figure}[ht!]
  \subfloat[Normalized spectrum (corresponding to the data in figure \ref{fig:VMD100/1000p} (a)) in the invariant mass of the $\eta^\prime\pi^-$ system is plotted.]{\includegraphics[scale=0.3,angle=-90]{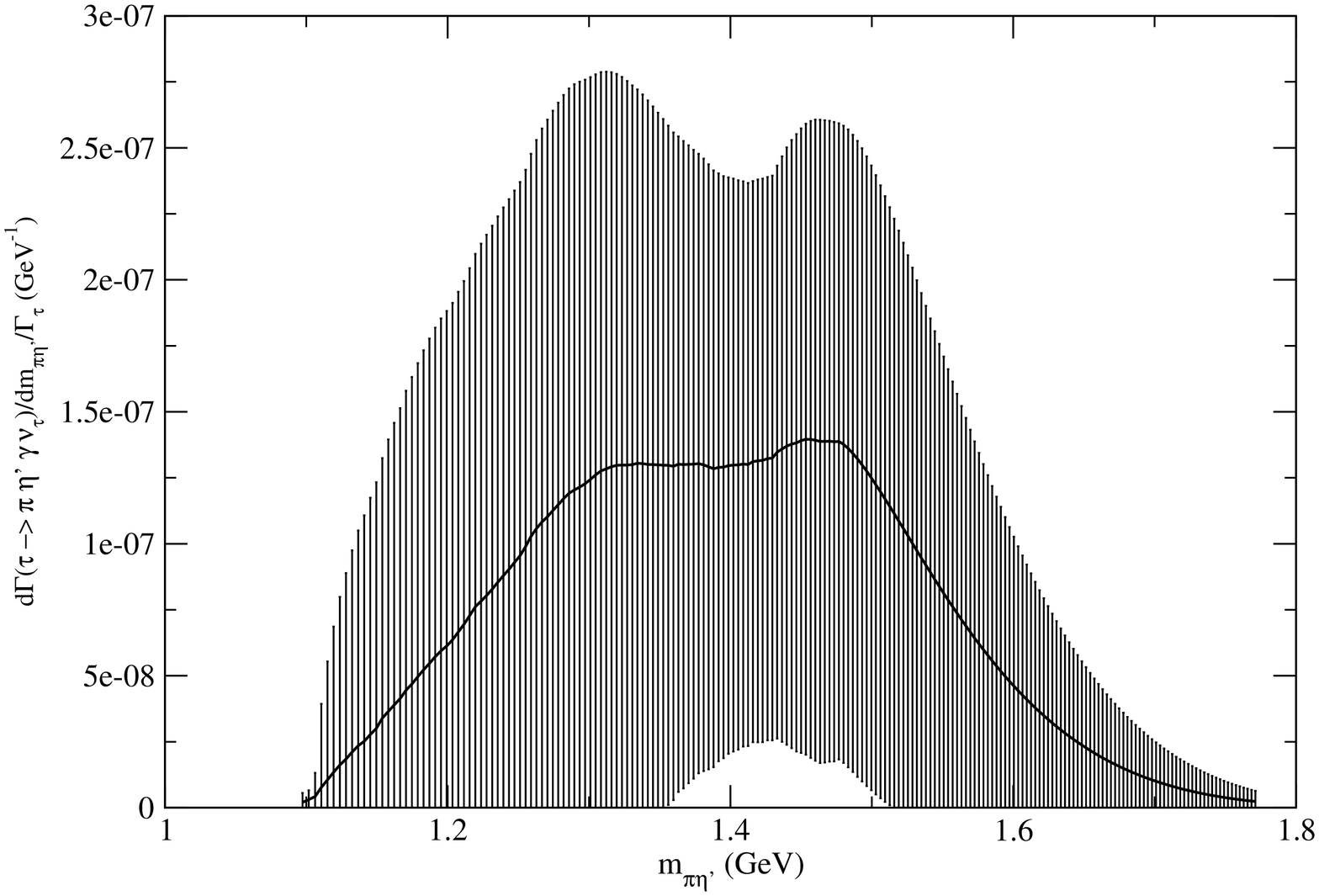}}\hspace*{0.2cm}
  \subfloat[For the same points as in (a), the normalized spectra in $E_\gamma$ is drawn.]{\includegraphics[scale=0.3,angle=-90]{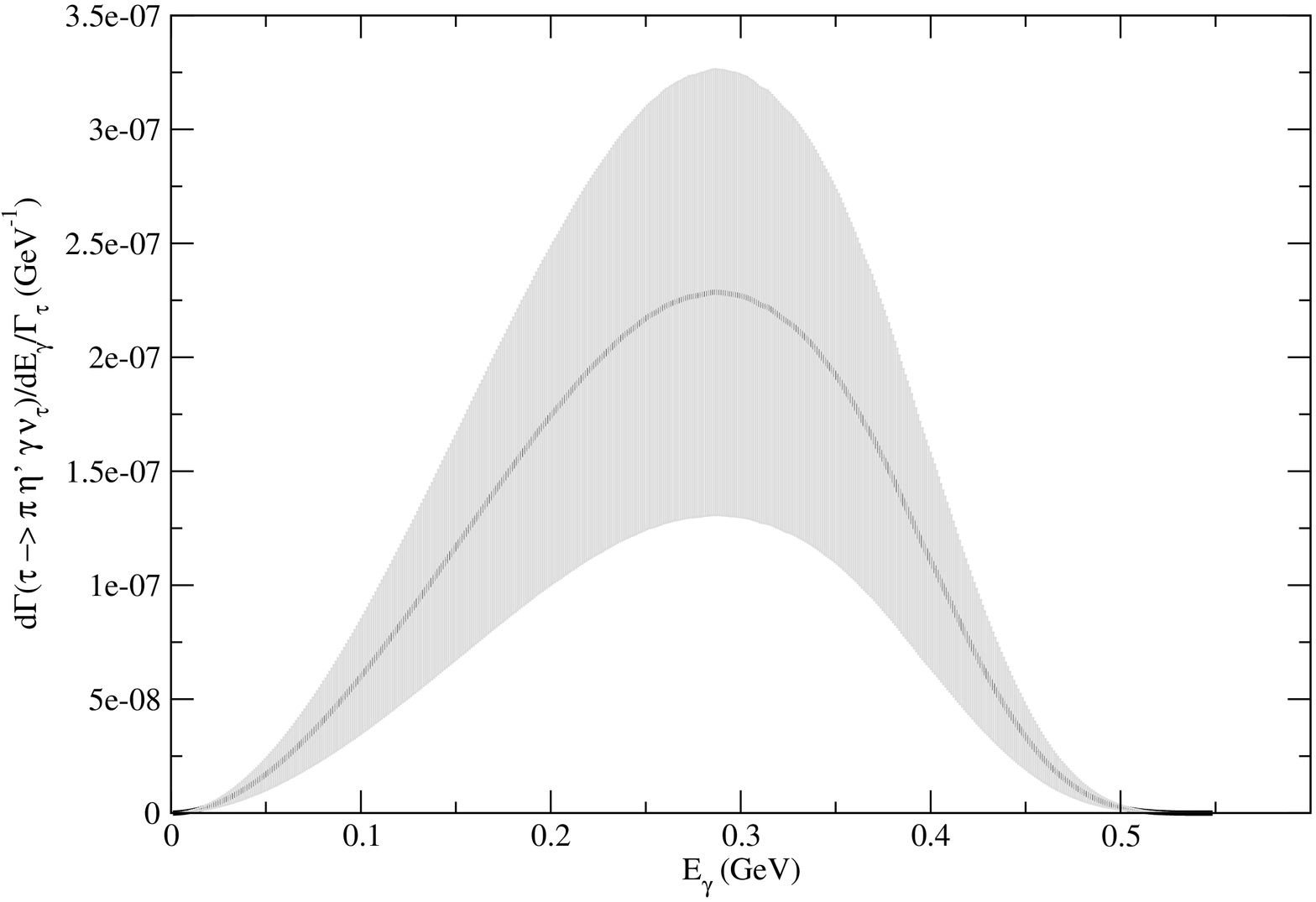}}
  \caption{Predictions for the normalized spectra of the $\tau^-\to\eta^\prime\pi^-\nu_\tau \gamma$ decays according to MDM.}\label{fig:VMDSpectra_etap}
  \end{figure}

For the previously simulated $100$ data points we have obtained their spectra in both the $\pi^-\eta$ invariant mass (left) and on the photon energy (right). In the first (second) case 200 (500) points constitute the spectra for every simulated point in the 
sampled parameters space. The corresponding normalized spectra (the differential decay distributions are divided by the tau full width) are plotted in figs.~\ref{fig:VMDSpectra_etap}. 
Again no hint of the underlying dynamics is seen and cutting medium- and high-energy photons appears promising to eliminate this 
background. Since the phase space does not allow for on-shell $a_0$ exchanges, no possible related substructure can arise~\footnote{We are neglecting excited resonance contributions, which 
specifically forbids any trace of the $a_0(1450)$ meson in this approach.}.\\ \\

\vspace*{1.5cm}
\vspace*{1.5cm}

\begin{figure}[ht!]
  \subfloat[When cutting photons with $E_\gamma>100$ MeV.]{\includegraphics[scale=0.3,angle=-90]{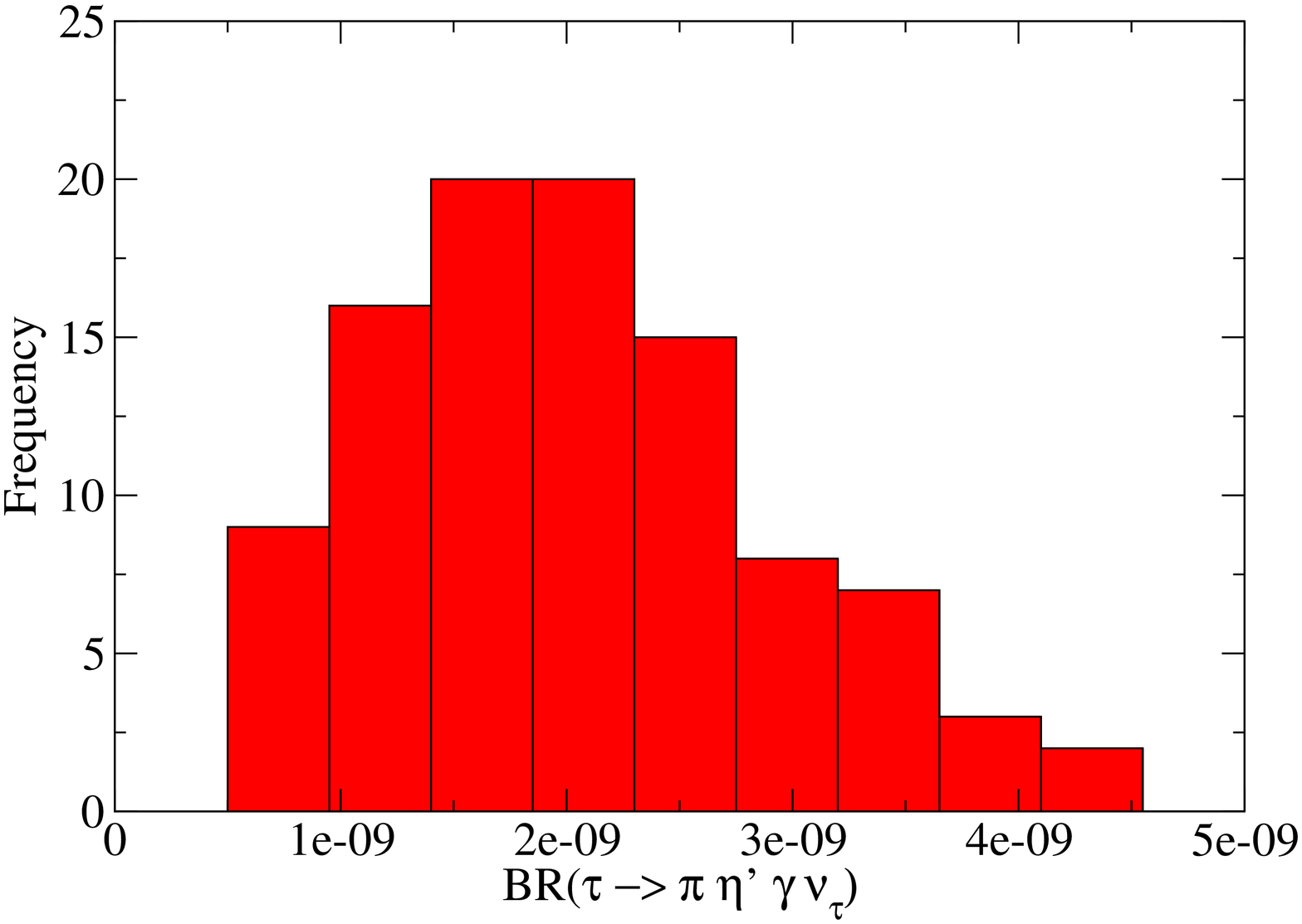}}
  \hspace*{0.2cm}
  \subfloat[When cutting photons with $E_\gamma>50$ MeV.]{\includegraphics[scale=0.3,angle=-90]{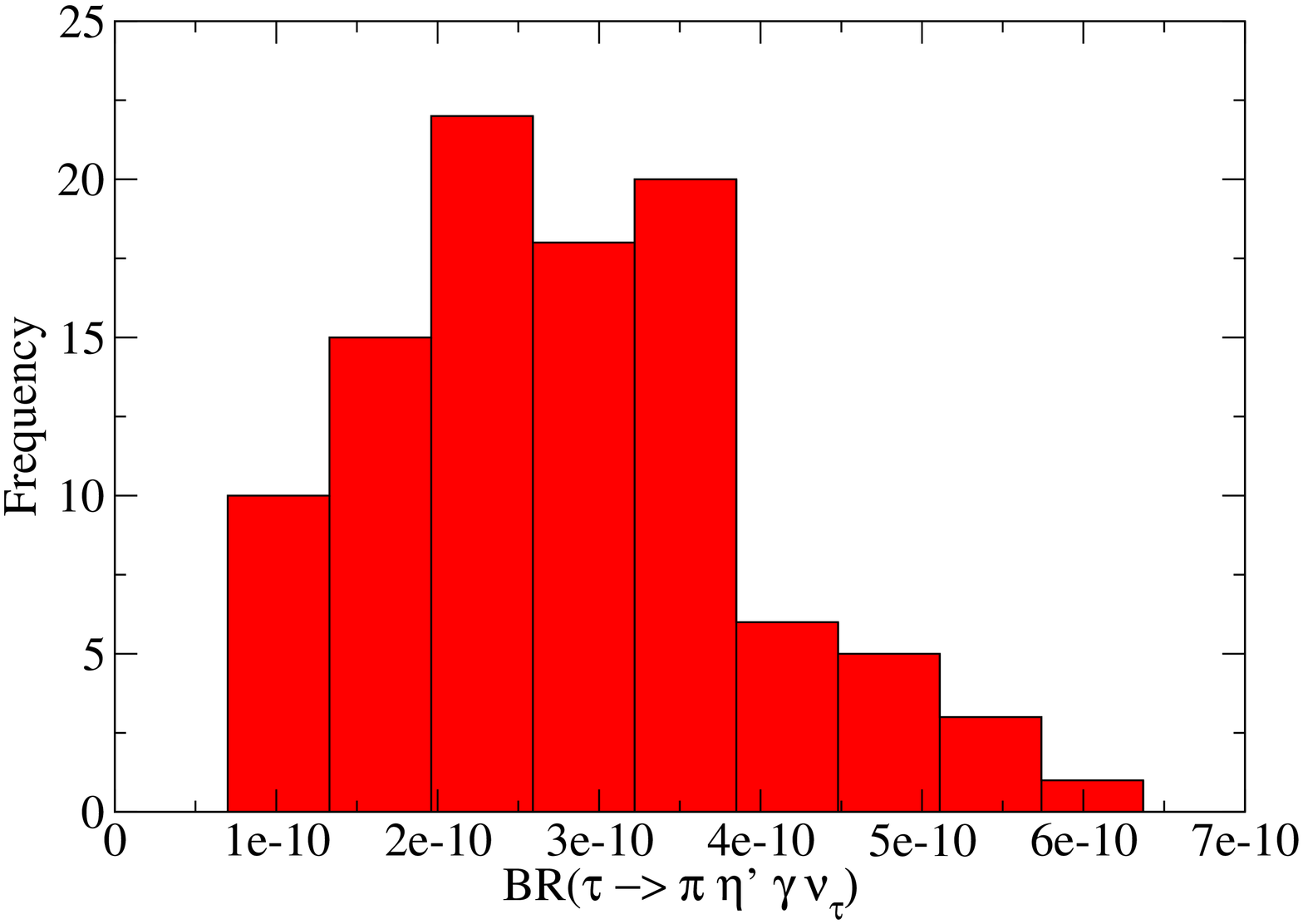}}
  \caption{$BR(\tau^-\to\eta\pi^-\nu_\tau \gamma)$ are represented as a function of the photon energy cut.}\label{fig:VMDpcuts}
  \end{figure}

  Finally, in figures \ref{fig:VMDpcuts} we present the decreased normalized decay rates, resulting from cutting photons above $100$ MeV (left) and $50$ MeV (right). The corresponding branching fractions are $\leq0.2\cdot10^{-8}$ 
and $\leq0.3\cdot10^{-9}$, respectively (obtained with the $1000$ data point samples not shown in the figure), at least a factor 50 smaller than their non-radiative counterparts, a feature that needs to be confronted to the 
results using $R\chi L$ presented in the next section. We also note that in MDM the bulk of the contribution to the branching ratio comes from the last diagram in the last line of figure \ref{fig:hadvertex}. Neglecting all other diagrams one 
gets $\sim 80\%$ of the branching ratio only from this diagram in the $\eta$ meson decay mode, while the $\eta^\prime$ channel is essentially saturated by this contribution.\\ \\

\vspace*{1.5cm}
\vspace*{1.5cm}
 
\subsection{$R\chi L$ predictions}\label{sec:radbkgRChT}
As we noted in section \ref{sec:MDM}, the MDM form factors are obtained from two-resonance mediated diagrams only. In the $R\chi L$ framework one has, in addition to the chiral (anomalous) contribution, one- and two-resonance mediated diagrams. Along 
this section we will be comparing the results obtained with/without the two-resonance exchange diagrams. Comparison of both will show that the main features of these decays are already captured without including the two-resonance contributions.\\ \\

\vspace*{1.5cm}
  \begin{figure}[ht!]
  \subfloat[All contributions included.]{\includegraphics[scale=0.3,angle=-90]{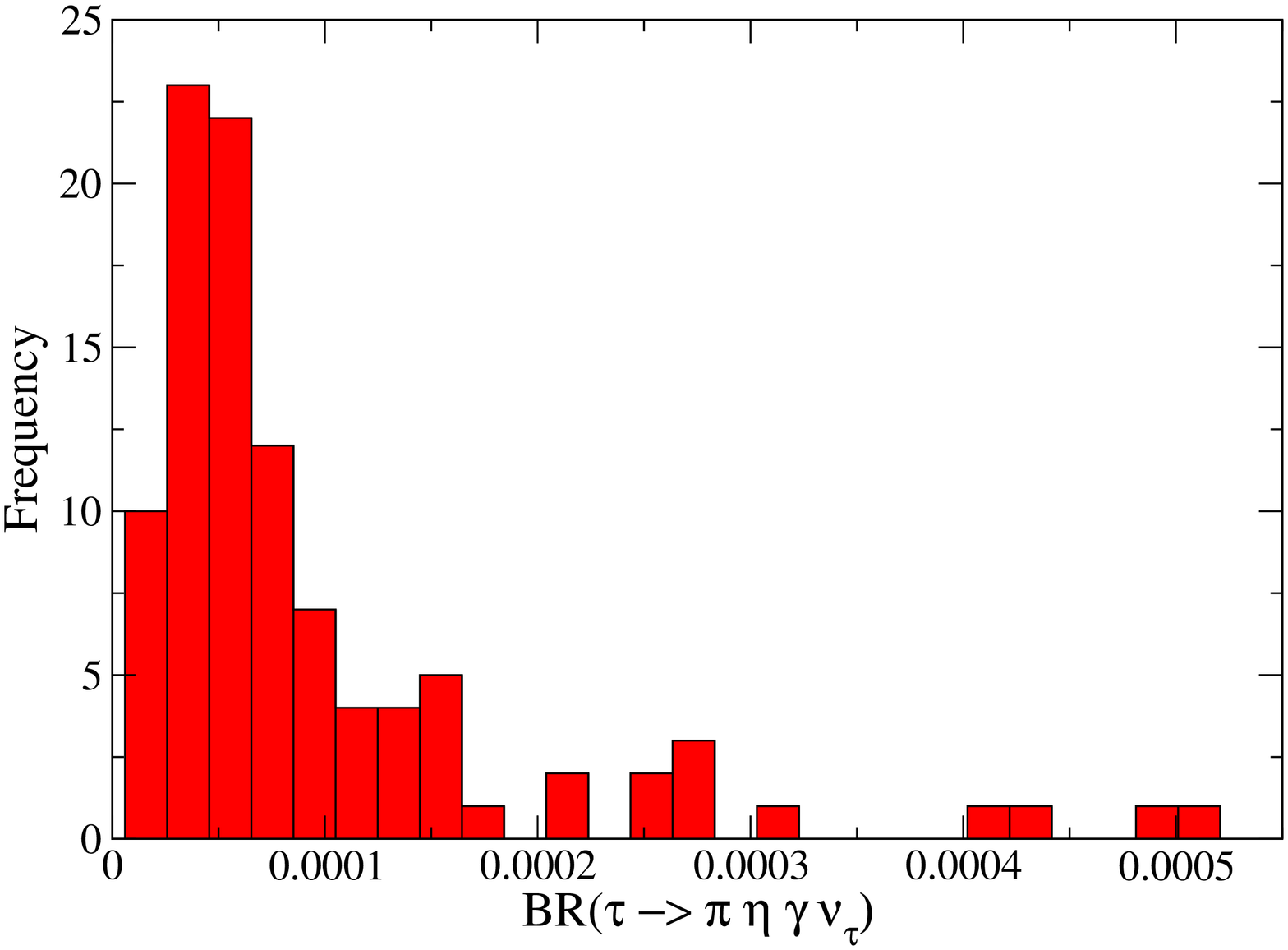}}
  \hspace*{0.2cm}
  \subfloat[Two-resonance contributions neglected.]{\includegraphics[scale=0.3,angle=-90]{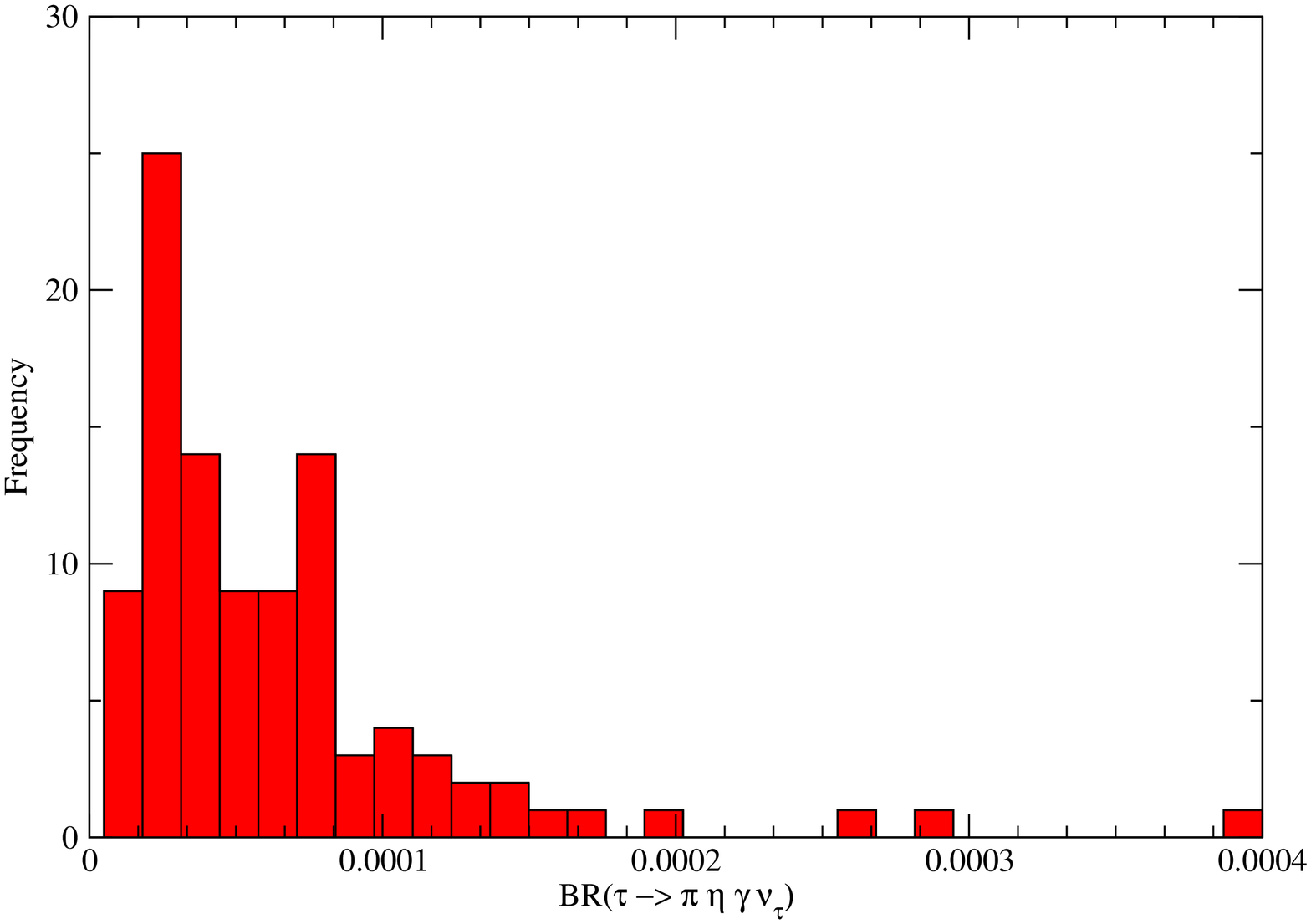}}
  \caption{Predictions for the $\tau^-\to\eta\pi^-\nu_\tau \gamma$ decays branching ratios: $100$ normally sampled points in the $R\chi L$ parameter space are plotted.}\label{fig:RChT100}
  \end{figure}
\vspace*{1.5cm}

After normally sampling $100$ points in the parameter space, the resulting branching ratio is $(1.0\pm0.2)\cdot10^{-4}$, which is plotted in fig.~\ref{fig:RChT100}(a) (the error is only statistical). This one can be reduced enlarging the simulation, but 
then the systematic theory error would saturate the total uncertainty. In particular, with $1000$ data points we find $(0.98\pm0.15)\cdot10^{-4}$, as mean and standard deviation of the sample. If, based on the large-$N_C$ expansion, we assign a $1/N_C$ error at 
the amplitude level, a $1/N_C^2$ error in branching fractions would become comparable to the previous statistical error. Still, our conservative educated guess on this branching ratio uncertainty would be $\sim 0.22\cdot10^{-4}$, accounting for a possible 
larger (double) theory error. In this way we quote $(0.98\pm0.27)\cdot10^{-4}$ as our predicted branching fraction for this decay channel when all (chiral and one and two-resonance mediated contributions) are included. This result is an order of 
magnitude larger than the MDM prediction. Part of it could be due to an statistical artifact caused by the sizable probability of having a significant number of couplings with magnitude outside the one-sigma error range, given the large number of couplings that 
are normally sampled in order to get our predictions. However, according to our previous simulations \cite{v1} where $R\chi L$ couplings were sampled uniformly within the one-sigma interval (with zero probability of lying outside of it), the different dynamics of the 
$MDM$ approach and of the $R\chi L$ has a similarly important effect in explaining this difference. We compare the results corresponding to fig.~\ref{fig:RChT100} (a) (including two-resonance mediated contributions) to the case where these are neglected 
(figure \ref{fig:RChT100} (b)). The predicted branching 
ratios do not vary much. Our previously quoted branching fraction, $(0.98\pm0.27)\cdot10^{-4}$, is reduced to $(0.65 \pm 0.17)\cdot10^{-4}$, where both numbers were obtained from the $1000$ data point simulations and the errors are dominated by the theory 
uncertainty.\\ \\

\begin{figure}[ht!]
   \subfloat[Normalized spectrum (corresponding to the data in figure \ref{fig:RChT100}) in the invariant mass of the $\eta\pi^-$ system is plotted.]{\includegraphics[scale=0.3,angle=-90]{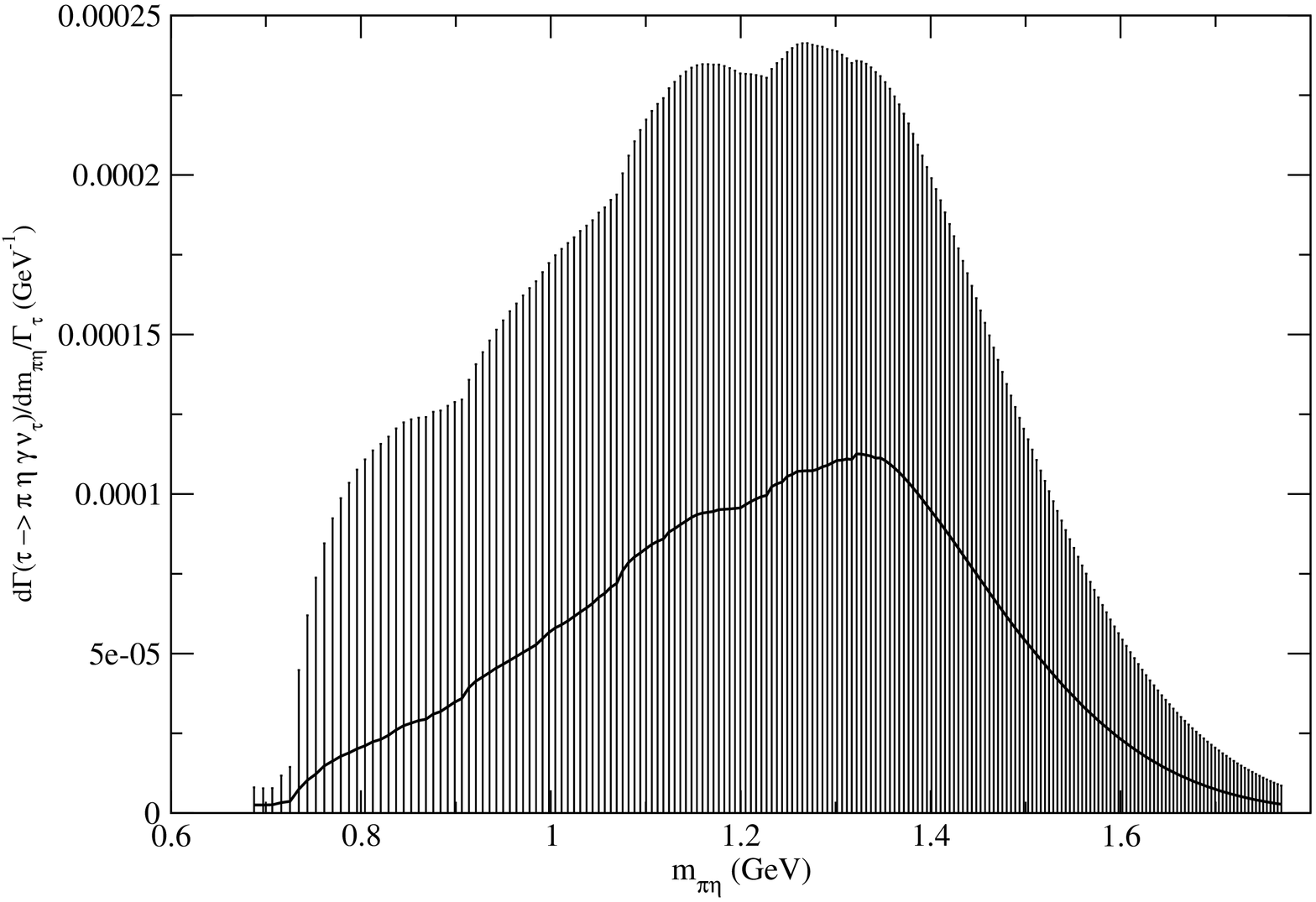}}
   \hspace*{0.2cm}
   \subfloat[For the same points as in (a), the normalized spectrum in $E_\gamma$ is drawn.]{\includegraphics[scale=0.3,angle=-90]{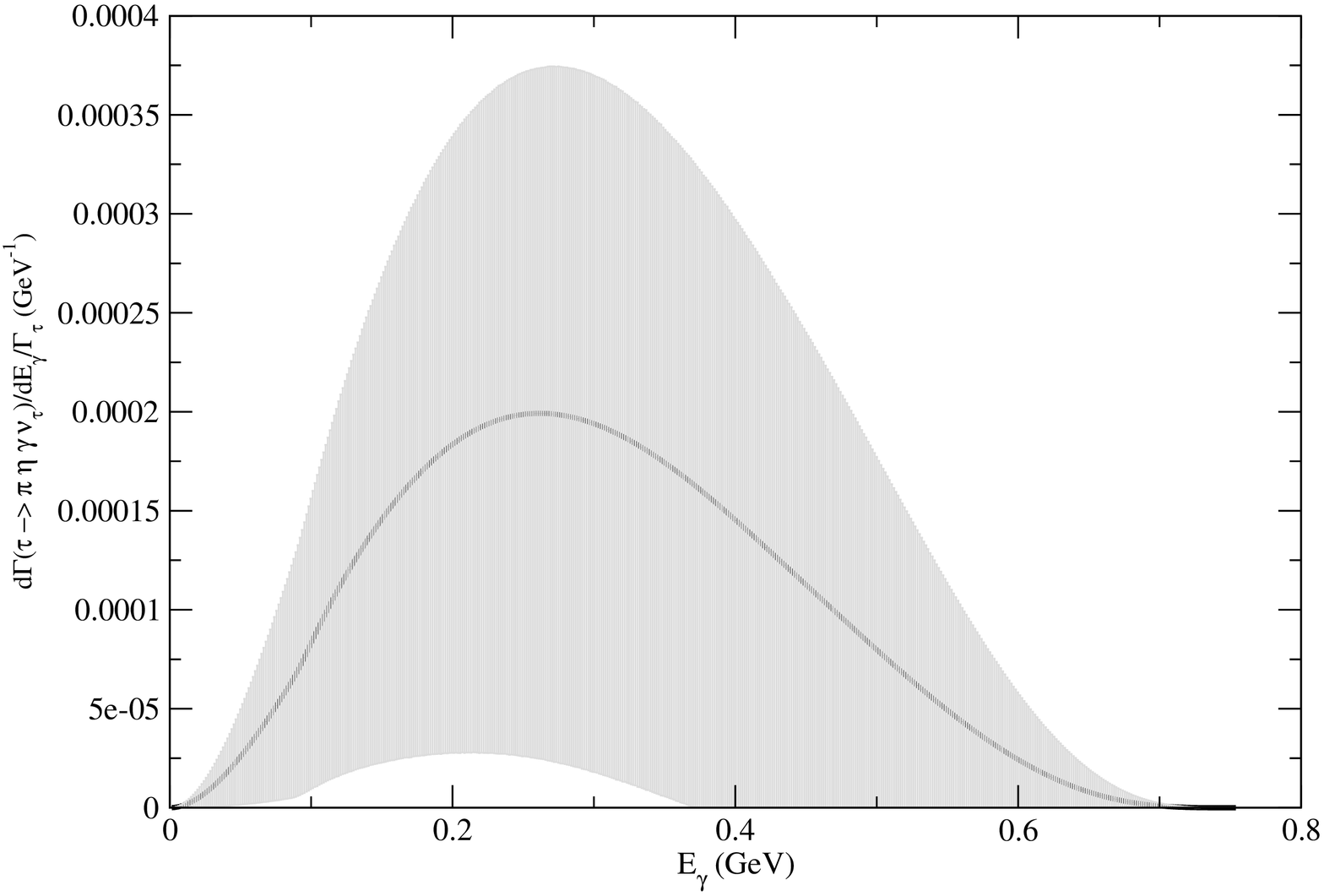}}
   \caption{Normalized spectra of the $\tau^-\to\eta\pi^-\nu_\tau \gamma$ decays according to $R\chi L$.}\label{fig:RChTSpectra}
   \end{figure}

In figures \ref{fig:RChTSpectra} (a) and (b) we plot the normalized spectra in $m_{\eta\pi}$ and $E_\gamma$, with 200 and 500 data points, respectively. In this case, as opposed to the MDM description, the spectra change appreciably depending on the precise 
values of the Lagrangian couplings (see fig. 16(a) in \cite{v1} for illustration).
In fig. \ref{fig:RChTSpectra} (a) we  see that the maximum of the spectra is distributed with significant probability in the $1.15-1.35$ GeV range, in agreement with the MDM prediction. The analysis of the photon energy spectrum (in fig.~\ref{fig:RChTSpectra} (b))
also confirms that, as suggested by the MDM analysis, it seems possible to suppress the bulk of this mode decay rate by cutting photons with energies above some $100$ GeV. These features stay basically the same when neglecting the two-resonance mediated 
contributions. In agreement with the MDM prediction, there is not any signature of the $a_0(980)$ meson in the $\eta\pi$ invariant mass distribution. Since, as in the MDM approach, we are disregarding excited meson multiplets, no possible trace of the 
$a_0(1450)$ meson can result in $R\chi L$ either.\\ \\

  \vspace*{1.5cm}
  \vspace*{1.5cm}

\begin{figure}[ht!]
  \subfloat[When cutting photons with $E_\gamma>100$ MeV.]{\includegraphics[scale=0.3,angle=-90]{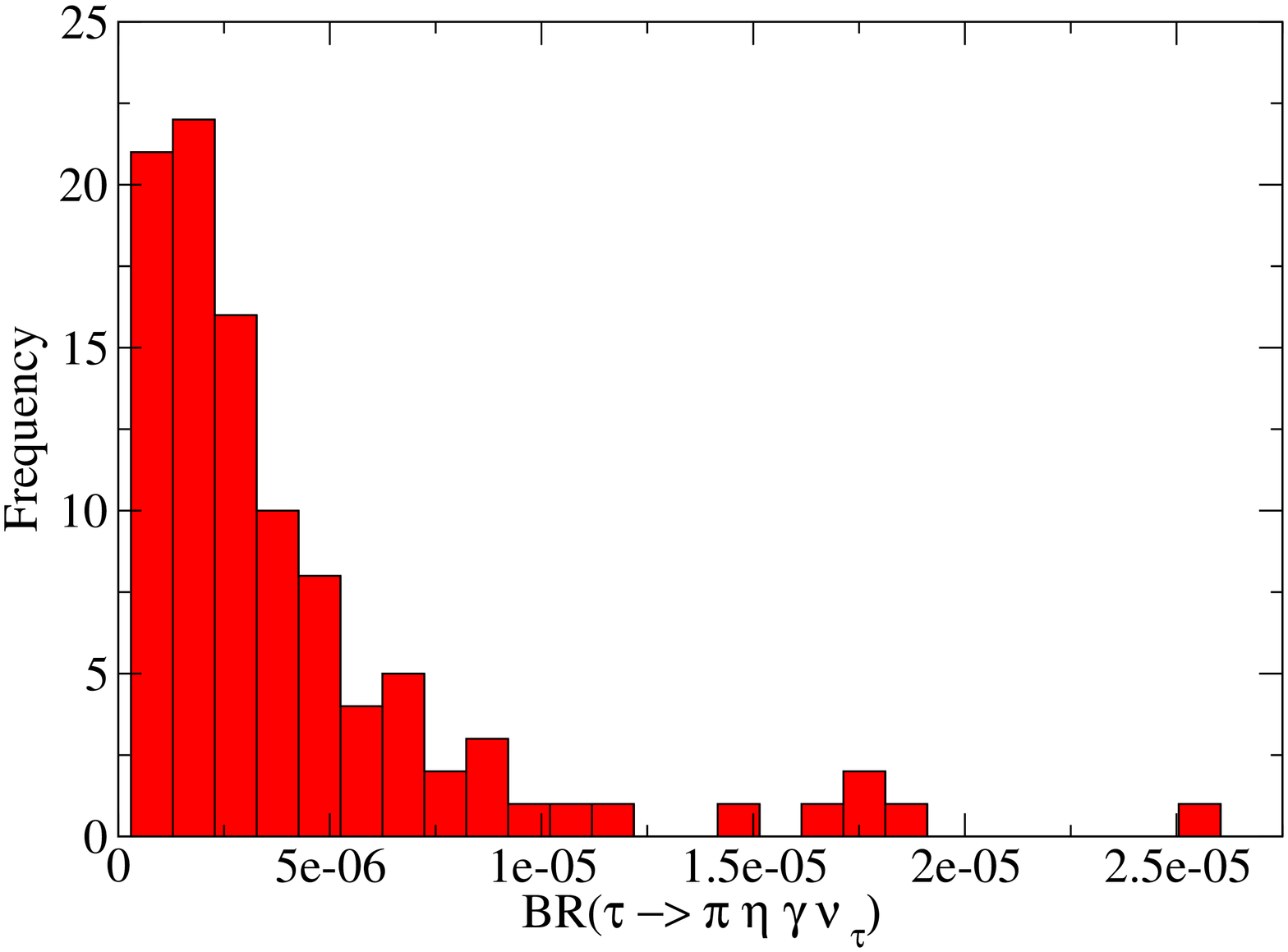}}
  \hspace*{0.2cm}
  \subfloat[When cutting photons with $E_\gamma>50$ MeV.]{\includegraphics[scale=0.3,angle=-90]{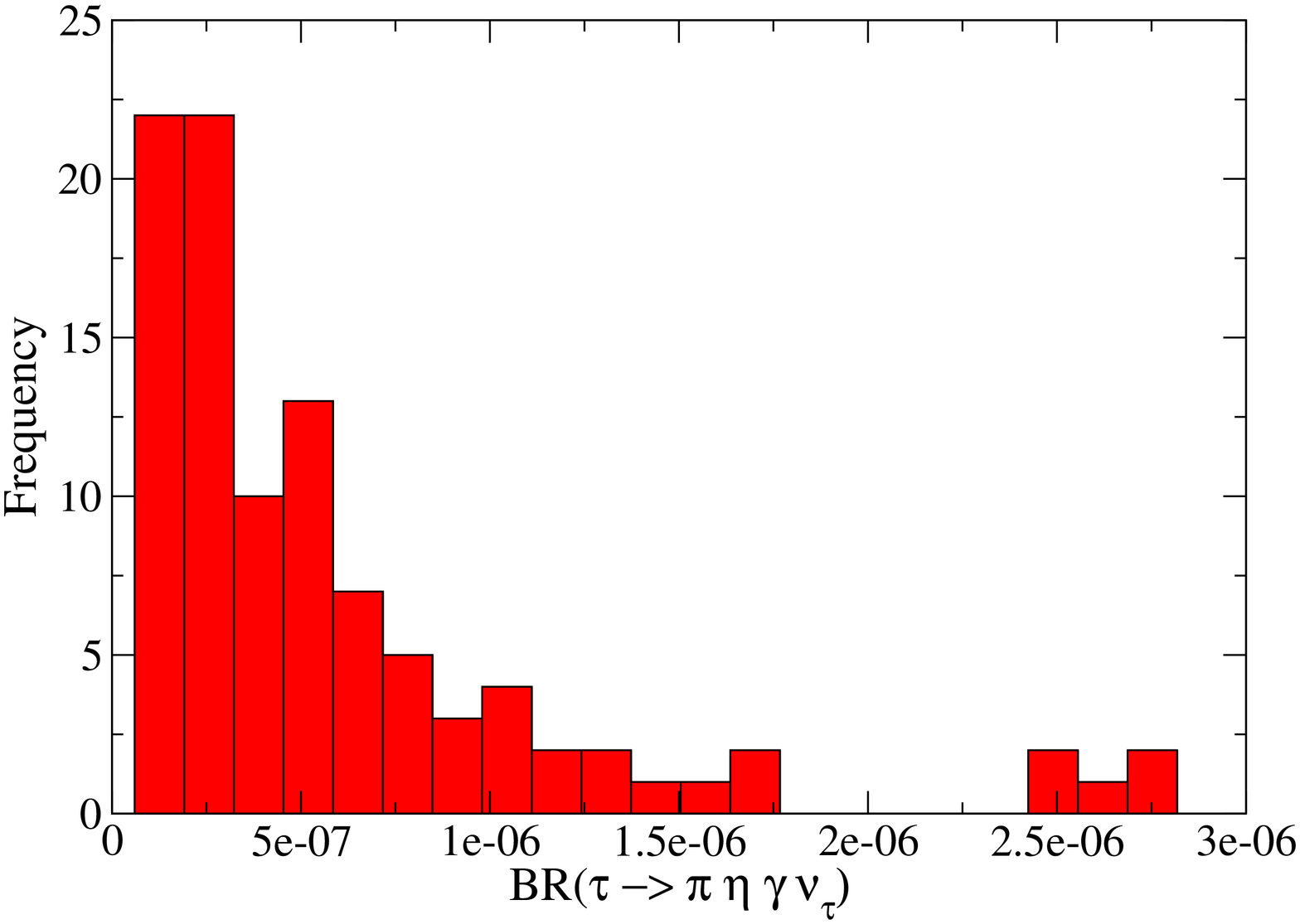}}
  \caption{$BR(\tau^-\to\eta\pi^-\nu_\tau \gamma)$ are represented as a function of the photon energy cut.}\label{fig:RChTcuts}
  \end{figure}

In figures \ref{fig:RChTcuts} we present the simulated branching fractions when photons above $100$ MeV are indeed cut (left plot), yielding $(0.44\pm0.06)\cdot10^{-5}$. If it was possible to cut above $50$ MeV photons, the branching ration would be shrinked 
to $(0.67\pm0.28)\cdot10^{-6}$ (right plot). Again, the quoted statistical errors have been obtained from the $1000$ data point sample, not shown in the figure. 
Vetoing photons with $E>100$ MeV should be able to reduce the number of background events to a 
fourth of the non-radiative decay, which 
should allow for a first detection of the $\tau^-\to\eta\pi^-\nu_\tau \gamma$ decays (supplemented by a phase-space discriminator, if needed). 

\vspace*{1.5cm}
\vspace*{1.5cm}

This conclusion does not change when we neglect the contributions from two-resonance mediated diagrams. The branching ratios obtained when cutting photons with with $E_\gamma > 50(100)$ MeV change from the previous values $(0.44\pm0.06)\cdot10^{-5}$ 
$((0.67\pm0.28)\cdot10^{-6})$ to 
$(0.30\pm0.04)\cdot10^{-5}$ $((0.45\pm0.16)\cdot10^{-6})$.\\ \\

\begin{figure}[ht!]
  \subfloat[All contributions included.]{\includegraphics[scale=0.3,angle=-90]{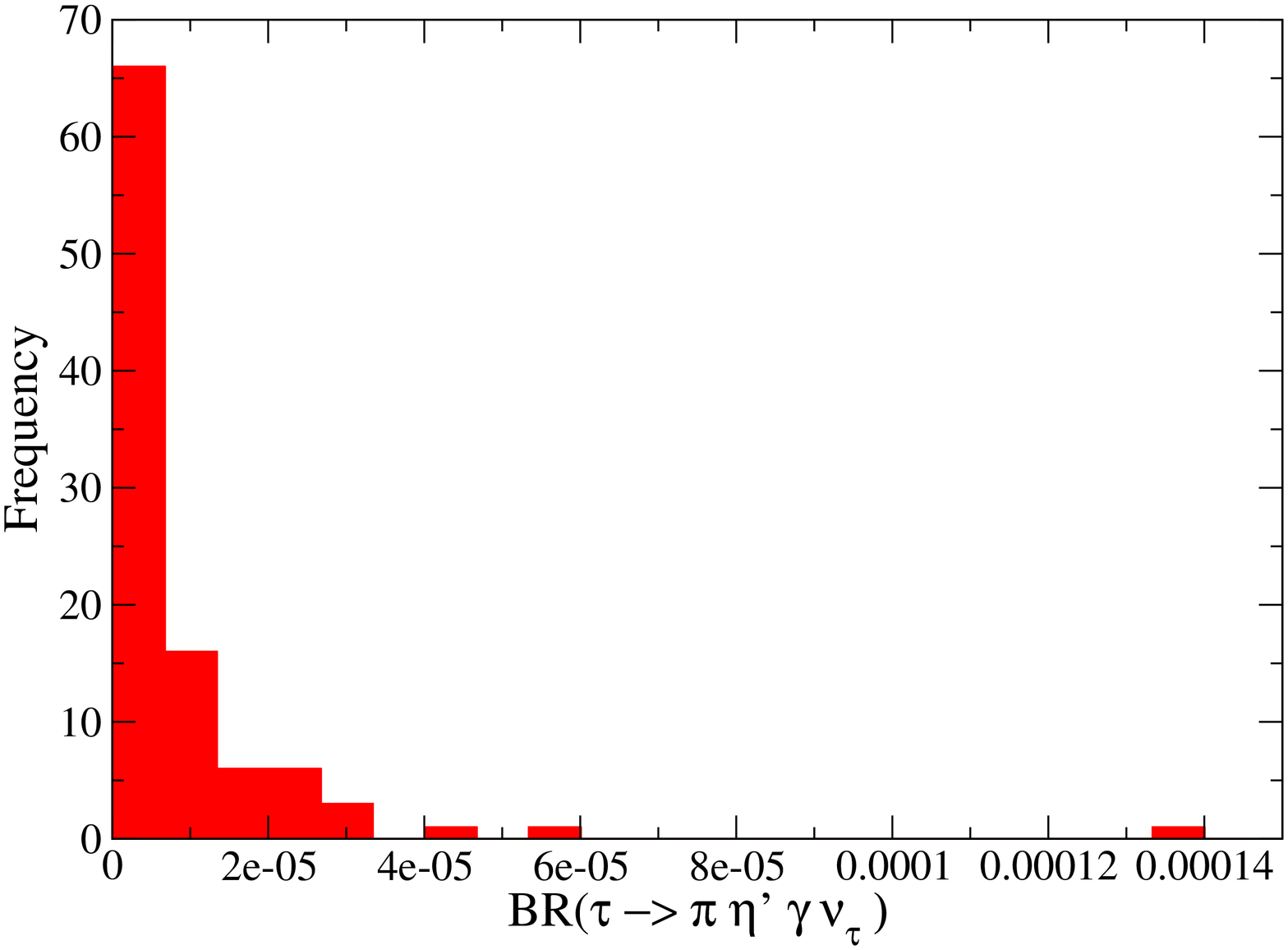}}
  \hspace*{0.2cm}
  \subfloat[Two-resonance contributions neglected.]{\includegraphics[scale=0.3,angle=-90]{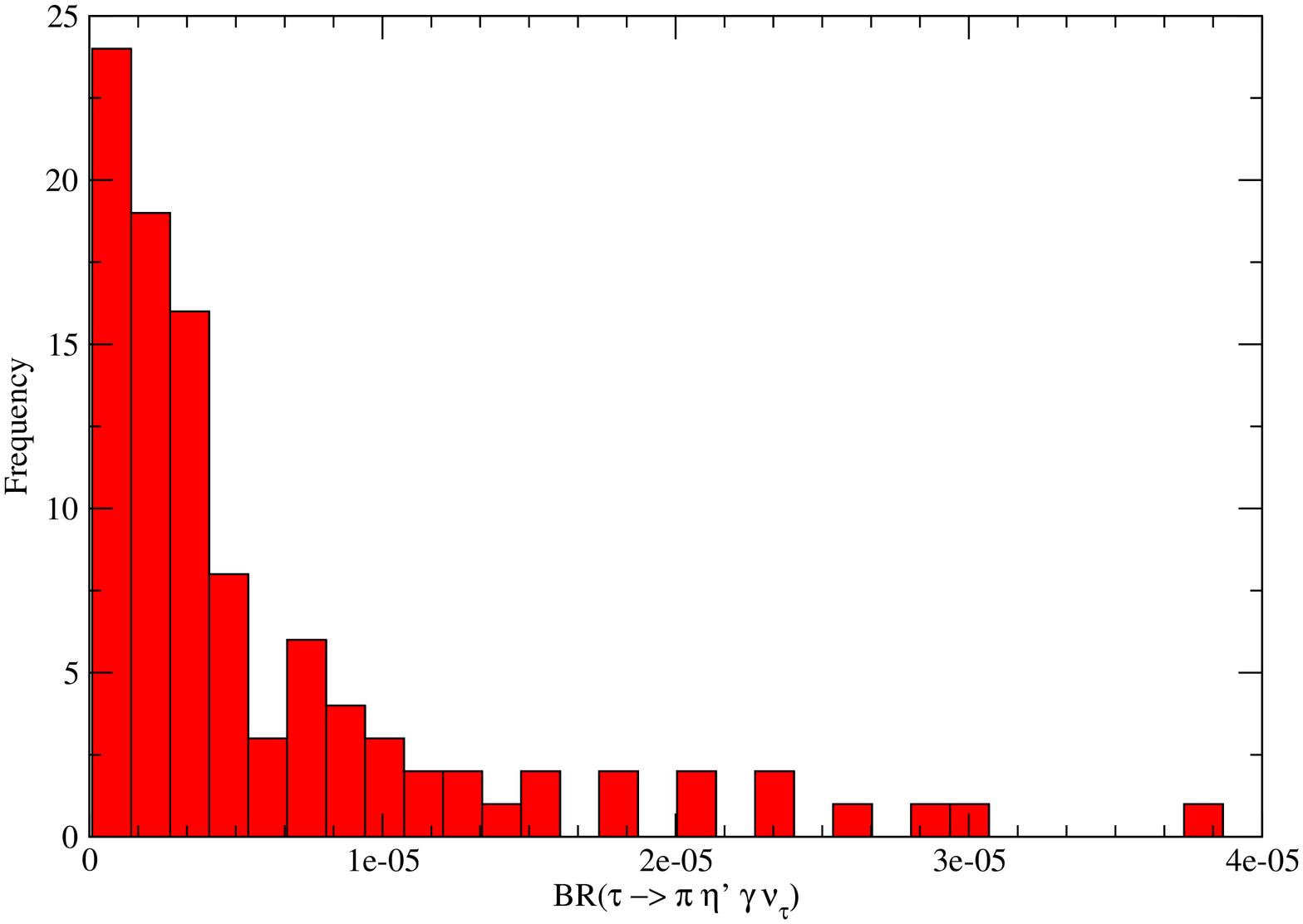}}
  \caption{Predictions for the $\tau^-\to\eta^\prime\pi^-\nu_\tau \gamma$ decays branching ratios: $100$ normally sampled points in the $R\chi L$ parameter space are plotted.}\label{fig:RChT100p}
  \end{figure}

 In figure \ref{fig:RChT100p} we show the plot analogous to fig.~\ref{fig:RChT100}, but for the $\eta^\prime$ mode. Using 100 sample data points, the predicted mean branching fraction is $(0.9\pm0.4)\cdot10^{-5}$, which is 
larger than the non-radiative decay. Once again, this feature remains when neglecting the two-resonance contributions, yielding $(0.7\pm0.3)\cdot10^{-5}$. If we now enlarge our sampling to
 $1000$ data points, our uncertainties become theory dominated. The corresponding results are $(0.84\pm0.06)\cdot10^{-5}$ (all contributions) and $(0.65\pm0.05)\cdot10^{-6}$ (without two-resonance contributions). 
 As we noted for the $\eta$ channel, the results of $R\chi L$ are noticeably larger than those of VMD (typically two orders of magnitude for the $\eta'$ channel). We again understand partly this discrepancy from the fact that, 
having so many parameters normally sampled for $R\chi L$, it becomes rather probable to have enough of them outside the one-sigma error band so as to increase substantially the predictions for the observables. In the $\eta'$ channel, 
however, the most of this difference comes from the richer dynamics of $R\chi L$ with respect to $MDM$, as confirmed by our earlier simulations~\cite{v1}.  
We compare the results corresponding to fig.~\ref{fig:RChT100p} (a) (all contributions included) to the case where the two-resonance contributions are neglected (figure \ref{fig:RChT100p} (b)). The predicted branching ratios remain basically 
constant, as just quoted.\\ \\

\vspace*{1.5cm}
\vspace*{1.5cm}

\begin{figure}[ht!]
  \subfloat[Normalized spectrum in the invariant mass of the $\eta^\prime\pi^-$ system (corresponding to the points in fig.~\ref{fig:RChT100p}) is plotted.]{\includegraphics[scale=0.3,angle=-90]{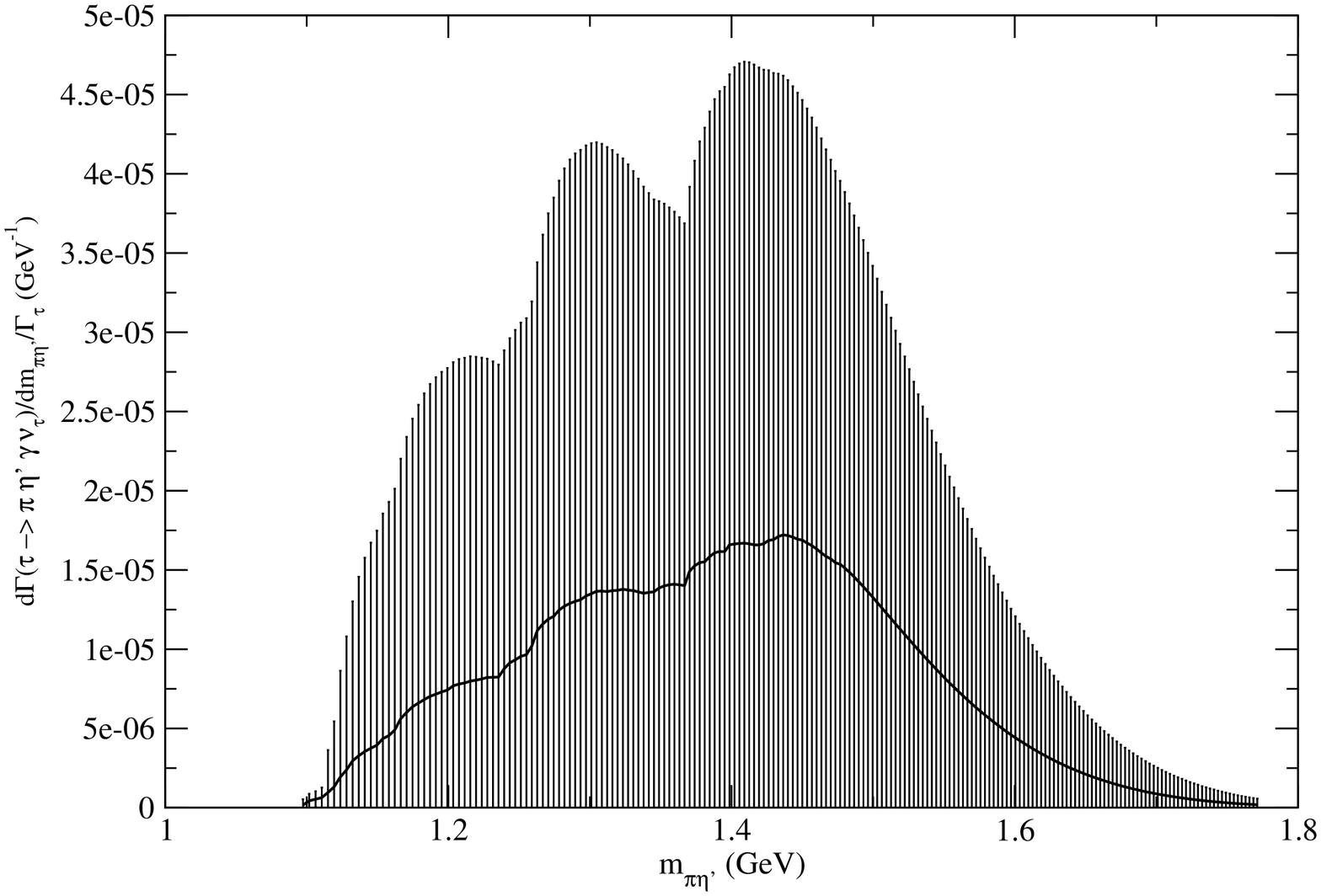}}
  \hspace*{0.2cm}
  \subfloat[For the same points as in (a), the normalized spectra in $E_\gamma$ are drawn.]{\includegraphics[scale=0.3,angle=-90]{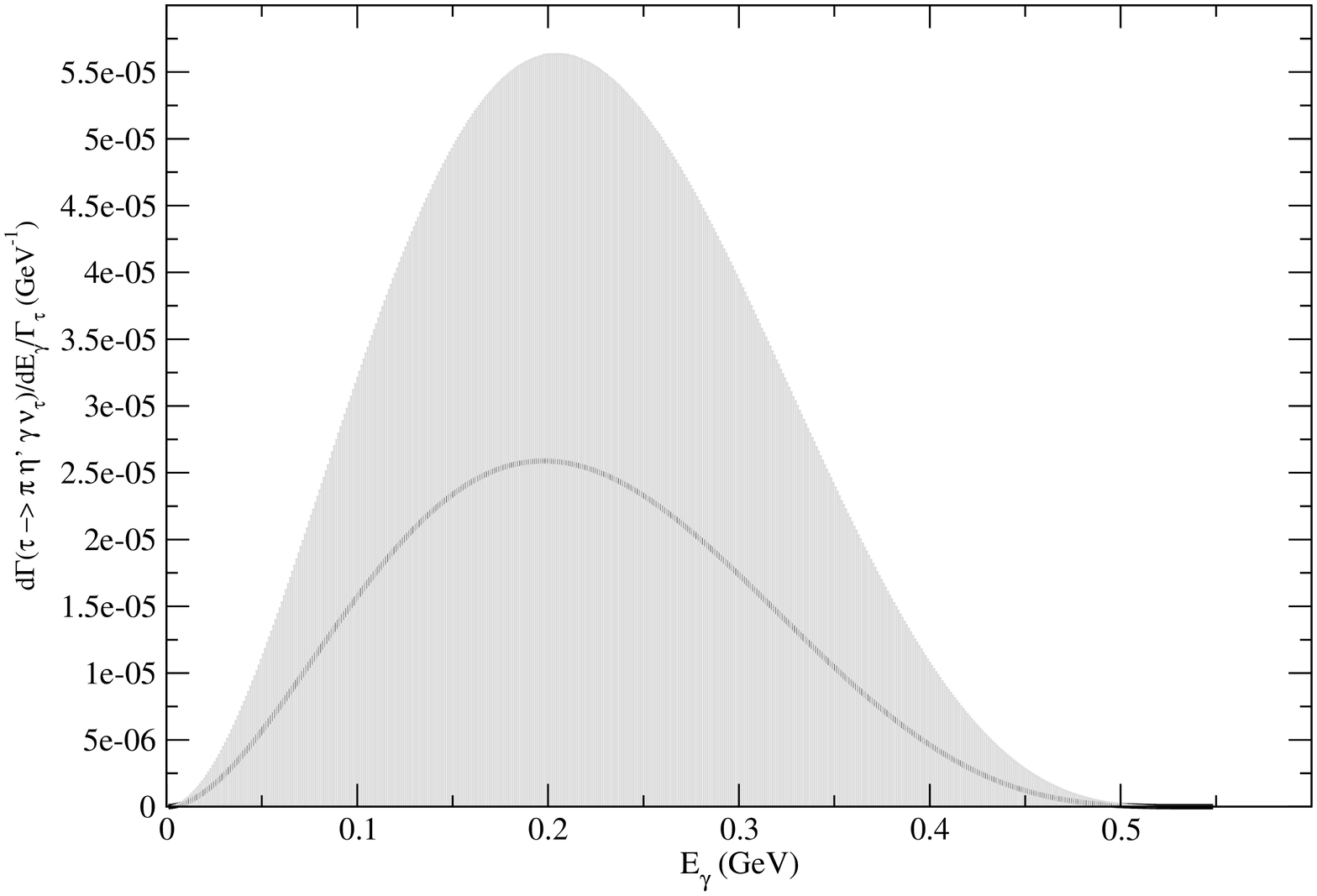}}
  \caption{Normalized spectra of the $\tau^-\to\eta^\prime\pi^-\nu_\tau \gamma$ decays according to $R\chi L$.}\label{fig:RChTSpectrap}
  \end{figure}

 In figs.~\ref{fig:RChTSpectrap} we show the normalized spectra of the $\tau^-\to\eta^\prime\pi^-\nu_\tau \gamma$ decays versus the meson system invariant mass (a) and the photon energy (b), with 200 and 500 data points, respectively. In this case 
for the $\eta^\prime\pi$ invariant mass distribution, a maximum is expected in the region $1.30-1.45$ GeV. 
The photon energy spectra 
suggests an $\mathcal{O}(100)$ MeV cut 
on $E_\gamma$ that we consider in the following. Again, we point out that the spectra change only very mildly when 
 neglecting the two-resonant contributions.\\ \\

 \vspace*{3.0cm}
  \vspace*{2.0cm}

\begin{figure}[ht!]
  \subfloat[When cutting photons with $E_\gamma>100$ MeV.]{\includegraphics[scale=0.3,angle=-90]{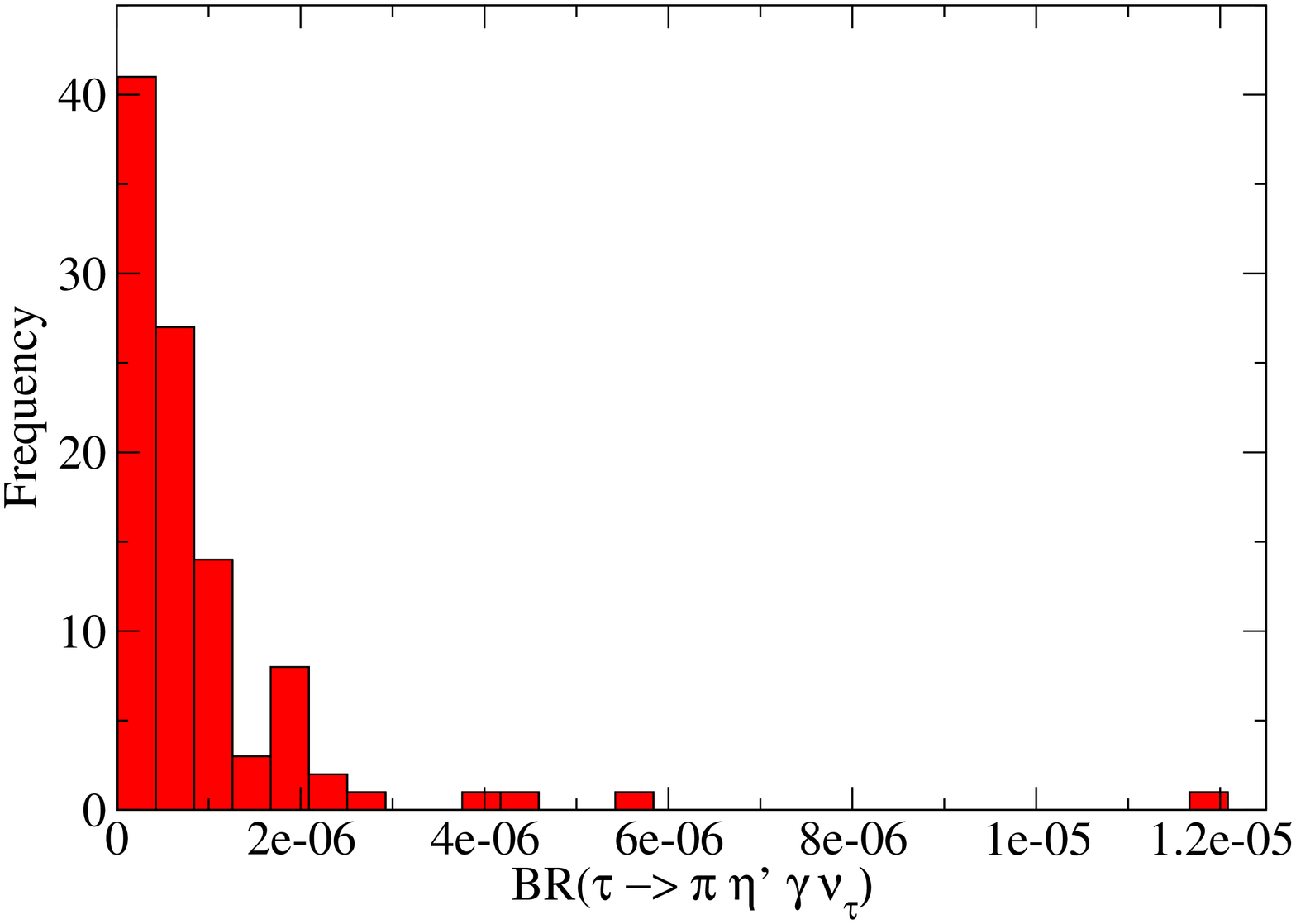}}
  \hspace*{0.2cm}
  \subfloat[When cutting photons with $E_\gamma>50$ MeV.]{\includegraphics[scale=0.3,angle=-90]{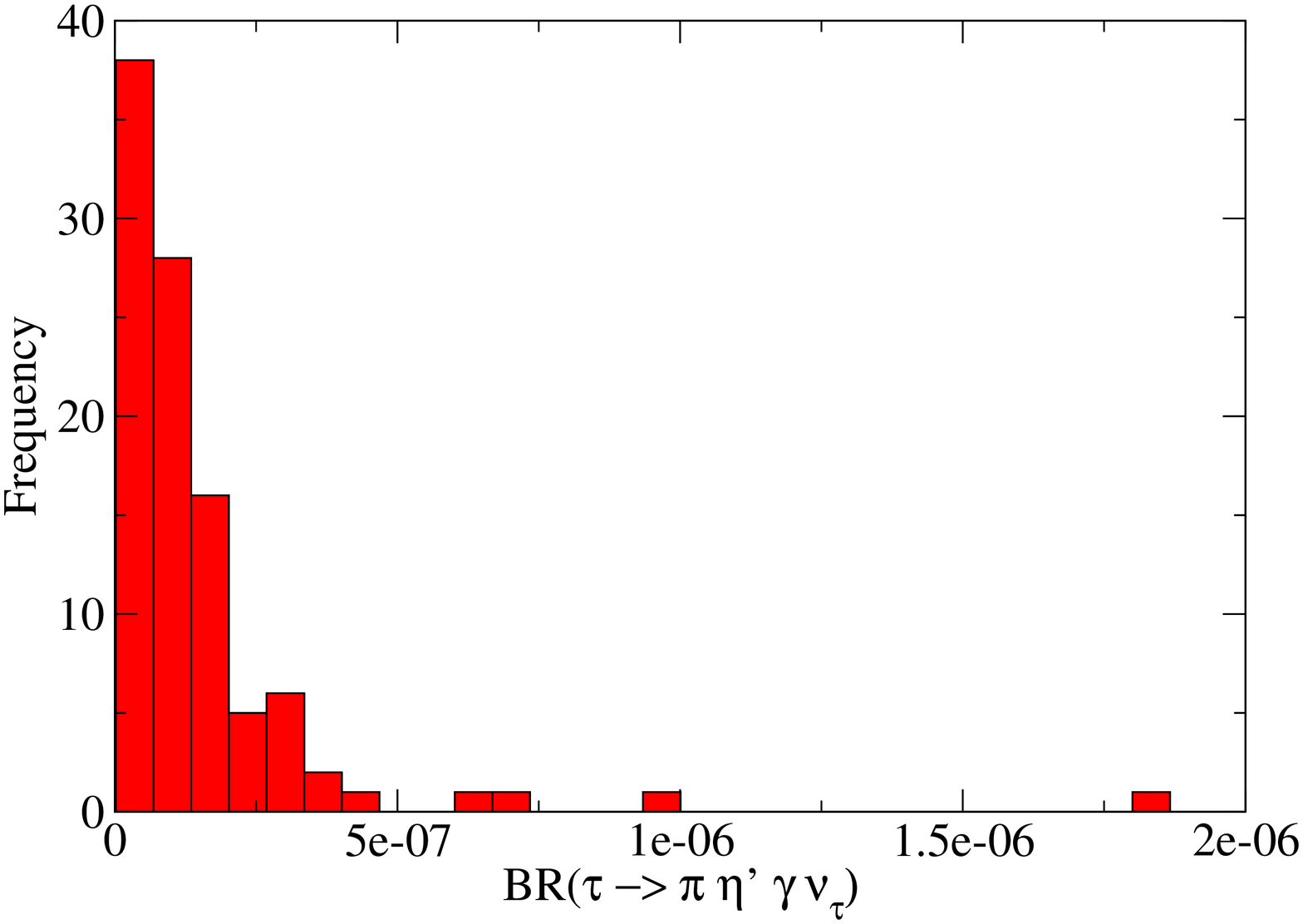}}
  \caption{$BR(\tau^-\to\eta\pi^-\nu_\tau \gamma)$ are represented as a function of the photon energy cut.}\label{fig:RChTpcuts}
  \end{figure}

The branching ratio into the $\pi^-\eta^\prime\gamma\nu_\tau$ decay channel, when cutting photons above $100$ MeV are $(0.9\pm0.2)\cdot10^{-6}$. It is reduced to $(1.5\pm1.5)\cdot10^{-7}$ when the cut is established from $50$ MeV on. 
These results are depicted in figure \ref{fig:RChTpcuts}. We recall that the 
systematic errors have been obtained from the $1000$ points data sample, not shown in the figure. We could reduce the errors by an extended sampling but this is not needed. It is already evident that if the corresponding non-radiative decay has a 
branching fraction which is close to 
minimum 
of the predicted interval, then one would require to use the different event topology of the three- and four-body decays to get rid of this background, which again looks feasible in the Belle-II environment (even if the branching ratio lies close to 
the predicted upper limit, using this additional information would be needed). Once more, these conclusions also 
apply when the two-resonance contributions are included, because the previous numbers barely change to $(0.7\pm0.2)\cdot10^{-6}$ and $(1\pm1)\cdot10^{-7}$, respectively.\\ \\

\vspace*{1.5cm}
\vspace*{1.5cm}

\section{Conclusions and outlook}\label{sec:Concl}
Induced SCC remain as yet undetected suppressed effects within the SM. In nuclear physics, the difficulty in splitting their signatures from ordinary CVC violation makes semileptonic tau decays at Belle-II 
the most promising arena for their discovery in an era of precision tau physics \cite{Pich:2013lsa}, where eventual departures of the corresponding rates from the expectations coming from $G$-parity violation may signal 
new physics providing genuine SCC. Actually, current upper limits \cite{PDG} on SCC searches lie close to the expected predictions according to isospin violating effects in the SM. With this motivation in mind, a number 
of theory papers and experimental analyses have been conducted in the last years in an effort that promises to continue with the start of Belle-II data taking.\\

In this paper, we point out for the first time the importance of the $\tau^-\to\pi^-\eta^{(\prime)}\nu_\tau\gamma$ decays as backgrounds in these searches. Within the framework of Resonance Chiral Lagrangians, we have found that 
their corresponding branching 
ratios are comparable to those of the non-radiative decays (in agreement with the expectations from $G$-parity violation as compared to electromagnetic suppression). Our main conclusion is that cutting photons above a 
realistic energy $E_\gamma\gtrsim 100$ MeV (leaving small windows for detecting $\pi^0$ and $\eta^{(\prime)}$ decays involving photons) should get rid of this background in the searches for SCC in $\tau^-\to\pi^-\eta\nu_\tau$ 
decays. On the other hand, given the theory errors in predicting $BR(\tau^-\to\pi^-\eta^\prime\nu_\tau)$, it is unclear if a feasible cut on photon energy will be able to eliminate this background. In that case, however, rejection appears 
possible taking advantage of the different kinematics of the three- (signal) and four-body (background) decays. Our most important results are summarized in table \ref{results}.\\

\begin{table}[h!]
 \centering
   \begin{tabular}{|c|c|c|c|c|} \hline
    SCC bkg & BR (no cuts) & BR ($E_\gamma^{\mathrm{cut}}>100$ MeV) & BR SCC signal & Bkg rejection\\ \hline
    $\tau^-\to\pi^-\eta\gamma\nu_\tau$ & $(1.0\pm0.3)\cdot10^{-4}$ & $(0.4\pm0.1)\cdot10^{-5}$ & $\sim1.7\cdot10^{-5}$ & Yes \\ \hline
    $\tau^-\to\pi^-\eta^\prime\gamma\nu_\tau$ & $(0.8\pm0.2)\cdot10^{-5}$ & $(0.9\pm0.3)\cdot10^{-6}$ & $[10^{-7},10^{-6}]$ & No \\ \hline
   \end{tabular}
 \caption{The main conclusions of our analysis are summarized: Our predicted branching ratios for the $\tau^-\to\pi^-\eta^{(\prime)}\gamma\nu_\tau$ decays and the corresponding results when the cut $E_\gamma>100$ MeV is 
    applied. We also compare the latter results to the prediction for the corresponding non-radiative decay (SCC signal) according to Ref.~\cite{Escribano:2016ntp} and conclude if this cut 
    is able to get rid of the 
    corresponding background in SCC searches.}
   \label{results}
\end{table}

It is also interesting to note our finding that, within the $R\chi L$ frame, a simplified description of these decays neglecting the two-resonance mediated contributions is a good approximation for branching ratios and decay spectra, 
which will ease the coding of the corresponding form factors in the Monte Carlo generators. Finally, in $\tau^-\to\pi^-\eta\gamma\nu_\tau$ decays, we do not find any signature corresponding to the $a_0(980)$ meson in the $\eta\pi$ invariant mass 
distribution. Therefore, an observation of such structure in the corresponding non-radiative decay would be in accord with the prediction of Ref.~\cite{Descotes-Genon:2014tla} and disagree with the one in Ref.~\cite{Escribano:2016ntp}. On 
the contrary, the sharp peak predicted in the same spectrum at $\sim 1.4$ GeV \cite{Escribano:2016ntp} should be a distinctive feature of a dynamically generated scalar resonance prominent contribution in the $\tau^-\to\pi^-\eta\nu_\tau$ decays, 
testable with early data.\\

 \section*{Acknowledgements}
We have benefited from discussions on this topic with Jean Pestieau, Jorge Portol\'es and Juanjo Sanz-Cillero. G.~L.~C.~ acknowledges the kind hospitality of IFIC (Valencia), where part of this work was done. P.~R.~ is indebted to Simon Eidelman, 
Kiyoshi Hayasaka and Benjamin Oberhof for clarifications concerning photon detection/rejection at Belle-II and discussions on other important backgrounds for SCC searches in $\tau^-\to\pi^-\eta^{(\prime)}\nu_\tau$ decays. Financial support from 
projects 296 ('Fronteras de la Ciencia'), 236394, 250628 ('Ciencia B\'asica') and SNI (Conacyt, Mexico) is acknowledged.

\section*{Appendix A: Inner bremsstrahlung contributions}
In this appendix we check that inner bremsstrahlung contributions can indeed be neglected in our study. As we argued in the introduction, radiation off the external lines will be doubly suppressed: by $\alpha$ (as it corresponds to the $\gamma$ emission) 
and by G-parity violation (as it happens with the non-radiative decay). Of course, this will no longer be true if photons with extremely low energy are considered because of the well-known infrared singularity (see for example
section 7 of Ref.~\cite{Denner:1991kt}). In order to study this question a threshold energy for photon detection ($E_{thr}$) needs to be specified. We consider that $10$ MeV is a realistic value for it in a B-factory. In this way, photons with 
$E_\gamma < E_{\mathrm{thr}}$ will not be resolved and will be included in the non-radiative decay rate (inclusive in low-energy photons). We want to quantify the impact of detected inner bremsstrahlung photons in the radiative decay rates.\\

According to Low's theorem, the expansion of the radiative amplitude at low photon energies ($E_\gamma=k$) reads
\begin{equation}
 \mathcal{M}_\gamma \, = \, \frac{A}{k}+B+\mathcal{O}(k)\,,\;
\end{equation}
where $A$ and $B$ are given in terms of the non-radiative amplitude, $\mathcal{M}_0$. In fact, one has
\begin{equation}
 \mathcal{M}_\gamma \, = \, -e \mathcal{M}_0 \left(\frac{P\cdot \epsilon}{P\cdot k}-\frac{p \cdot \epsilon}{p\cdot k}\right)+...
\end{equation}
In the previous equation, $P(p)$ are the momenta of the charged particles $\tau^-(\pi^-)$ and only the coupling to the electric charge is given (higher electromagnetic multipoles and $\mathcal{O}(k^0)$ terms are to be understood in '...' and are 
neglected since they will be subleading in the infrared limit). In this approximation, one can estimate the leading Low contribution to the radiative decay as (a bar over the matrix element stands for sum over polarizations)
\begin{equation}\label{eq:LeadingLow}
 |\overline{\mathcal{M}_\gamma}|^2 \, = \, e^2 |\overline{\mathcal{M}_0}|^2 \sum_{\gamma \;\mathrm{pols.}} \Big|\frac{P\cdot \epsilon}{P\cdot k}-\frac{p \cdot \epsilon}{p\cdot k}\Big|^2\,,
\end{equation}
where $|\overline{\mathcal{M}_0}|^2$ has to be evaluated using the kinematics of the radiative decay.\\

Using the $\mathcal{M}_0$ worked out in Ref.~\cite{Escribano:2016ntp}, we have evaluated the leading Low contribution -as given by eq.~(\ref{eq:LeadingLow})- to the radiative decay rates. The corresponding spectra ($\eta$ and $\eta^\prime$ decay modes) are 
given in figs.~22. 
The respective branching ratios (for $E_\gamma > 10$ MeV) are $\sim 2.5\cdot 10^{-8}$ and $\sim 4.6\cdot 10^{-12}$, respectively. This suppression is larger than it could be expected according only to the $\alpha$ and isospin suppressions mentioned at the 
beginning. The additional suppression comes from the fact that the scalar form factors are very peaked around $m_{\eta^{(\prime)}\pi}\sim1.4$ GeV~\cite{Escribano:2016ntp}, which dilutes the effect of the 
$\Big|\frac{P\cdot \epsilon}{P\cdot k}-\frac{p \cdot \epsilon}{p\cdot k}\Big|^2$ factor increasing for low photon energies. Moreover, in the case of the $\eta^\prime$ decay channel, also the vector form factor is very suppressed, as the $\rho(770)$ 
contribution is below the kinematical threshold for the $\eta^\prime \pi^-$ form factor. As a result of this, we see the characteristic damping of the inner-bremsstrahlung spectra corresponding to a very smooth variation of the integrated effect of the 
meson form factors.

\vspace*{1.5cm}
  \begin{figure}[ht!]
  \subfloat[$\eta$ channel.]{\includegraphics[scale=0.3,angle=-90]{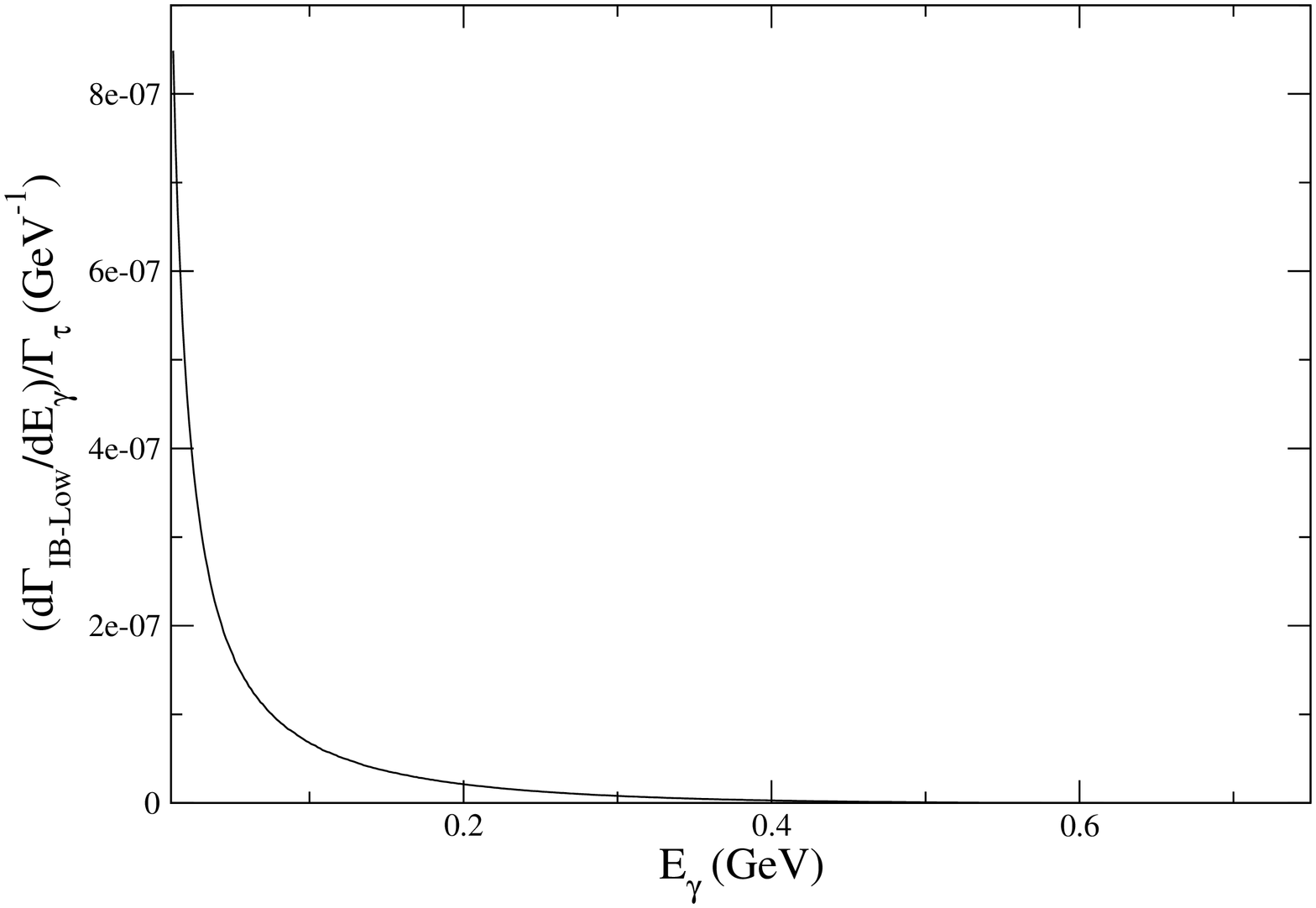}}
  \hspace*{0.2cm}
  \subfloat[$\eta^\prime$ channel.]{\includegraphics[scale=0.3,angle=-90]{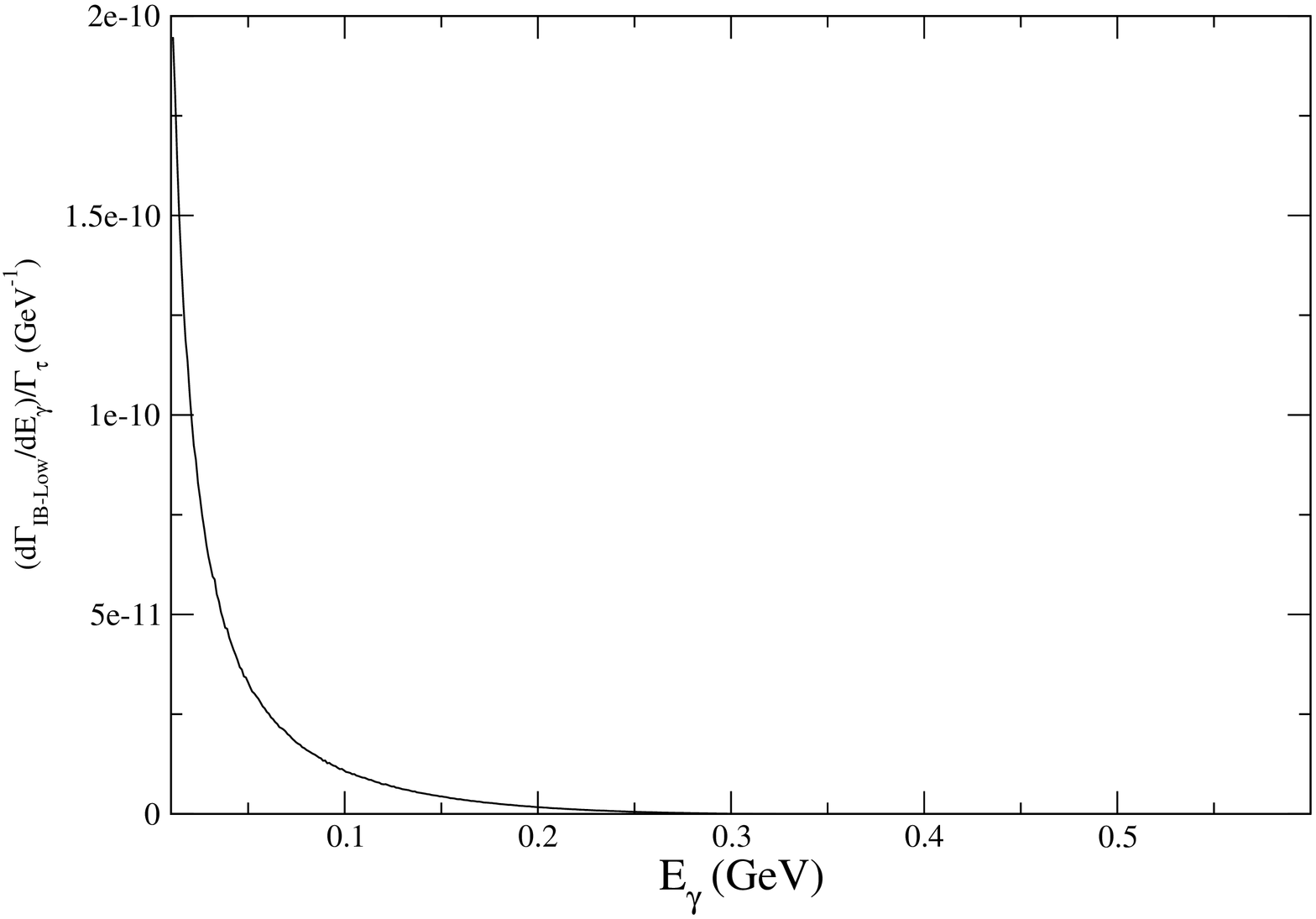}}
  \caption{The normalized photon spectra of the leading Low contributions to the $BR(\tau^-\to\eta^{(\prime)}\pi^-\nu_\tau \gamma)$ are plotted.}
  \end{figure}\label{fig:LeadingLow}
\vspace*{1.5cm}

\section*{Appendix B: Form factors results according to Resonance Chiral Lagrangians}
\footnotesize{
In this appendix we include the different contributions to the (axial-)vector form factors obtained using $R\chi L$. Only the anomalous contribution was included in section \ref{sec:FFsRChT}. Here we explicitly quote the analytic expressions for the 
model-dependent (resonant-mediated) contributions to these form factors following the order in the figures. We start with fig.~\ref{fig:1R-FA}, giving rise to $a_{i=1,2,3,4}^{1R}$ in $R\chi L$:\\

\begin{multline}\label{a1-1R-RChL}
 a_1^{1R}=
   -\frac{4C_q}{F^2M_Vm_\rho^2D_\rho[(p_0+k)^2]}\left(k\cdot p_0 \left(\left(F_V-2 G_V\right) \left(-\left(8 c_3+2
   c_5+3 c_7\right) m_{\eta }^2+8 c_3 m_{\pi }^2+2 \left(c_5+c_7\right)
   p\cdot p_0\right)\right.\right.\\\left.
   +m_{\rho }^2 \left(\left(c_7+c_{1256}\right) F_V-2
   c_7 G_V\right)\right)-\frac{1}{2} \left(F_V-2 G_V\right) k\cdot p
   \left(4 c_5 k\cdot p_0+2 \left(c_5-4 c_3\right) m_{\eta }^2-\left(2
   c_5+c_{1256}\right) m_{\rho }^2+8 c_3 m_{\pi }^2\right)\\
   -2 c_7
   \left(F_V-2 G_V\right) \left(k\cdot p_0\right){}^2+\frac{1}{2}
   m_{\rho }^2 \left(m_{\eta }^2 \left(\left(2 \left(8
   c_3+c_5+c_7\right)+c_{1256}\right) F_V-4 \left(4 c_3+c_5+c_7\right)
   G_V\right)\right.\\\left.\left.
   +16 c_3 m_{\pi }^2 \left(G_V-F_V\right)-\left(2
   \left(c_5+c_7\right)-c_{1256}\right) p\cdot p_0 \left(F_V-2
   G_V\right)\right)\right.\\\left.+\left(\left(4 c_3+c_5+c_7\right) m_{\eta }^2-4 c_3
   m_{\pi }^2\right) \left(F_V-2 G_V\right) \left(p\cdot p_0-m_{\eta
   }^2\right)\right)\\
 +\frac{4F_A}{F^2D_{a_1}[(p+p_0+k)^2]}\left(-C_q \left(-\left(\kappa _5^A+\kappa _6^A-\kappa
   _7^A+\kappa _{16}^A\right) \left(k\cdot p+m_\pi^2\right)-\left(\kappa
   _3^A+2 \kappa _8^A+\kappa _{15}^A\right) k\cdot p_0\right.\right.\\\left.
+2 m_{\pi }^2
   \left(2 \kappa _9^A-\kappa _{10}^A+2 \kappa _{11}^A+\kappa
   _{14}^A\right)+\left(-3 \kappa _3^A+4 \kappa _4^A+\kappa _5^A+\kappa
   _6^A+\kappa _7^A-2 \kappa _8^A-\kappa _{15}^A\right) p\cdot
   p_0\right)\\
   -C_q \left(-\left(\kappa _5^A+\kappa _6^A+\kappa
   _7^A\right) k\cdot p_0+2 m_{\pi }^2 \left(\kappa _{14}^A+2
   \left(\kappa _9^A+\kappa _{11}^A\right)\right)+m_\eta^2 \left(\kappa
   _3^A-\kappa _5^A-\kappa _6^A-\kappa _7^A+2 \kappa _8^A+\kappa
   _{15}^A\right)\right.\\\left.\left.+\left(-2 \kappa _3^A+4 \kappa _4^A+\kappa _5^A+\kappa
   _6^A-\kappa _7^A+\kappa _{16}^A\right) p\cdot p_0\right)-2 \sqrt{2}
   \kappa _9^A C_s \left(m_{\pi }^2-2 m_K^2\right)+\sqrt{2} \left(2
   \kappa _9^A-\kappa _{10}^A\right) C_s \left(2 m_K^2-m_{\pi
   }^2\right)\right)\\
   -\frac{4F_A}{F^2m_{a_1}^2D_{a_1}[(p+k)^2]}\left(\left(m_{a_1}^2-k\cdot p\right) \left(C_q \left(\kappa
   _{16}^A \left(2 k\cdot p+m_\pi^2\right)+\kappa _{16}^A \left(2
   \left(k\cdot p_0+p\cdot p_0\right)+m_\eta^2\right)\right.\right.\right.\\\left.\left.-\left(\kappa _3^A+2
   \kappa _8^A+\kappa _{15}^A\right) \left(k\cdot p_0+m_\eta^2+p\cdot
   p_0\right)-2 m_{\pi }^2 \left(2 \kappa _9^A+\kappa _{10}^A+2 \kappa
   _{11}^A+\kappa _{12}^A\right)\right)+\sqrt{2} \left(2 \kappa
   _9^A+\kappa _{10}^A\right) C_s \left(2 m_K^2-m_{\pi
   }^2\right)\right)\\
   +\sqrt{2} C_s \left(2 m_K^2-m_{\pi }^2\right)
   \left(-2 \kappa _9^A \left(-m_{a_1}^2+2 k\cdot p+m_\pi^2+p\cdot
   p_0\right)-\kappa _{10}^A p\cdot p_0\right)\\
   +C_q \left(p\cdot p_0
   \left(-m_{a_1}^2 \left(\kappa _3^A+\kappa
   _{15}^A\right)+\left(\kappa _3^A+\kappa _{15}^A-\kappa
   _{16}^A\right) \left(2 k\cdot p+m_\pi^2\right)+2 m_{\pi }^2 \left(2
   \kappa _9^A+\kappa _{10}^A+2 \kappa _{11}^A+\kappa
   _{12}^A\right)\right.\right.\\\left.
   +m_\eta^2 \left(\kappa _3^A+2 \kappa _8^A+\kappa
   _{15}^A-\kappa _{16}^A\right)\right)-\left(4 m_{\pi }^2 \left(\kappa
   _9^A+\kappa _{11}^A\right)+m_\eta^2 \left(\kappa _3^A+\kappa
   _{15}^A\right)\right) \left(m_{a_1}^2-2 k\cdot p-m_\pi^2\right)\\\left.\left.
   +k\cdot
   p_0 \left(\left(\kappa _3^A+\kappa _{15}^A\right) \left(-m_{a_1}^2+2
   k\cdot p+m_\pi^2\right)+\left(\kappa _3^A+2 \kappa _8^A+\kappa _{15}^A-2
   \kappa _{16}^A\right) p\cdot p_0\right)+\left(\kappa _3^A+2 \kappa
   _8^A+\kappa _{15}^A-2 \kappa _{16}^A\right) \left(p\cdot
   p_0\right){}^2\right)\right)\\
  -\frac{4F_VC_q}{F^2 m_\rho^2}\left(\frac{2 \sqrt{2} C_s \left(2 m_K^2-m_{\pi }^2\right)
   \kappa _{13}^V}{C_q}+\frac{2 \sqrt{2} C_s \left(2 m_K^2-m_{\pi
   }^2\right) \left(\kappa _{13}^V+\kappa
   _{18}^V\right)}{C_q}\right.\\
   -\left(\kappa _1^V+\kappa _2^V+\kappa
   _3^V+\kappa _6^V+\kappa _7^V+\kappa _8^V\right) k\cdot
   p_0-\left(\kappa _6^V+\kappa _8^V+2 \kappa _{12}^V+\kappa _{16}^V-2
   \kappa _{17}^V\right) k\cdot p_0\\
   +\left(-\kappa _1^V+\kappa
   _2^V-\kappa _3^V+\kappa _7^V+\kappa _{17}^V\right) k\cdot p-m_{\eta
   }^2 \kappa _1^V-m_{\pi }^2 \left(\kappa _1^V+2 \left(-\kappa
   _4^V+\kappa _{10}^V+2 \left(\kappa _{13}^V+\kappa _{14}^V+\kappa
   _{15}^V\right)\right)\right)\\
   -2 m_{\pi }^2 \left(-\kappa _4^V+2
   \kappa _9^V+\kappa _{10}^V+2 \left(\kappa _{13}^V+\kappa
   _{14}^V+\kappa _{15}^V+\kappa _{18}^V\right)\right)+m_\pi^2 \left(\kappa
   _2^V-\kappa _3^V+\kappa _7^V\right)-m_\eta^2 \left(\kappa _2^V+\kappa
   _3^V+\kappa _6^V+\kappa _7^V+\kappa _8^V\right)\\
   -m_\eta^2 \left(\kappa
   _6^V+\kappa _8^V+2 \kappa _{12}^V+\kappa _{16}^V-\kappa
   _{17}^V\right)+p\cdot p_0 \kappa _1^V+p\cdot p_0 \left(\kappa
   _1^V+\kappa _2^V+\kappa _3^V-4 \kappa _5^V-\kappa _6^V-\kappa
   _7^V-\kappa _8^V\right)\\\left.
   -p\cdot p_0 \left(-3 \kappa _2^V+3 \kappa
   _3^V+4 \kappa _5^V+\kappa _6^V-\kappa _7^V+\kappa _8^V+2 \kappa
   _{12}^V+\kappa _{16}^V-\kappa _{17}^V\right)\frac{}{}\right)\,,
   \\
\end{multline}
where $c_{1256}\equiv c_1-c_2-c_5+2c_6$ was used. Its value is fixed by eqs.~(─\ref{eq: Consistent set of relations}).\\

\begin{multline}\label{a2-1R-RChL}
 a_2^{1R}=
  \frac{4C_q\left(F_V-2 G_V\right)}{F^2M_Vm_\rho^2D_\rho[(p_0+k)^2]} \left(c_7 k\cdot p_0+\left(4
   c_3+c_5+c_7\right) m_{\eta }^2-4 c_3 m_{\pi }^2\right) \left(-2
   k\cdot p_0-m_{\eta }^2+m_{\rho }^2\right)\\
 +\frac{4F_A}{F^2D_{a_1}[(p+p_0+k)^2]}\left(C_q \left(-\left(\kappa _5^A+\kappa _6^A+\kappa
   _7^A\right) k\cdot p_0+2 m_{\pi }^2 \left(\kappa _{14}^A+2
   \left(\kappa _9^A+\kappa _{11}^A\right)\right)\right.\right.\\\left.\left.+m_\eta^2 \left(\kappa
   _3^A-\kappa _5^A-\kappa _6^A-\kappa _7^A+2 \kappa _8^A+\kappa
   _{15}^A\right)+\left(-2 \kappa _3^A+4 \kappa _4^A+\kappa _5^A+\kappa
   _6^A-\kappa _7^A+\kappa _{16}^A\right) p\cdot p_0\right)+2 \sqrt{2}
   \kappa _9^A C_s \left(m_{\pi }^2-2 m_K^2\right)\right)\\
   +\frac{4F_A}{F^2m_{a_1}^2D_{a_1}[(p+k)^2]}\left(\sqrt{2} C_s \left(2 m_K^2-m_{\pi }^2\right) \left(2 \kappa _9^A \left(m_{a_1}^2-2 k\cdot
   p-m_\pi^2\right)-\left(2 \kappa _9^A+\kappa _{10}^A\right) p\cdot p_0\right)\right.\\
   +C_q \left(p\cdot p_0 \left(-m_{a_1}^2
   \left(\kappa _3^A+\kappa _{15}^A\right)+\left(\kappa _3^A+\kappa _{15}^A-\kappa _{16}^A\right) \left(2 k\cdot
   p+m_\pi^2\right)+2 m_{\pi }^2 \left(2 \kappa _9^A+\kappa _{10}^A+2 \kappa _{11}^A+\kappa _{12}^A\right)\right.\right.\\\left.
   +m_\eta^2 \left(\kappa
   _3^A+2 \kappa _8^A+\kappa _{15}^A-\kappa _{16}^A\right)\right)-\left(4 m_{\pi }^2 \left(\kappa _9^A+\kappa
   _{11}^A\right)+m_\eta^2 \left(\kappa _3^A+\kappa _{15}^A\right)\right) \left(m_{a_1}^2-2 k\cdot p-m_\pi^2\right)\\\left.\left.
   +k\cdot p_0
   \left(\left(\kappa _3^A+2 \kappa _8^A+\kappa _{15}^A-2 \kappa _{16}^A\right) p\cdot p_0-\left(\kappa _3^A+\kappa
   _{15}^A\right) \left(m_{a_1}^2-2 k\cdot p-m_\pi^2\right)\right)+\left(\kappa _3^A+2 \kappa _8^A+\kappa _{15}^A-2 \kappa
   _{16}^A\right) \left(p\cdot p_0\right){}^2\right)\right)\\
  +\frac{4F_VC_q}{F^2 m_\rho^2}\left(\frac{2 \sqrt{2} C_s \left(2 m_K^2-m_{\pi }^2\right)
   \left(\kappa _{13}^V+\kappa _{18}^V\right)}{C_q}-\left(\kappa
   _1^V+\kappa _2^V+\kappa _3^V+\kappa _6^V+\kappa _7^V+\kappa
   _8^V\right) k\cdot p_0-m_{\eta }^2 \kappa _1^V\right.\\
   -2 m_{\pi }^2
   \left(-\kappa _4^V+2 \kappa _9^V+\kappa _{10}^V+2 \left(\kappa
   _{13}^V+\kappa _{14}^V+\kappa _{15}^V+\kappa
   _{18}^V\right)\right)-m_\eta^2 \left(\kappa _2^V+\kappa _3^V+\kappa
   _6^V+\kappa _7^V+\kappa _8^V\right)\\\left.
   +p\cdot p_0 \left(\kappa
   _1^V+\kappa _2^V+\kappa _3^V-4 \kappa _5^V-\kappa _6^V-\kappa
   _7^V-\kappa _8^V\right)\frac{}{}\right)\,,
  \\
\end{multline}

\begin{multline}\label{a3-1R-RChL}
 a_3^{1R}=
 \frac{16G_VC_q}{F^2M_V[m_\eta^2+2(k\cdot p + k\cdot p_0 + p\cdot p_0)]D_\rho[(p_0+k)^2]}\left[-(c_{1256}+8c_3)\frac{m_\eta^2}{2}+c_{1256} k\cdot p_0
 +4c_3m_\pi^2\right]\\
 +\frac{4F_A}{F^2m^2_{a_1}D_{a_1}[(p+p_0+k)^2]} \left(\kappa _{10}^A \left(\sqrt{2} C_s \left(2 m_K^2-m_{\pi
   }^2\right)-2 m_{\pi }^2 C_q\right)-m_{a_1}^2 \left(\kappa
   _5^A+\kappa _6^A-\kappa _7^A+\kappa _{16}^A\right)
   C_q\right)\\
   +\frac{4F_A}{F^2m_{a_1}^2D_{a_1}[(p+k)^2]}\left(C_q \left(\kappa _{16}^A \left(m_{a_1}^2+2 k\cdot
   p_0+m_\eta^2+2 p\cdot p_0\right)-\left(\kappa _3^A+2 \kappa _8^A+\kappa
   _{15}^A\right) \left(k\cdot p_0+m_\eta^2+p\cdot p_0\right)\right.\right.\\\left.\left.
   -2 m_{\pi }^2
   \left(2 \kappa _9^A+\kappa _{10}^A+2 \kappa _{11}^A+\kappa
   _{12}^A\right)\right)+\sqrt{2} \left(2 \kappa _9^A+\kappa
   _{10}^A\right) C_s \left(2 m_K^2-m_{\pi }^2\right)\right)\\
  +\frac{4F_VC_q}{F^2 m_\rho^2}\left(3 \kappa _1^V-3 \kappa _2^V+3 \kappa _3^V-\kappa
   _6^V+\kappa _7^V-\kappa _8^V\right)\,,
   \\
\end{multline}

\begin{multline}\label{a4-1R-RChL}
 a_4^{1R}= 
  \frac{4C_q\left(F_V-2 G_V\right) }{F^2M_Vm_\rho^2D_\rho[(p_0+k)^2]}\left(\left(c_5+c_7\right) \left(2
   k\cdot p_0+m_{\eta }^2\right)+4 c_3 \left(m_{\eta }^2-m_{\pi }^2\right)+\left(-c_5-c_7+\frac{c_{1256}}{2}\right) m_{\rho }^2\right)
   \\
 -\frac{4F_A C_q}{F^2D_{a_1}[(p+p_0+k)^2]}\left(\kappa _3^A-2 \kappa _5^A-2 \kappa _6^A+2 \kappa
   _8^A+\kappa _{15}^A-\kappa _{16}^A\right)\\
   -\frac{4F_A}{F^2m_{a_1}^2D_{a_1}[(p+k)^2]}\left(C_q \left(\kappa _{16}^A \left(m_{a_1}^2+2 k\cdot
   p_0+m_\eta^2+2 p\cdot p_0\right)-\left(\kappa _3^A+2 \kappa _8^A+\kappa
   _{15}^A\right) \left(k\cdot p_0+m_\eta^2+p\cdot p_0\right)\right.\right.\\\left.\left.-2 m_{\pi }^2
   \left(2 \kappa _9^A+\kappa _{10}^A+2 \kappa _{11}^A+\kappa
   _{12}^A\right)\right)+\sqrt{2} \left(2 \kappa _9^A+\kappa
   _{10}^A\right) C_s \left(2 m_K^2-m_{\pi }^2\right)-C_q \left(\kappa _{16}^A-2 \kappa _8^A\right)
   \left(m_{a_1}^2-2 k\cdot p-m_\pi^2\right)\right)\\
  -\frac{4F_VC_q}{F^2 m_\rho^2}\left(2 \kappa _1^V-4 \kappa _2^V+6 \kappa _3^V-2 \kappa _6^V-2
   \kappa _8^V+2 \kappa _{12}^V-\kappa _{16}^V-\kappa _{17}^V\right)\,.\\
\end{multline}

The two-resonance mediated contributions to the axial-vector form factors, corresponding to figs.~\ref{fig:2R-FA}, are given in the following:\\

\begin{multline}\label{a1-2R-RChL}
  a_1^{2R}=
   -\frac{8F_AC_q}{F^2M_Vm_\rho^2D_{a_1}[(p+p_0+k)^2]D_\rho[(p_0+k)^2]}\\
   \left(-2 \sqrt{2} (k\cdot p)^2 \left(\frac{1}{2} \left(2
   c_5+c_{1256}\right) m_{\rho }^2-c_5 \left(m_{\eta }^2+2 k\cdot
   p_0\right)+4 c_3 \left(m_{\eta }^2-m_{\pi }^2\right)\right) \lambda
   ^{\prime\prime}\right.\\
   -\left(\left(c_5+c_7\right) m_{\eta }^2+k\cdot p_0 c_7-4
   c_3 \left(m_{\pi }^2+m_{\eta }^2\right)\right) \left(m_{\eta
   }^2-m_{\rho }^2+2 k\cdot p_0\right) \left(-2 \sqrt{2} \left(k\cdot
   p+p\cdot p_0\right) \lambda ^{\prime\prime}-2 m_{\pi }^2 \left(2 \lambda
   _1+\lambda _2\right)\right.\\\left.
   +\left(m_{\eta }^2+2 k\cdot p_0\right) \lambda
   _4\right)-\frac{1}{2} k\cdot p \left(-4 \sqrt{2} p\cdot p_0 \left(8
   c_3 m_{\pi }^2-c_7 m_{\eta }^2+\left(c_7-c_{1256}\right) m_{\rho
   }^2\right) \lambda ^{\prime\prime}\right.\\
   +2 c_5 \left(4 \lambda _4 \left(k\cdot
   p_0\right){}^2+\left(m_{\eta }^2-m_{\rho }^2\right) \left(m_{\eta }^2
   \lambda _4-2 m_{\pi }^2 \left(2 \lambda _1+\lambda
   _2\right)\right)\right)\\
   -8 c_3 \left(m_{\pi }^2-m_{\eta }^2\right)
   \left(2 \left(2 \lambda _1+\lambda _2\right) m_{\pi }^2-m_{\eta }^2
   \lambda _4+m_{\rho }^2 \left(\lambda _3-2 \lambda
   _5\right)\right)+c_{1256} m_{\rho }^2 \left(2 \left(2 \lambda
   _1+\lambda _2\right) m_{\pi }^2+m_{\eta }^2 \left(\lambda _3-\lambda
   _4-2 \lambda _5\right)\right)\\
   +2 k\cdot p_0 \left(4 \sqrt{2} p\cdot
   p_0 c_7 \lambda ^{\prime\prime}+2 c_5 \left(\left(2 m_{\eta }^2-m_{\rho
   }^2\right) \lambda _4-2 m_{\pi }^2 \left(2 \lambda _1+\lambda
   _2\right)\right)+8 c_3 \left(\left(4 \lambda _1+\lambda _4\right)
   m_{\pi }^2+m_{\rho }^2 \lambda _3-2 m_{\eta }^2 \lambda
   _4\right)\right.\\\left.\left.
   +c_{1256} m_{\rho }^2 \left(\lambda _3-\lambda _4-2 \lambda
   _5\right)\right)\right)+p\cdot p_0 \left(4 c_3 \left(m_{\pi
   }^2+m_{\eta }^2\right) \left(2 \left(2 \lambda _1+\lambda _2\right)
   m_{\pi }^2-m_{\eta }^2 \lambda _4+p\cdot p_0 \left(2 \lambda
   _2-\lambda _4-2 \lambda _5\right)\right)\right.\\\left.
   -\left(c_5+c_7\right)
   \left(-4 \lambda _4 \left(k\cdot p_0\right){}^2-m_{\eta }^2 \left(-2
   \left(2 \lambda _1+\lambda _2\right) m_{\pi }^2+m_{\eta }^2 \lambda
   _4+p\cdot p_0 \left(-2 \lambda _2+\lambda _4+2 \lambda
   _5\right)\right)\right)\right)\\
   +k\cdot p_0 \left(\left(-c_{1256}
   \lambda _3 m_{\rho }^2-4 p\cdot p_0 \left(c_5+c_7\right) \lambda
   _2\right) m_{\pi }^2\right.\\
   +p\cdot p_0 \left(-4 \sqrt{2} p\cdot p_0
   \left(c_5+c_7\right) \lambda ^{\prime\prime}+4 \left(-4
   c_3+c_5+c_7\right) m_{\eta }^2 \lambda _4+8 m_{\pi }^2
   \left(-\left(-4 c_3+c_5+c_7\right) \lambda _1-c_3 \lambda
   _4\right)\right)\\\left.
   -m_{\rho }^2 \left(c_{1256} \left(4 \lambda _1 m_{\pi
   }^2+p\cdot p_0 \left(\lambda _3-\lambda _4-2 \lambda
   _5\right)\right)+2 p\cdot p_0 \left(\left(c_5+c_7\right) \lambda _4+4
   c_3 \left(\lambda _3-2 \left(\lambda _4+\lambda
   _5\right)\right)\right)\right)\right)\\
   -m_{\rho }^2 \left(-\sqrt{2}
   \left(p\cdot p_0\right){}^2 \left(2
   \left(c_5+c_7\right)-c_{1256}\right) \lambda ^{\prime\prime}+4 m_{\pi
   }^2 \left(\frac{1}{2} \left(8 c_3+c_{1256}\right) m_{\eta }^2-4 c_3
   m_{\pi }^2\right) \lambda _1\right.\\
   +\frac{1}{2} m_{\pi }^2 \left(2 p\cdot
   p_0 \left(c_{1256}-2 \left(c_5+c_7\right)\right) \lambda
   _2+\left(\left(8 c_3+c_{1256}\right) m_{\eta }^2-8 c_3 m_{\pi
   }^2\right) \lambda _3\right)\\
   +p\cdot p_0 \left(\frac{1}{2} m_{\eta }^2
   \left(2 \left(c_5+c_7\right) \lambda _4+c_{1256} \left(\lambda
   _3-\lambda _4-2 \lambda _5\right)+8 c_3 \left(\lambda _3-2
   \left(\lambda _4+\lambda _5\right)\right)\right)\right.\\\left.\left.\left.
   \left(\left(c_5+c_7-\frac{c_{1256}}{2}\right) \lambda _1+c_3
   \left(\lambda _3-2 \lambda _5\right)\right)\right)\right)\frac{}{}\right)\\
 +\frac{8 F_A C_q \kappa _1^{SA} \left(c_d p\cdot
    p_0+m_{\pi }^2 c_m\right)}{F^2D_{a_1}[(p+p_0+k)^2]D_{a_0}[(p+p_0)^2]}\,,
  \\
\end{multline}

\begin{multline}\label{a2-2R-RChL}
 a_2^{2R}=
  -\frac{8F_AC_q}{F^2M_Vm_\rho^2D_{a_1}[(p+p_0+k)^2]D_\rho[(p_0+k)^2]}\\
   \left(c_7 k\cdot p_0+\left(c_5+c_7\right) m_{\eta }^2-4 c_3
   \left(m_{\eta }^2+m_{\pi }^2\right)\right) \left(2 k\cdot p_0+m_{\eta
   }^2-m_{\rho }^2\right)\\
   \left(\lambda _4 \left(2 k\cdot p_0+m_{\eta
   }^2\right)-2 \sqrt{2} \lambda ^{\prime\prime} k\cdot p-2 \left(2 \lambda
   _1+\lambda _2\right) m_{\pi}^2-2 \sqrt{2} p\cdot p_0 \lambda
   ^{\prime\prime}\right)\\
   -\frac{4 F_A C_q \kappa _1^{SA} \left(c_d p\cdot
   p_0+m_{\pi }^2 c_m\right)}{F^2D_{a_1}[(p+p_0+k)^2]D_{a_0}[(p+p_0)^2]}\,,
   \\
\end{multline}

\begin{multline}\label{a3-2R-RChL}
 a_3^{2R}=
   -\frac{4\sqrt{2}F_AC_q\left(\lambda ^\prime+\lambda^{\prime\prime}\right)}{F^2M_VD_{a_1}[(p+p_0+k)^2]D_\rho[(p_0+k)^2]}
    \left(2 \left(8 c_3+c_{1256}\right) k\cdot
   p_0+\left(8 c_3+c_{1256}\right) m_{\eta }^2-8 c_3 m_{\pi }^2\right)\,,
 \\
\end{multline}

\begin{multline}\label{a4-2R-RChL}
 a_4^{2R}=
 -\frac{8F_AC_q}{F^2M_Vm_\rho^2D_{a_1}[(p+p_0+k)^2]D_\rho[(p_0+k)^2]}\left(\frac{1}{2} \left(2 \lambda _2-\lambda _3\right)
 m_{\rho }^2 \left(c_{1256} \left(2 k\cdot p_0+m_{\eta
   }^2\right)+8 c_3 \left(m_{\eta }^2-m_{\pi }^2\right)\right)\right.\\
   -2 \sqrt{2} \lambda ^{\prime\prime} k\cdot p
   \left(-\left(c_5+c_7\right) \left(2 k\cdot p_0+m_{\eta }^2\right)+4 c_3 \left(m_{\eta }^2+m_{\pi }^2\right)+\frac{1}{2}
   \left(2 \left(c_5+c_7\right)-c_{1256}\right) m_{\rho }^2\right)\\
   +2 k\cdot p_0 \left(\left(c_5+c_7\right) \left(\lambda _4
   \left(m_{\rho }^2-2 m_{\eta }^2\right)+2 \left(2 \lambda _1+\lambda _2\right) m_{\pi }^2\right)+4 c_3 \left(2 \lambda _4
   m_{\eta }^2+\left(\lambda _4-4 \lambda _1\right) m_{\pi }^2\right)+2 \sqrt{2} \left(c_5+c_7\right) p\cdot p_0 \lambda
   ^{\prime\prime}\right)\\
   -4 \left(c_5+c_7\right) \lambda _4 \left(k\cdot p_0\right){}^2-\left(4 c_3 \left(m_{\eta }^2+m_{\pi
   }^2\right)-\left(c_5+c_7\right) m_{\eta }^2\right) \left(-\lambda _4 m_{\eta }^2+2 \left(2 \lambda _1+\lambda _2\right)
   m_{\pi }^2+\left(2 \lambda _2-\lambda _4-2 \lambda _5\right) p\cdot p_0\right)\\\left.
   -m_{\rho }^2 \left(\lambda _4 \left(4 c_3
   \left(m_{\eta }^2+m_{\pi }^2\right)-\left(c_5+c_7\right) m_{\eta }^2\right)+\left(2 \left(c_5+c_7\right)-c_{1256}\right)
   \left(2 \lambda _1+\lambda _2\right) m_{\pi }^2+\sqrt{2} \left(2 \left(c_5+c_7\right)-c_{1256}\right) p\cdot p_0 \lambda
   ^{\prime\prime}\right)\right)\,.
 \\
\end{multline}

We will display separately the contributions from the last diagram in the first line of figure \ref{fig:2R-FA}, due to the length of the corresponding expressions.\\

\begin{multline}
 a_1^{W^-\to (a_1^-)\eta\to\pi^-(\rho^0)\eta\to\pi^-\gamma\eta}=   
  +\frac{8F_V}{F^2m_{a_1}^2m_\rho^2D_{a_1}[(p+k)^2]}\left(8 m_\pi^2 C_q m_{\pi }^2 m_\eta^2 \lambda _1 \kappa
   _3^A\right.\\
   +16 k\cdot p C_q m_{\pi }^2 m_\eta^2 \lambda _1 \kappa _3^A-8 C_q
   m_{\pi }^2 m_{a_1}^2 m_\eta^2 \lambda _1 \kappa _3^A-2 m_\pi^2 C_q
   m_{a_1}^2 m_\eta^2 \lambda _2 \kappa _3^A+2 k\cdot p C_q m_{a_1}^2
   m_\eta^2 \lambda _2 \kappa _3^A+2 p^4 C_q m_\eta^2 \lambda _2 \kappa
   _3^A\\
   -4 (k\cdot p)^2 C_q m_\eta^2 \lambda _2 \kappa _3^A+2 m_\pi^2 k\cdot p
   C_q m_\eta^2 \lambda _2 \kappa _3^A+2 k\cdot p C_q m_{a_1}^2 m_\eta^2
   \lambda _4 \kappa _3^A-4 (k\cdot p)^2 C_q m_\eta^2 \lambda _4 \kappa
   _3^A-2 m_\pi^2 k\cdot p C_q m_\eta^2 \lambda _4 \kappa _3^A\\
   +4 k\cdot p C_q
   m_{a_1}^2 m_\eta^2 \lambda _5 \kappa _3^A-8 (k\cdot p)^2 C_q m_\eta^2
   \lambda _5 \kappa _3^A-4 m_\pi^2 k\cdot p C_q m_\eta^2 \lambda _5 \kappa
   _3^A+8 C_q m_{\pi }^2 p_0^4 \kappa _8^A \lambda _1+8 k\cdot p C_q
   m_{\pi }^2 m_\eta^2 \kappa _8^A \lambda _1\\
   -8 C_q m_{\pi }^2 m_{a_1}^2
   m_\eta^2 \kappa _8^A \lambda _1+32 m_\pi^2 C_q m_{\pi }^4 \kappa _9^A
   \lambda _1+64 k\cdot p C_q m_{\pi }^4 \kappa _9^A \lambda _1+16
   \sqrt{2} m_\pi^2 C_s m_{\pi }^4 \kappa _9^A \lambda _1+32 \sqrt{2}
   k\cdot p C_s m_{\pi }^4 \kappa _9^A \lambda _1\\
   -32 \sqrt{2} m_\pi^2 C_s
   m_K^2 m_{\pi }^2 \kappa _9^A \lambda _1-64 \sqrt{2} k\cdot p C_s
   m_K^2 m_{\pi }^2 \kappa _9^A \lambda _1-32 C_q m_{\pi }^4 m_{a_1}^2
   \kappa _9^A \lambda _1-16 \sqrt{2} C_s m_{\pi }^4 m_{a_1}^2 \kappa
   _9^A \lambda _1\\
   +32 \sqrt{2} C_s m_K^2 m_{\pi }^2 m_{a_1}^2 \kappa
   _9^A \lambda _1+8 k\cdot p C_q m_{\pi }^4 \kappa _{10}^A \lambda
   _1+4 \sqrt{2} k\cdot p C_s m_{\pi }^4 \kappa _{10}^A \lambda _1-8
   \sqrt{2} k\cdot p C_s m_K^2 m_{\pi }^2 \kappa _{10}^A \lambda _1\\
   -8
   C_q m_{\pi }^4 m_{a_1}^2 \kappa _{10}^A \lambda _1-4 \sqrt{2} C_s
   m_{\pi }^4 m_{a_1}^2 \kappa _{10}^A \lambda _1+8 \sqrt{2} C_s m_K^2
   m_{\pi }^2 m_{a_1}^2 \kappa _{10}^A \lambda _1+8 C_q m_{\pi }^4
   m_\eta^2 \kappa _{10}^A \lambda _1+4 \sqrt{2} C_s m_{\pi }^4 m_\eta^2
   \kappa _{10}^A \lambda _1\\
   -8 \sqrt{2} C_s m_K^2 m_{\pi }^2 m_\eta^2
   \kappa _{10}^A \lambda _1+32 m_\pi^2 C_q m_{\pi }^4 \kappa _{11}^A
   \lambda _1+64 k\cdot p C_q m_{\pi }^4 \kappa _{11}^A \lambda _1-32
   C_q m_{\pi }^4 m_{a_1}^2 \kappa _{11}^A \lambda _1+8 k\cdot p C_q
   m_{\pi }^4 \kappa _{12}^A \lambda _1\\
   -8 C_q m_{\pi }^4 m_{a_1}^2
   \kappa _{12}^A \lambda _1+8 C_q m_{\pi }^4 m_\eta^2 \kappa _{12}^A
   \lambda _1+8 m_\pi^2 C_q m_{\pi }^2 m_\eta^2 \kappa _{15}^A \lambda _1+16
   k\cdot p C_q m_{\pi }^2 m_\eta^2 \kappa _{15}^A \lambda _1-8 C_q m_{\pi
   }^2 m_{a_1}^2 m_\eta^2 \kappa _{15}^A \lambda _1\\
   -4 C_q m_{\pi }^2 p_0^4
   \kappa _{16}^A \lambda _1-8 (k\cdot p)^2 C_q m_{\pi }^2 \kappa
   _{16}^A \lambda _1-4 m_\pi^2 k\cdot p C_q m_{\pi }^2 \kappa _{16}^A
   \lambda _1+4 m_\pi^2 C_q m_{\pi }^2 m_{a_1}^2 \kappa _{16}^A \lambda
   _1+8 k\cdot p C_q m_{\pi }^2 m_{a_1}^2 \kappa _{16}^A \lambda _1\\
   -4
   m_\pi^2 C_q m_{\pi }^2 m_\eta^2 \kappa _{16}^A \lambda _1-12 k\cdot p C_q
   m_{\pi }^2 m_\eta^2 \kappa _{16}^A \lambda _1+4 C_q m_{\pi }^2
   m_{a_1}^2 m_\eta^2 \kappa _{16}^A \lambda _1-4 k\cdot p C_q p_0^4
   \kappa _8^A \lambda _2-4 m_\pi^2 C_q m_{a_1}^2 m_\eta^2 \kappa _8^A \lambda
   _2\\
   -4 (k\cdot p)^2 C_q m_\eta^2 \kappa _8^A \lambda _2-8 \sqrt{2} p^4
   C_s m_K^2 \kappa _9^A \lambda _2+16 \sqrt{2} (k\cdot p)^2 C_s m_K^2
   \kappa _9^A \lambda _2-8 \sqrt{2} m_\pi^2 k\cdot p C_s m_K^2 \kappa _9^A
   \lambda _2+8 p^4 C_q m_{\pi }^2 \kappa _9^A \lambda _2\\
   -16 (k\cdot
   p)^2 C_q m_{\pi }^2 \kappa _9^A \lambda _2+8 m_\pi^2 k\cdot p C_q m_{\pi
   }^2 \kappa _9^A \lambda _2+4 \sqrt{2} p^4 C_s m_{\pi }^2 \kappa _9^A
   \lambda _2-8 \sqrt{2} (k\cdot p)^2 C_s m_{\pi }^2 \kappa _9^A
   \lambda _2+4 \sqrt{2} m_\pi^2 k\cdot p C_s m_{\pi }^2 \kappa _9^A
   \lambda _2\\
   +8 \sqrt{2} m_\pi^2 C_s m_K^2 m_{a_1}^2 \kappa _9^A \lambda
   _2-8 \sqrt{2} k\cdot p C_s m_K^2 m_{a_1}^2 \kappa _9^A \lambda _2-8
   m_\pi^2 C_q m_{\pi }^2 m_{a_1}^2 \kappa _9^A \lambda _2+8 k\cdot p C_q
   m_{\pi }^2 m_{a_1}^2 \kappa _9^A \lambda _2\\
   -4 \sqrt{2} m_\pi^2 C_s
   m_{\pi }^2 m_{a_1}^2 \kappa _9^A \lambda _2+4 \sqrt{2} k\cdot p C_s
   m_{\pi }^2 m_{a_1}^2 \kappa _9^A \lambda _2+4 \sqrt{2} (k\cdot p)^2
   C_s m_K^2 \kappa _{10}^A \lambda _2-4 (k\cdot p)^2 C_q m_{\pi }^2
   \kappa _{10}^A \lambda _2\\
   -2 \sqrt{2} (k\cdot p)^2 C_s m_{\pi }^2
   \kappa _{10}^A \lambda _2+4 \sqrt{2} m_\pi^2 C_s m_K^2 m_{a_1}^2 \kappa
   _{10}^A \lambda _2-4 m_\pi^2 C_q m_{\pi }^2 m_{a_1}^2 \kappa _{10}^A
   \lambda _2-2 \sqrt{2} m_\pi^2 C_s m_{\pi }^2 m_{a_1}^2 \kappa _{10}^A
   \lambda _2\\
   +4 \sqrt{2} k\cdot p C_s m_K^2 m_\eta^2 \kappa _{10}^A
   \lambda _2-4 k\cdot p C_q m_{\pi }^2 m_\eta^2 \kappa _{10}^A \lambda
   _2-2 \sqrt{2} k\cdot p C_s m_{\pi }^2 m_\eta^2 \kappa _{10}^A \lambda
   _2+8 p^4 C_q m_{\pi }^2 \kappa _{11}^A \lambda _2\\
   -16 (k\cdot p)^2
   C_q m_{\pi }^2 \kappa _{11}^A \lambda _2+8 m_\pi^2 k\cdot p C_q m_{\pi
   }^2 \kappa _{11}^A \lambda _2-8 m_\pi^2 C_q m_{\pi }^2 m_{a_1}^2 \kappa
   _{11}^A \lambda _2+8 k\cdot p C_q m_{\pi }^2 m_{a_1}^2 \kappa
   _{11}^A \lambda _2-4 (k\cdot p)^2 C_q m_{\pi }^2 \kappa _{12}^A
   \lambda _2\\
   -4 m_\pi^2 C_q m_{\pi }^2 m_{a_1}^2 \kappa _{12}^A \lambda
   _2-4 k\cdot p C_q m_{\pi }^2 m_\eta^2 \kappa _{12}^A \lambda _2-2 m_\pi^2
   C_q m_{a_1}^2 m_\eta^2 \kappa _{15}^A \lambda _2+2 k\cdot p C_q
   m_{a_1}^2 m_\eta^2 \kappa _{15}^A \lambda _2+2 p^4 C_q m_\eta^2 \kappa
   _{15}^A \lambda _2\\
   -4 (k\cdot p)^2 C_q m_\eta^2 \kappa _{15}^A \lambda
   _2+2 m_\pi^2 k\cdot p C_q m_\eta^2 \kappa _{15}^A \lambda _2+2 k\cdot p C_q
   p_0^4 \kappa _{16}^A \lambda _2+2 p^4 C_q m_{a_1}^2 \kappa _{16}^A
   \lambda _2+4 m_\pi^2 k\cdot p C_q m_{a_1}^2 \kappa _{16}^A \lambda _2+2
   m_\pi^2 C_q m_{a_1}^2 m_\eta^2 \kappa _{16}^A \lambda _2\\
   +6 (k\cdot p)^2 C_q
   m_\eta^2 \kappa _{16}^A \lambda _2+2 m_\pi^2 k\cdot p C_q m_\eta^2 \kappa
   _{16}^A \lambda _2+4 (k\cdot p)^3 C_q \kappa _{16}^A \lambda _2+2
   m_\pi^2 (k\cdot p)^2 C_q \kappa _{16}^A \lambda _2+4 \left(p\cdot
   p_0\right){}^3 C_q \left(\kappa _8^A-\kappa _{16}^A\right) \lambda
   _2\\
   -2 \left(p\cdot p_0\right){}^2 \left(\sqrt{2} C_s \left(2
   m_K^2-m_{\pi }^2\right) \kappa _{10}^A+C_q
   \left(\left(m_\pi^2-m_{a_1}^2+2 k\cdot p\right) \kappa _3^A+2
   \left(m_\pi^2-m_{a_1}^2-m_\eta^2+k\cdot p\right) \kappa _8^A-2 m_{\pi }^2
   \kappa _{10}^A\right.\right.\\\left.\left.
   -2 m_{\pi }^2 \kappa _{12}^A+m_\pi^2 \kappa
   _{15}^A-m_{a_1}^2 \kappa _{15}^A+2 k\cdot p \kappa _{15}^A-m_\pi^2
   \kappa _{16}^A+2 m_{a_1}^2 \kappa _{16}^A+m_\eta^2 \kappa
   _{16}^A\right)\right) \lambda _2-2 k\cdot p C_q p_0^4 \kappa _8^A
   \lambda _4+2 k\cdot p C_q m_{a_1}^2 m_\eta^2 \kappa _8^A \lambda _4\\
   -2
   (k\cdot p)^2 C_q m_\eta^2 \kappa _8^A \lambda _4+16 \sqrt{2} (k\cdot
   p)^2 C_s m_K^2 \kappa _9^A \lambda _4+8 \sqrt{2} m_\pi^2 k\cdot p C_s
   m_K^2 \kappa _9^A \lambda _4-16 (k\cdot p)^2 C_q m_{\pi }^2 \kappa
   _9^A \lambda _4\\
   -8 m_\pi^2 k\cdot p C_q m_{\pi }^2 \kappa _9^A \lambda
   _4-8 \sqrt{2} (k\cdot p)^2 C_s m_{\pi }^2 \kappa _9^A \lambda _4-4
   \sqrt{2} m_\pi^2 k\cdot p C_s m_{\pi }^2 \kappa _9^A \lambda _4-8
   \sqrt{2} k\cdot p C_s m_K^2 m_{a_1}^2 \kappa _9^A \lambda _4\\
   +8
   k\cdot p C_q m_{\pi }^2 m_{a_1}^2 \kappa _9^A \lambda _4+4 \sqrt{2}
   k\cdot p C_s m_{\pi }^2 m_{a_1}^2 \kappa _9^A \lambda _4+2 \sqrt{2}
   (k\cdot p)^2 C_s m_K^2 \kappa _{10}^A \lambda _4-2 (k\cdot p)^2 C_q
   m_{\pi }^2 \kappa _{10}^A \lambda _4\\
   -\sqrt{2} (k\cdot p)^2 C_s
   m_{\pi }^2 \kappa _{10}^A \lambda _4-2 \sqrt{2} k\cdot p C_s m_K^2
   m_{a_1}^2 \kappa _{10}^A \lambda _4+2 k\cdot p C_q m_{\pi }^2
   m_{a_1}^2 \kappa _{10}^A \lambda _4+\sqrt{2} k\cdot p C_s m_{\pi }^2
   m_{a_1}^2 \kappa _{10}^A \lambda _4\\
   +2 \sqrt{2} k\cdot p C_s m_K^2
   m_\eta^2 \kappa _{10}^A \lambda _4-2 k\cdot p C_q m_{\pi }^2 m_\eta^2
   \kappa _{10}^A \lambda _4-\sqrt{2} k\cdot p C_s m_{\pi }^2 m_\eta^2
   \kappa _{10}^A \lambda _4-16 (k\cdot p)^2 C_q m_{\pi }^2 \kappa
   _{11}^A \lambda _4-8 m_\pi^2 k\cdot p C_q m_{\pi }^2 \kappa _{11}^A
   \lambda _4\\
   +8 k\cdot p C_q m_{\pi }^2 m_{a_1}^2 \kappa _{11}^A
   \lambda _4-2 (k\cdot p)^2 C_q m_{\pi }^2 \kappa _{12}^A \lambda _4+2
   k\cdot p C_q m_{\pi }^2 m_{a_1}^2 \kappa _{12}^A \lambda _4-2 k\cdot
   p C_q m_{\pi }^2 m_\eta^2 \kappa _{12}^A \lambda _4\\
   +2 k\cdot p C_q
   m_{a_1}^2 m_\eta^2 \kappa _{15}^A \lambda _4-4 (k\cdot p)^2 C_q m_\eta^2
   \kappa _{15}^A \lambda _4-2 m_\pi^2 k\cdot p C_q m_\eta^2 \kappa _{15}^A
   \lambda _4+k\cdot p C_q p_0^4 \kappa _{16}^A \lambda _4-2 (k\cdot
   p)^2 C_q m_{a_1}^2 \kappa _{16}^A \lambda _4\\
   -m_\pi^2 k\cdot p C_q
   m_{a_1}^2 \kappa _{16}^A \lambda _4-k\cdot p C_q m_{a_1}^2 m_\eta^2
   \kappa _{16}^A \lambda _4+3 (k\cdot p)^2 C_q m_\eta^2 \kappa _{16}^A
   \lambda _4+m_\pi^2 k\cdot p C_q m_\eta^2 \kappa _{16}^A \lambda _4+2
   (k\cdot p)^3 C_q \kappa _{16}^A \lambda _4\\
   +m_\pi^2 (k\cdot p)^2 C_q
   \kappa _{16}^A \lambda _4-4 k\cdot p C_q p_0^4 \kappa _8^A \lambda
   _5+4 k\cdot p C_q m_{a_1}^2 m_\eta^2 \kappa _8^A \lambda _5-4 (k\cdot
   p)^2 C_q m_\eta^2 \kappa _8^A \lambda _5+32 \sqrt{2} (k\cdot p)^2 C_s
   m_K^2 \kappa _9^A \lambda _5\nonumber
   \end{multline}
   
   \begin{multline}
   +16 \sqrt{2} m_\pi^2 k\cdot p C_s m_K^2
   \kappa _9^A \lambda _5-32 (k\cdot p)^2 C_q m_{\pi }^2 \kappa _9^A
   \lambda _5-16 m_\pi^2 k\cdot p C_q m_{\pi }^2 \kappa _9^A \lambda _5-16
   \sqrt{2} (k\cdot p)^2 C_s m_{\pi }^2 \kappa _9^A \lambda _5\\
   -8
   \sqrt{2} m_\pi^2 k\cdot p C_s m_{\pi }^2 \kappa _9^A \lambda _5-16
   \sqrt{2} k\cdot p C_s m_K^2 m_{a_1}^2 \kappa _9^A \lambda _5+16
   k\cdot p C_q m_{\pi }^2 m_{a_1}^2 \kappa _9^A \lambda _5+8 \sqrt{2}
   k\cdot p C_s m_{\pi }^2 m_{a_1}^2 \kappa _9^A \lambda _5\\
   +4 \sqrt{2}
   (k\cdot p)^2 C_s m_K^2 \kappa _{10}^A \lambda _5-4 (k\cdot p)^2 C_q
   m_{\pi }^2 \kappa _{10}^A \lambda _5-2 \sqrt{2} (k\cdot p)^2 C_s
   m_{\pi }^2 \kappa _{10}^A \lambda _5-4 \sqrt{2} k\cdot p C_s m_K^2
   m_{a_1}^2 \kappa _{10}^A \lambda _5\\
   +4 k\cdot p C_q m_{\pi }^2
   m_{a_1}^2 \kappa _{10}^A \lambda _5+2 \sqrt{2} k\cdot p C_s m_{\pi
   }^2 m_{a_1}^2 \kappa _{10}^A \lambda _5+4 \sqrt{2} k\cdot p C_s
   m_K^2 m_\eta^2 \kappa _{10}^A \lambda _5-4 k\cdot p C_q m_{\pi }^2
   m_\eta^2 \kappa _{10}^A \lambda _5\\
   -2 \sqrt{2} k\cdot p C_s m_{\pi }^2
   m_\eta^2 \kappa _{10}^A \lambda _5-32 (k\cdot p)^2 C_q m_{\pi }^2
   \kappa _{11}^A \lambda _5-16 m_\pi^2 k\cdot p C_q m_{\pi }^2 \kappa
   _{11}^A \lambda _5+16 k\cdot p C_q m_{\pi }^2 m_{a_1}^2 \kappa
   _{11}^A \lambda _5\\
   -4 (k\cdot p)^2 C_q m_{\pi }^2 \kappa _{12}^A
   \lambda _5+4 k\cdot p C_q m_{\pi }^2 m_{a_1}^2 \kappa _{12}^A
   \lambda _5-4 k\cdot p C_q m_{\pi }^2 m_\eta^2 \kappa _{12}^A \lambda
   _5+4 k\cdot p C_q m_{a_1}^2 m_\eta^2 \kappa _{15}^A \lambda _5-8
   (k\cdot p)^2 C_q m_\eta^2 \kappa _{15}^A \lambda _5\\
   -4 m_\pi^2 k\cdot p C_q
   m_\eta^2 \kappa _{15}^A \lambda _5+2 k\cdot p C_q p_0^4 \kappa _{16}^A
   \lambda _5-4 (k\cdot p)^2 C_q m_{a_1}^2 \kappa _{16}^A \lambda _5-2
   m_\pi^2 k\cdot p C_q m_{a_1}^2 \kappa _{16}^A \lambda _5-2 k\cdot p C_q
   m_{a_1}^2 m_\eta^2 \kappa _{16}^A \lambda _5\\
   +6 (k\cdot p)^2 C_q m_\eta^2
   \kappa _{16}^A \lambda _5+2 m_\pi^2 k\cdot p C_q m_\eta^2 \kappa _{16}^A
   \lambda _5+4 (k\cdot p)^3 C_q \kappa _{16}^A \lambda _5+2 m_\pi^2
   (k\cdot p)^2 C_q \kappa _{16}^A \lambda _5\\
   -2 \left(k\cdot
   p_0\right){}^2 C_q \left(\kappa _8^A-\kappa _{16}^A\right) \left(4
   \lambda _1 m_{\pi }^2+2 \left(m_\pi^2-p\cdot p_0\right) \lambda
   _2-k\cdot p \left(\lambda _4+2 \lambda _5\right)\right)\\
   +2 p\cdot p_0
   \left(\sqrt{2} C_s \left(2 m_K^2-m_{\pi }^2\right) \left(2
   \left(m_\pi^2-m_{a_1}^2+2 k\cdot p\right) \kappa
   _9^A+\left(m_\pi^2-m_{a_1}^2+k\cdot p\right) \kappa _{10}^A\right)
   \lambda _2\right.\\
   +C_q \left(\left(m_\pi^2-m_{a_1}^2+2 k\cdot p\right)
   \left(\left(m_\pi^2-m_\eta^2-k\cdot p\right) \lambda _2-k\cdot p
   \left(\lambda _4+2 \lambda _5\right)\right) \kappa _3^A-2 (k\cdot
   p)^2 \kappa _8^A \lambda _2-2 m_\pi^2 m_{a_1}^2 \kappa _8^A \lambda _2-2
   m_\pi^2 m_\eta^2 \kappa _8^A \lambda _2\right.\\
   +2 m_{a_1}^2 m_\eta^2 \kappa _8^A
   \lambda _2-4 k\cdot p m_\eta^2 \kappa _8^A \lambda _2+p^4 \kappa
   _{15}^A \lambda _2-2 (k\cdot p)^2 \kappa _{15}^A \lambda _2-m_\pi^2
   m_{a_1}^2 \kappa _{15}^A \lambda _2+k\cdot p m_{a_1}^2 \kappa
   _{15}^A \lambda _2-m_\pi^2 m_\eta^2 \kappa _{15}^A \lambda _2\\
   +m_{a_1}^2
   m_\eta^2 \kappa _{15}^A \lambda _2-2 k\cdot p m_\eta^2 \kappa _{15}^A
   \lambda _2+m_\pi^2 k\cdot p \kappa _{15}^A \lambda _2+p^4 \kappa _{16}^A
   \lambda _2+4 (k\cdot p)^2 \kappa _{16}^A \lambda _2+m_\pi^2 m_{a_1}^2
   \kappa _{16}^A \lambda _2-2 k\cdot p m_{a_1}^2 \kappa _{16}^A
   \lambda _2\\
   +m_\pi^2 m_\eta^2 \kappa _{16}^A \lambda _2-m_{a_1}^2 m_\eta^2
   \kappa _{16}^A \lambda _2+3 k\cdot p m_\eta^2 \kappa _{16}^A \lambda
   _2+3 m_\pi^2 k\cdot p \kappa _{16}^A \lambda _2+2 m_{\pi }^2 \left(2
   \left(m_\pi^2-m_{a_1}^2+2 k\cdot p\right) \lambda _1 \kappa _3^A-2
   m_{a_1}^2 \kappa _8^A \lambda _1\right.\\
   +2 m_\eta^2 \kappa _8^A \lambda _1+2
   m_\pi^2 \kappa _{15}^A \lambda _1-2 m_{a_1}^2 \kappa _{15}^A \lambda
   _1+2 m_{a_1}^2 \kappa _{16}^A \lambda _1-2 m_\eta^2 \kappa _{16}^A
   \lambda _1-2 m_\pi^2 \kappa _9^A \lambda _2+2 m_{a_1}^2 \kappa _9^A
   \lambda _2-m_\pi^2 \kappa _{10}^A \lambda _2+m_{a_1}^2 \kappa _{10}^A
   \lambda _2-2 m_\pi^2 \kappa _{11}^A \lambda _2\\\left.
   +2 m_{a_1}^2 \kappa
   _{11}^A \lambda _2-m_\pi^2 \kappa _{12}^A \lambda _2+m_{a_1}^2 \kappa
   _{12}^A \lambda _2+k\cdot p \left(2 \lambda _1 \kappa _8^A+4 \kappa
   _{15}^A \lambda _1-2 \kappa _{16}^A \lambda _1-4 \kappa _9^A \lambda
   _2-\kappa _{10}^A \lambda _2-4 \kappa _{11}^A \lambda _2-\kappa
   _{12}^A \lambda _2\right)\right)\\
   -(k\cdot p)^2 \kappa _8^A \lambda
   _4+k\cdot p m_{a_1}^2 \kappa _8^A \lambda _4-k\cdot p m_\eta^2 \kappa
   _8^A \lambda _4-2 (k\cdot p)^2 \kappa _{15}^A \lambda _4+k\cdot p
   m_{a_1}^2 \kappa _{15}^A \lambda _4-m_\pi^2 k\cdot p \kappa _{15}^A
   \lambda _4+(k\cdot p)^2 \kappa _{16}^A \lambda _4\\
   -k\cdot p m_{a_1}^2
   \kappa _{16}^A \lambda _4+k\cdot p m_\eta^2 \kappa _{16}^A \lambda _4-2
   (k\cdot p)^2 \kappa _8^A \lambda _5+2 k\cdot p m_{a_1}^2 \kappa _8^A
   \lambda _5-2 k\cdot p m_\eta^2 \kappa _8^A \lambda _5-4 (k\cdot p)^2
   \kappa _{15}^A \lambda _5\\\left.\left.
   +2 k\cdot p m_{a_1}^2 \kappa _{15}^A
   \lambda _5-2 m_\pi^2 k\cdot p \kappa _{15}^A \lambda _5+2 (k\cdot p)^2
   \kappa _{16}^A \lambda _5-2 k\cdot p m_{a_1}^2 \kappa _{16}^A
   \lambda _5+2 k\cdot p m_\eta^2 \kappa _{16}^A \lambda
   _5\right)\right)\\
   +k\cdot p_0 \left(\sqrt{2} C_s \left(2 m_K^2-m_{\pi
   }^2\right) \left(4 \lambda _1 m_{\pi }^2+2 \left(m_\pi^2-p\cdot
   p_0\right) \lambda _2-k\cdot p \left(\lambda _4+2 \lambda
   _5\right)\right) \kappa _{10}^A\right.\\
   +C_q \left(2 \left(m_\pi^2-m_{a_1}^2+2
   k\cdot p\right) \left(\left(m_\pi^2-k\cdot p-p\cdot p_0\right) \lambda
   _2-k\cdot p \left(\lambda _4+2 \lambda _5\right)\right) \kappa
   _3^A-8 m_{\pi }^4 \left(\kappa _{10}^A+\kappa _{12}^A\right) \lambda
   _1-4 (k\cdot p)^2 \kappa _8^A \lambda _2\right.\\
   +8 \left(p\cdot
   p_0\right){}^2 \kappa _8^A \lambda _2-4 m_\pi^2 m_{a_1}^2 \kappa _8^A
   \lambda _2+4 p\cdot p_0 m_{a_1}^2 \kappa _8^A \lambda _2-4 m_\pi^2 m_\eta^2
   \kappa _8^A \lambda _2-4 k\cdot p m_\eta^2 \kappa _8^A \lambda _2+4
   p\cdot p_0 m_\eta^2 \kappa _8^A \lambda _2-8 m_\pi^2 p\cdot p_0 \kappa _8^A
   \lambda _2\\
   -4 k\cdot p p\cdot p_0 \kappa _8^A \lambda _2+2 p^4 \kappa
   _{15}^A \lambda _2-4 (k\cdot p)^2 \kappa _{15}^A \lambda _2-2 m_\pi^2
   m_{a_1}^2 \kappa _{15}^A \lambda _2+2 k\cdot p m_{a_1}^2 \kappa
   _{15}^A \lambda _2+2 p\cdot p_0 m_{a_1}^2 \kappa _{15}^A \lambda
   _2\\
   +2 m_\pi^2 k\cdot p \kappa _{15}^A \lambda _2-2 m_\pi^2 p\cdot p_0 \kappa
   _{15}^A \lambda _2-4 k\cdot p p\cdot p_0 \kappa _{15}^A \lambda _2+2
   p^4 \kappa _{16}^A \lambda _2+4 (k\cdot p)^2 \kappa _{16}^A \lambda
   _2-8 \left(p\cdot p_0\right){}^2 \kappa _{16}^A \lambda _2\\
   +4 m_\pi^2
   m_{a_1}^2 \kappa _{16}^A \lambda _2-4 p\cdot p_0 m_{a_1}^2 \kappa
   _{16}^A \lambda _2+2 m_\pi^2 m_\eta^2 \kappa _{16}^A \lambda _2+4 k\cdot p
   m_\eta^2 \kappa _{16}^A \lambda _2-2 p\cdot p_0 m_\eta^2 \kappa _{16}^A
   \lambda _2+4 m_\pi^2 k\cdot p \kappa _{16}^A \lambda _2\\
   +6 m_\pi^2 p\cdot p_0
   \kappa _{16}^A \lambda _2-2 (k\cdot p)^2 \kappa _8^A \lambda _4+2
   k\cdot p m_{a_1}^2 \kappa _8^A \lambda _4+2 k\cdot p p\cdot p_0
   \kappa _8^A \lambda _4-4 (k\cdot p)^2 \kappa _{15}^A \lambda _4+2
   k\cdot p m_{a_1}^2 \kappa _{15}^A \lambda _4\\
   -2 m_\pi^2 k\cdot p \kappa
   _{15}^A \lambda _4-2 k\cdot p m_{a_1}^2 \kappa _{16}^A \lambda
   _4+k\cdot p m_\eta^2 \kappa _{16}^A \lambda _4-m_\pi^2 k\cdot p \kappa
   _{16}^A \lambda _4-2 k\cdot p p\cdot p_0 \kappa _{16}^A \lambda _4-4
   (k\cdot p)^2 \kappa _8^A \lambda _5\\
   +4 k\cdot p m_{a_1}^2 \kappa _8^A
   \lambda _5+4 k\cdot p p\cdot p_0 \kappa _8^A \lambda _5-8 (k\cdot
   p)^2 \kappa _{15}^A \lambda _5+4 k\cdot p m_{a_1}^2 \kappa _{15}^A
   \lambda _5-4 m_\pi^2 k\cdot p \kappa _{15}^A \lambda _5-4 k\cdot p
   m_{a_1}^2 \kappa _{16}^A \lambda _5\\
   +2 k\cdot p m_\eta^2 \kappa _{16}^A
   \lambda _5-2 m_\pi^2 k\cdot p \kappa _{16}^A \lambda _5-4 k\cdot p
   p\cdot p_0 \kappa _{16}^A \lambda _5\\
   +2 m_{\pi }^2 \left(4
   \left(m_\pi^2-m_{a_1}^2+2 k\cdot p\right) \lambda _1 \kappa _3^A+2
   \left(-m_\pi^2 \lambda _2 \kappa _{10}^A+2 m_\pi^2 \kappa _{15}^A \lambda
   _1+m_\pi^2 \kappa _{16}^A \lambda _1-m_\eta^2 \kappa _{16}^A \lambda _1-2
   m_{a_1}^2 \left(\kappa _8^A+\kappa _{15}^A-\kappa _{16}^A\right)
   \lambda _1\right.\right.\\\left.\left.\left.\left.\left.
   -m_\pi^2 \kappa _{12}^A \lambda _2+p\cdot p_0 \left(-2 \lambda
   _1 \kappa _8^A+2 \kappa _{16}^A \lambda _1+\left(\kappa
   _{10}^A+\kappa _{12}^A\right) \lambda _2\right)\right)+k\cdot p
   \left(4 \lambda _1 \kappa _8^A+8 \kappa _{15}^A \lambda
   _1+\left(\kappa _{10}^A+\kappa _{12}^A\right) \left(\lambda _4+2
   \lambda _5\right)\right)\right)\right)\right)\right)\,,
 \nonumber \\
\end{multline}

\begin{multline}
 a_2^{W^-\to (a_1^-)\eta\to\pi^-(\rho^0)\eta\to\pi^-\gamma\eta}=   
  +\frac{8F_V}{F^2m_{a_1}^2m_\rho^2D_{a_1}[(p+k)^2]}\left(2 \left(C_q m_\eta^2 \kappa _3^A-4 \sqrt{2} C_s m_K^2 \kappa _9^A
  +4 C_q m_{\pi }^2 \kappa _9^A+2
   \sqrt{2} C_s m_{\pi }^2 \kappa _9^A\right.\right.\\\left.
   +4 C_q m_{\pi }^2 \kappa _{11}^A+C_q m_\eta^2 \kappa _{15}^A-C_q m_\eta^2 \kappa
   _{16}^A+k\cdot p_0 C_q \left(\kappa _3^A+\kappa _{15}^A\right)+p\cdot p_0 C_q \left(\kappa _3^A+\kappa _{15}^A-\kappa
   _{16}^A\right)\right) \left(2 \lambda _2+\lambda _4+2 \lambda _5\right) (k\cdot p)^2\\
   +\left(-8 p\cdot p_0 C_q m_{\pi }^2
   \lambda _1 \kappa _3^A-8 C_q m_{\pi }^2 m_\eta^2 \lambda _1 \kappa _3^A-2 p\cdot p_0 C_q m_{a_1}^2 \lambda _2 \kappa _3^A-2
   C_q m_{a_1}^2 m_\eta^2 \lambda _2 \kappa _3^A+4 p\cdot p_0 C_q m_\eta^2 \lambda _2 \kappa _3^A\right.\\
   +4 \left(p\cdot p_0\right){}^2 C_q
   \lambda _2 \kappa _3^A-p\cdot p_0 C_q m_{a_1}^2 \lambda _4 \kappa _3^A-C_q m_{a_1}^2 m_\eta^2 \lambda _4 \kappa _3^A-2 p\cdot
   p_0 C_q m_{a_1}^2 \lambda _5 \kappa _3^A-2 C_q m_{a_1}^2 m_\eta^2 \lambda _5 \kappa _3^A\\
   -32 C_q m_{\pi }^4 \kappa _9^A \lambda
   _1-16 \sqrt{2} C_s m_{\pi }^4 \kappa _9^A \lambda _1+32 \sqrt{2} C_s m_K^2 m_{\pi }^2 \kappa _9^A \lambda _1-32 C_q m_{\pi
   }^4 \kappa _{11}^A \lambda _1-8 p\cdot p_0 C_q m_{\pi }^2 \kappa _{15}^A \lambda _1\\
   -8 C_q m_{\pi }^2 m_\eta^2 \kappa _{15}^A
   \lambda _1+8 p\cdot p_0 C_q m_{\pi }^2 \kappa _{16}^A \lambda _1+8 C_q m_{\pi }^2 m_\eta^2 \kappa _{16}^A \lambda _1+4 C_q
   p_0^4 \kappa _8^A \lambda _2+8 p\cdot p_0 C_q m_\eta^2 \kappa _8^A \lambda _2+4 \left(p\cdot p_0\right){}^2 C_q \kappa _8^A
   \lambda _2\\
   -16 \sqrt{2} p\cdot p_0 C_s m_K^2 \kappa _9^A \lambda _2+16 p\cdot p_0 C_q m_{\pi }^2 \kappa _9^A \lambda _2+8
   \sqrt{2} p\cdot p_0 C_s m_{\pi }^2 \kappa _9^A \lambda _2+8 \sqrt{2} C_s m_K^2 m_{a_1}^2 \kappa _9^A \lambda _2-8 C_q
   m_{\pi }^2 m_{a_1}^2 \kappa _9^A \lambda _2\\
   -4 \sqrt{2} C_s m_{\pi }^2 m_{a_1}^2 \kappa _9^A \lambda _2-4 \sqrt{2} p\cdot
   p_0 C_s m_K^2 \kappa _{10}^A \lambda _2+4 p\cdot p_0 C_q m_{\pi }^2 \kappa _{10}^A \lambda _2+2 \sqrt{2} p\cdot p_0 C_s
   m_{\pi }^2 \kappa _{10}^A \lambda _2-4 \sqrt{2} C_s m_K^2 m_\eta^2 \kappa _{10}^A \lambda _2\\
   +4 C_q m_{\pi }^2 m_\eta^2 \kappa
   _{10}^A \lambda _2+2 \sqrt{2} C_s m_{\pi }^2 m_\eta^2 \kappa _{10}^A \lambda _2+16 p\cdot p_0 C_q m_{\pi }^2 \kappa _{11}^A
   \lambda _2-8 C_q m_{\pi }^2 m_{a_1}^2 \kappa _{11}^A \lambda _2+4 p\cdot p_0 C_q m_{\pi }^2 \kappa _{12}^A \lambda _2+4 C_q
   m_{\pi }^2 m_\eta^2 \kappa _{12}^A \lambda _2\\
   -2 p\cdot p_0 C_q m_{a_1}^2 \kappa _{15}^A \lambda _2-2 C_q m_{a_1}^2 m_\eta^2
   \kappa _{15}^A \lambda _2+4 p\cdot p_0 C_q m_\eta^2 \kappa _{15}^A \lambda _2+4 \left(p\cdot p_0\right){}^2 C_q \kappa _{15}^A
   \lambda _2-2 C_q p_0^4 \kappa _{16}^A \lambda _2\\
   +4 p\cdot p_0 C_q m_{a_1}^2 \kappa _{16}^A \lambda _2-6 p\cdot p_0 C_q
   m_\eta^2 \kappa _{16}^A \lambda _2+2 C_q p_0^4 \kappa _8^A \lambda _4+4 p\cdot p_0 C_q m_\eta^2 \kappa _8^A \lambda _4+2
   \left(p\cdot p_0\right){}^2 C_q \kappa _8^A \lambda _4\\
   +4 \sqrt{2} C_s m_K^2 m_{a_1}^2 \kappa _9^A \lambda _4-4 C_q m_{\pi
   }^2 m_{a_1}^2 \kappa _9^A \lambda _4-2 \sqrt{2} C_s m_{\pi }^2 m_{a_1}^2 \kappa _9^A \lambda _4-2 \sqrt{2} p\cdot p_0 C_s
   m_K^2 \kappa _{10}^A \lambda _4+2 p\cdot p_0 C_q m_{\pi }^2 \kappa _{10}^A \lambda _4\\
   +\sqrt{2} p\cdot p_0 C_s m_{\pi }^2
   \kappa _{10}^A \lambda _4-2 \sqrt{2} C_s m_K^2 m_\eta^2 \kappa _{10}^A \lambda _4+2 C_q m_{\pi }^2 m_\eta^2 \kappa _{10}^A
   \lambda _4+\sqrt{2} C_s m_{\pi }^2 m_\eta^2 \kappa _{10}^A \lambda _4-4 C_q m_{\pi }^2 m_{a_1}^2 \kappa _{11}^A \lambda _4\\
   +2
   p\cdot p_0 C_q m_{\pi }^2 \kappa _{12}^A \lambda _4+2 C_q m_{\pi }^2 m_\eta^2 \kappa _{12}^A \lambda _4-p\cdot p_0 C_q
   m_{a_1}^2 \kappa _{15}^A \lambda _4-C_q m_{a_1}^2 m_\eta^2 \kappa _{15}^A \lambda _4-C_q p_0^4 \kappa _{16}^A \lambda _4-3
   p\cdot p_0 C_q m_\eta^2 \kappa _{16}^A \lambda _4\\
   -2 \left(p\cdot p_0\right){}^2 C_q \kappa _{16}^A \lambda _4+4 C_q p_0^4
   \kappa _8^A \lambda _5+8 p\cdot p_0 C_q m_\eta^2 \kappa _8^A \lambda _5+4 \left(p\cdot p_0\right){}^2 C_q \kappa _8^A \lambda
   _5+8 \sqrt{2} C_s m_K^2 m_{a_1}^2 \kappa _9^A \lambda _5\\
   -8 C_q m_{\pi }^2 m_{a_1}^2 \kappa _9^A \lambda _5-4 \sqrt{2} C_s
   m_{\pi }^2 m_{a_1}^2 \kappa _9^A \lambda _5-4 \sqrt{2} p\cdot p_0 C_s m_K^2 \kappa _{10}^A \lambda _5+4 p\cdot p_0 C_q
   m_{\pi }^2 \kappa _{10}^A \lambda _5+2 \sqrt{2} p\cdot p_0 C_s m_{\pi }^2 \kappa _{10}^A \lambda _5\\
   -4 \sqrt{2} C_s m_K^2
   m_\eta^2 \kappa _{10}^A \lambda _5+4 C_q m_{\pi }^2 m_\eta^2 \kappa _{10}^A \lambda _5+2 \sqrt{2} C_s m_{\pi }^2 m_\eta^2 \kappa
   _{10}^A \lambda _5-8 C_q m_{\pi }^2 m_{a_1}^2 \kappa _{11}^A \lambda _5+4 p\cdot p_0 C_q m_{\pi }^2 \kappa _{12}^A \lambda
   _5\\
   +4 C_q m_{\pi }^2 m_\eta^2 \kappa _{12}^A \lambda _5-2 p\cdot p_0 C_q m_{a_1}^2 \kappa _{15}^A \lambda _5-2 C_q m_{a_1}^2
   m_\eta^2 \kappa _{15}^A \lambda _5-2 C_q p_0^4 \kappa _{16}^A \lambda _5-6 p\cdot p_0 C_q m_\eta^2 \kappa _{16}^A \lambda _5-4
   \left(p\cdot p_0\right){}^2 C_q \kappa _{16}^A \lambda _5\\
   +m_\pi^2 \left(-4 \sqrt{2} C_s m_K^2 \kappa _9^A+4 C_q m_{\pi }^2
   \kappa _9^A+2 \sqrt{2} C_s m_{\pi }^2 \kappa _9^A+4 C_q m_{\pi }^2 \kappa _{11}^A+C_q m_\eta^2 \left(\kappa _3^A+\kappa
   _{15}^A-\kappa _{16}^A\right)+p\cdot p_0 C_q \left(\kappa _3^A+\kappa _{15}^A-\kappa _{16}^A\right)\right) \\
   \left(2 \lambda
   _2+\lambda _4+2 \lambda _5\right)+k\cdot p_0 C_q \left(-8 m_{\pi }^2 \lambda _1 \kappa _3^A-2 m_{a_1}^2 \lambda _2 \kappa
   _3^A-m_{a_1}^2 \lambda _4 \kappa _3^A-2 m_{a_1}^2 \lambda _5 \kappa _3^A-8 m_{\pi }^2 \kappa _{15}^A \lambda _1+4 m_\eta^2
   \kappa _8^A \lambda _2\right.\\
   -2 m_{a_1}^2 \kappa _{15}^A \lambda _2-4 m_\eta^2 \kappa _{16}^A \lambda _2+2 m_\eta^2 \kappa _8^A \lambda
   _4-m_{a_1}^2 \kappa _{15}^A \lambda _4-2 m_\eta^2 \kappa _{16}^A \lambda _4+4 m_\eta^2 \kappa _8^A \lambda _5-2 m_{a_1}^2 \kappa
   _{15}^A \lambda _5-4 m_\eta^2 \kappa _{16}^A \lambda _5\\\left.\left.
   +m_\pi^2 \left(\kappa _3^A+\kappa _{15}^A\right) \left(2 \lambda _2+\lambda
   _4+2 \lambda _5\right)+2 p\cdot p_0 \left(2 \lambda _2 \kappa _3^A+2 \kappa _8^A \lambda _2+2 \kappa _{15}^A \lambda
   _2+\kappa _8^A \lambda _4-\kappa _{16}^A \lambda _4+2 \kappa _8^A \lambda _5-2 \kappa _{16}^A \lambda
   _5\right)\right)\right) k\cdot p\\
   +2 \left(2 p\cdot p_0 C_q m_{\pi }^2 m_{a_1}^2 \lambda _1 \kappa _3^A+2 C_q m_{\pi }^2
   m_{a_1}^2 m_\eta^2 \lambda _1 \kappa _3^A-\left(p\cdot p_0\right){}^2 C_q m_{a_1}^2 \lambda _2 \kappa _3^A-p\cdot p_0 C_q
   m_{a_1}^2 m_\eta^2 \lambda _2 \kappa _3^A-4 C_q m_{\pi }^2 p_0^4 \kappa _8^A \lambda _1\right.\\
   -4 \left(p\cdot p_0\right){}^2 C_q
   m_{\pi }^2 \kappa _8^A \lambda _1-8 p\cdot p_0 C_q m_{\pi }^2 m_\eta^2 \kappa _8^A \lambda _1+8 C_q m_{\pi }^4 m_{a_1}^2
   \kappa _9^A \lambda _1+4 \sqrt{2} C_s m_{\pi }^4 m_{a_1}^2 \kappa _9^A \lambda _1\\
   -8 \sqrt{2} C_s m_K^2 m_{\pi }^2 m_{a_1}^2
   \kappa _9^A \lambda _1-4 p\cdot p_0 C_q m_{\pi }^4 \kappa _{10}^A \lambda _1-2 \sqrt{2} p\cdot p_0 C_s m_{\pi }^4 \kappa
   _{10}^A \lambda _1+4 \sqrt{2} p\cdot p_0 C_s m_K^2 m_{\pi }^2 \kappa _{10}^A \lambda _1\\
   -4 C_q m_{\pi }^4 m_\eta^2 \kappa
   _{10}^A \lambda _1-2 \sqrt{2} C_s m_{\pi }^4 m_\eta^2 \kappa _{10}^A \lambda _1+4 \sqrt{2} C_s m_K^2 m_{\pi }^2 m_\eta^2 \kappa
   _{10}^A \lambda _1+8 C_q m_{\pi }^4 m_{a_1}^2 \kappa _{11}^A \lambda _1-4 p\cdot p_0 C_q m_{\pi }^4 \kappa _{12}^A \lambda
   _1\\
   -4 C_q m_{\pi }^4 m_\eta^2 \kappa _{12}^A \lambda _1+2 p\cdot p_0 C_q m_{\pi }^2 m_{a_1}^2 \kappa _{15}^A \lambda _1+2 C_q
   m_{\pi }^2 m_{a_1}^2 m_\eta^2 \kappa _{15}^A \lambda _1+2 C_q m_{\pi }^2 p_0^4 \kappa _{16}^A \lambda _1+4 \left(p\cdot
   p_0\right){}^2 C_q m_{\pi }^2 \kappa _{16}^A \lambda _1\\
   +6 p\cdot p_0 C_q m_{\pi }^2 m_\eta^2 \kappa _{16}^A \lambda _1-2
   \left(p\cdot p_0\right){}^2 C_q m_{a_1}^2 \kappa _8^A \lambda _2-2 p\cdot p_0 C_q m_{a_1}^2 m_\eta^2 \kappa _8^A \lambda _2-2
   \left(p\cdot p_0\right){}^2 C_q m_\eta^2 \kappa _8^A \lambda _2-2 \left(p\cdot p_0\right){}^3 C_q \kappa _8^A \lambda _2\\
   +4
   \sqrt{2} p\cdot p_0 C_s m_K^2 m_{a_1}^2 \kappa _9^A \lambda _2-4 p\cdot p_0 C_q m_{\pi }^2 m_{a_1}^2 \kappa _9^A \lambda
   _2-2 \sqrt{2} p\cdot p_0 C_s m_{\pi }^2 m_{a_1}^2 \kappa _9^A \lambda _2+2 \sqrt{2} \left(p\cdot p_0\right){}^2 C_s m_K^2
   \kappa _{10}^A \lambda _2\\
   -2 \left(p\cdot p_0\right){}^2 C_q m_{\pi }^2 \kappa _{10}^A \lambda _2-\sqrt{2} \left(p\cdot
   p_0\right){}^2 C_s m_{\pi }^2 \kappa _{10}^A \lambda _2+2 \sqrt{2} p\cdot p_0 C_s m_K^2 m_{a_1}^2 \kappa _{10}^A \lambda
   _2-2 p\cdot p_0 C_q m_{\pi }^2 m_{a_1}^2 \kappa _{10}^A \lambda _2\\
   -\sqrt{2} p\cdot p_0 C_s m_{\pi }^2 m_{a_1}^2 \kappa
   _{10}^A \lambda _2-4 p\cdot p_0 C_q m_{\pi }^2 m_{a_1}^2 \kappa _{11}^A \lambda _2-2 \left(p\cdot p_0\right){}^2 C_q m_{\pi
   }^2 \kappa _{12}^A \lambda _2-2 p\cdot p_0 C_q m_{\pi }^2 m_{a_1}^2 \kappa _{12}^A \lambda _2\\
   -\left(p\cdot p_0\right){}^2
   C_q m_{a_1}^2 \kappa _{15}^A \lambda _2-p\cdot p_0 C_q m_{a_1}^2 m_\eta^2 \kappa _{15}^A \lambda _2+2 \left(p\cdot
   p_0\right){}^2 C_q m_{a_1}^2 \kappa _{16}^A \lambda _2+p\cdot p_0 C_q m_{a_1}^2 m_\eta^2 \kappa _{16}^A \lambda
   _2\\
   +\left(p\cdot p_0\right){}^2 C_q m_\eta^2 \kappa _{16}^A \lambda _2+2 \left(p\cdot p_0\right){}^3 C_q \kappa _{16}^A \lambda
   _2+2 \left(k\cdot p_0\right){}^2 p\cdot p_0 C_q \left(\kappa _{16}^A-\kappa _8^A\right) \lambda _2\\
   +m_\pi^2 \left(C_q
   \left(\kappa _3^A+\kappa _{15}^A+\kappa _{16}^A\right) \lambda _2 \left(p\cdot p_0\right){}^2+\left(2 \sqrt{2} C_s
   \left(m_{\pi }^2-2 m_K^2\right) \lambda _2 \kappa _9^A+C_q m_\eta^2 \left(\kappa _3^A+\kappa _{15}^A\right) \lambda _2\right.\right.\\\left.
   +C_q
   \left(m_{a_1}^2 \lambda _2 \kappa _{16}^A+m_{\pi }^2 \left(-2 \lambda _1 \kappa _3^A-2 \kappa _{15}^A \lambda _1+2 \kappa
   _{16}^A \lambda _1+4 \kappa _9^A \lambda _2+4 \kappa _{11}^A \lambda _2\right)\right)\right) p\cdot p_0\\\left.
   -2 m_{\pi }^2
   \left(2 \sqrt{2} C_s \left(m_{\pi }^2-2 m_K^2\right) \kappa _9^A+4 C_q m_{\pi }^2 \left(\kappa _9^A+\kappa
   _{11}^A\right)+C_q m_\eta^2 \left(\kappa _3^A+\kappa _{15}^A-\kappa _{16}^A\right)\right) \lambda _1\right)\nonumber
   \end{multline}
   
   \begin{multline}
   +k\cdot p_0
   \left(C_q \left(p\cdot p_0 \left(\kappa _3^A+\kappa _{15}^A+\kappa _{16}^A\right) \lambda _2-2 m_{\pi }^2 \left(\kappa
   _3^A+\kappa _{15}^A\right) \lambda _1\right) m_\pi^2+2 C_q m_{\pi }^2 \left(\left(\kappa _3^A+\kappa _{15}^A\right) m_{a_1}^2+2
   m_\eta^2 \left(\kappa _{16}^A-\kappa _8^A\right)\right) \lambda _1\right.\\
   +4 \left(p\cdot p_0\right){}^2 C_q \left(\kappa
   _{16}^A-\kappa _8^A\right) \lambda _2+p\cdot p_0 \left(\sqrt{2} C_s \left(2 m_K^2-m_{\pi }^2\right) \lambda _2 \kappa
   _{10}^A+C_q m_\eta^2 \left(\kappa _{16}^A-2 \kappa _8^A\right) \lambda _2\right.\\\left.\left.\left.\left.
   +C_q \left(-2 \left(2 \lambda _1 \kappa _8^A-2 \kappa
   _{16}^A \lambda _1+\left(\kappa _{10}^A+\kappa _{12}^A\right) \lambda _2\right) m_{\pi }^2-m_{a_1}^2 \left(\kappa _3^A+2
   \kappa _8^A+\kappa _{15}^A-2 \kappa _{16}^A\right) \lambda _2\right)\right)\right)\right)\right)\,,
\end{multline}

\begin{multline}
 a_3^{W^-\to (a_1^-)\eta\to\pi^-(\rho^0)\eta\to\pi^-\gamma\eta}=   
  +\frac{8F_V}{F^2m_{a_1}^2m_\rho^2D_{a_1}[(p+k)^2]}\left(2 C_q m_{a_1}^2 m_\eta^2 \lambda _2 \kappa
   _3^A-2 m_\pi^2 C_q m_\eta^2 \lambda _2 \kappa _3^A-4 k\cdot p C_q m_\eta^2
   \lambda _2 \kappa _3^A\right.\\
   +8 C_q m_{\pi }^4 \kappa _{10}^A \lambda _1+4 \sqrt{2} C_s m_{\pi
   }^4 \kappa _{10}^A \lambda _1-8 \sqrt{2} C_s m_K^2 m_{\pi }^2 \kappa
   _{10}^A \lambda _1+8 C_q m_{\pi }^4 \kappa _{12}^A \lambda _1-4 C_q
   m_{\pi }^2 m_{a_1}^2 \kappa _{16}^A \lambda _1-4 C_q m_{\pi }^2
   m_\eta^2 \kappa _{16}^A \lambda _1\\
   +4 C_q m_{a_1}^2 m_\eta^2 \kappa _8^A
   \lambda _2-4 k\cdot p C_q m_\eta^2 \kappa _8^A \lambda _2+8 \sqrt{2}
   m_\pi^2 C_s m_K^2 \kappa _9^A \lambda _2+16 \sqrt{2} k\cdot p C_s m_K^2
   \kappa _9^A \lambda _2-8 m_\pi^2 C_q m_{\pi }^2 \kappa _9^A \lambda
   _2\\
   -16 k\cdot p C_q m_{\pi }^2 \kappa _9^A \lambda _2-4 \sqrt{2} m_\pi^2
   C_s m_{\pi }^2 \kappa _9^A \lambda _2-8 \sqrt{2} k\cdot p C_s m_{\pi
   }^2 \kappa _9^A \lambda _2-8 \sqrt{2} C_s m_K^2 m_{a_1}^2 \kappa
   _9^A \lambda _2+8 C_q m_{\pi }^2 m_{a_1}^2 \kappa _9^A \lambda _2\\
   +4
   \sqrt{2} C_s m_{\pi }^2 m_{a_1}^2 \kappa _9^A \lambda _2+4 \sqrt{2}
   k\cdot p C_s m_K^2 \kappa _{10}^A \lambda _2-4 k\cdot p C_q m_{\pi
   }^2 \kappa _{10}^A \lambda _2-2 \sqrt{2} k\cdot p C_s m_{\pi }^2
   \kappa _{10}^A \lambda _2-4 \sqrt{2} C_s m_K^2 m_{a_1}^2 \kappa
   _{10}^A \lambda _2\\
   +4 C_q m_{\pi }^2 m_{a_1}^2 \kappa _{10}^A \lambda
   _2+2 \sqrt{2} C_s m_{\pi }^2 m_{a_1}^2 \kappa _{10}^A \lambda _2-8
   m_\pi^2 C_q m_{\pi }^2 \kappa _{11}^A \lambda _2-16 k\cdot p C_q m_{\pi
   }^2 \kappa _{11}^A \lambda _2+8 C_q m_{\pi }^2 m_{a_1}^2 \kappa
   _{11}^A \lambda _2\\
   -4 k\cdot p C_q m_{\pi }^2 \kappa _{12}^A \lambda
   _2+4 C_q m_{\pi }^2 m_{a_1}^2 \kappa _{12}^A \lambda _2+2 C_q
   m_{a_1}^2 m_\eta^2 \kappa _{15}^A \lambda _2-2 m_\pi^2 C_q m_\eta^2 \kappa
   _{15}^A \lambda _2-4 k\cdot p C_q m_\eta^2 \kappa _{15}^A \lambda _2-2
   m_\pi^2 C_q m_{a_1}^2 \kappa _{16}^A \lambda _2\\
   -2 k\cdot p C_q m_{a_1}^2
   \kappa _{16}^A \lambda _2-2 C_q m_{a_1}^2 m_\eta^2 \kappa _{16}^A
   \lambda _2+2 k\cdot p C_q m_\eta^2 \kappa _{16}^A \lambda _2+4
   \left(k\cdot p_0\right){}^2 C_q \left(\kappa _8^A-\kappa
   _{16}^A\right) \lambda _2\\
   +4 \left(p\cdot p_0\right){}^2 C_q
   \left(\kappa _8^A-\kappa _{16}^A\right) \lambda _2-2 k\cdot p C_q
   m_\eta^2 \kappa _8^A \lambda _4+2 \sqrt{2} k\cdot p C_s m_K^2 \kappa
   _{10}^A \lambda _4-2 k\cdot p C_q m_{\pi }^2 \kappa _{10}^A \lambda
   _4-\sqrt{2} k\cdot p C_s m_{\pi }^2 \kappa _{10}^A \lambda _4\\
   -2
   k\cdot p C_q m_{\pi }^2 \kappa _{12}^A \lambda _4+k\cdot p C_q
   m_{a_1}^2 \kappa _{16}^A \lambda _4+k\cdot p C_q m_\eta^2 \kappa
   _{16}^A \lambda _4-4 k\cdot p C_q m_\eta^2 \kappa _8^A \lambda _5+4
   \sqrt{2} k\cdot p C_s m_K^2 \kappa _{10}^A \lambda _5\\
   -4 k\cdot p C_q
   m_{\pi }^2 \kappa _{10}^A \lambda _5-2 \sqrt{2} k\cdot p C_s m_{\pi
   }^2 \kappa _{10}^A \lambda _5-4 k\cdot p C_q m_{\pi }^2 \kappa
   _{12}^A \lambda _5+2 k\cdot p C_q m_{a_1}^2 \kappa _{16}^A \lambda
   _5+2 k\cdot p C_q m_\eta^2 \kappa _{16}^A \lambda _5\\
   +8 C_q m_{\pi }^2 m_\eta^2 \kappa _8^A \lambda
   _1+2 p\cdot p_0
   \left(C_q \left(-\left(m_\pi^2-m_{a_1}^2+2 k\cdot p\right) \lambda _2
   \kappa _3^A+2 m_{a_1}^2 \kappa _8^A \lambda _2+2 m_\eta^2 \kappa _8^A
   \lambda _2\right.\right.\\
   -2 k\cdot p \kappa _8^A \lambda _2-m_\pi^2 \kappa _{15}^A
   \lambda _2+m_{a_1}^2 \kappa _{15}^A \lambda _2-2 k\cdot p \kappa
   _{15}^A \lambda _2-m_\pi^2 \kappa _{16}^A \lambda _2-2 m_{a_1}^2 \kappa
   _{16}^A \lambda _2-m_\eta^2 \kappa _{16}^A \lambda _2\\\left.
   +2 m_{\pi }^2
   \left(2 \lambda _1 \kappa _8^A-2 \kappa _{16}^A \lambda
   _1+\left(\kappa _{10}^A+\kappa _{12}^A\right) \lambda
   _2\right)-k\cdot p \kappa _8^A \lambda _4+k\cdot p \kappa _{16}^A
   \lambda _4-2 k\cdot p \kappa _8^A \lambda _5+2 k\cdot p \kappa
   _{16}^A \lambda _5\right)\\\left.
   -\sqrt{2} C_s \left(2 m_K^2-m_{\pi
   }^2\right) \kappa _{10}^A \lambda _2\right)+2 k\cdot p_0 \left(C_q
   \left(-\left(m_\pi^2-m_{a_1}^2+2 k\cdot p\right) \lambda _2 \kappa
   _3^A+2 m_{a_1}^2 \kappa _8^A \lambda _2+2 m_\eta^2 \kappa _8^A \lambda
   _2-2 k\cdot p \kappa _8^A \lambda _2\right.\right.\\
   +4 p\cdot p_0 \kappa _8^A
   \lambda _2-m_\pi^2 \kappa _{15}^A \lambda _2+m_{a_1}^2 \kappa _{15}^A
   \lambda _2-2 k\cdot p \kappa _{15}^A \lambda _2-m_\pi^2 \kappa _{16}^A
   \lambda _2-2 m_{a_1}^2 \kappa _{16}^A \lambda _2-m_\eta^2 \kappa
   _{16}^A \lambda _2-4 p\cdot p_0 \kappa _{16}^A \lambda _2\\\left.
   +2 m_{\pi
   }^2 \left(2 \lambda _1 \kappa _8^A-2 \kappa _{16}^A \lambda
   _1+\left(\kappa _{10}^A+\kappa _{12}^A\right) \lambda
   _2\right)-k\cdot p \kappa _8^A \lambda _4+k\cdot p \kappa _{16}^A
   \lambda _4-2 k\cdot p \kappa _8^A \lambda _5+2 k\cdot p \kappa
   _{16}^A \lambda _5\right)\\\left.\left.
   -\sqrt{2} C_s \left(2 m_K^2-m_{\pi
   }^2\right) \kappa _{10}^A \lambda _2\right)\right)\,,
\end{multline}

\begin{multline}
 a_4^{W^-\to (a_1^-)\eta\to\pi^-(\rho^0)\eta\to\pi^-\gamma\eta}=   
  +\frac{16F_V}{F^2m_{a_1}^2m_\rho^2D_{a_1}[(p+k)^2]}\left(C_q \left(-\left(p\cdot p_0\right) m_{a_1}^2
   \lambda _2 \kappa _3^A+m_\pi^2 m_\eta^2 \lambda _2 \kappa _3^A-m_{a_1}^2
   m_\eta^2 \lambda _2 \kappa _3^A\right.\right.\\
   +2 k\cdot p m_\eta^2 \lambda _2 \kappa
   _3^A+m_\pi^2 p\cdot p_0 \lambda _2 \kappa _3^A+2 k\cdot p p\cdot p_0
   \lambda _2 \kappa _3^A+4 (k\cdot p)^2 \kappa _8^A \lambda _2-2
   \left(p\cdot p_0\right){}^2 \kappa _8^A \lambda _2-2 k\cdot p
   m_{a_1}^2 \kappa _8^A \lambda _2\\
   -2 p\cdot p_0 m_{a_1}^2 \kappa _8^A
   \lambda _2-2 m_{a_1}^2 m_\eta^2 \kappa _8^A \lambda _2-2 p\cdot p_0
   m_\eta^2 \kappa _8^A \lambda _2+2 m_\pi^2 k\cdot p \kappa _8^A \lambda
   _2-p\cdot p_0 m_{a_1}^2 \kappa _{15}^A \lambda _2+m_\pi^2 m_\eta^2 \kappa
   _{15}^A \lambda _2\\
   -m_{a_1}^2 m_\eta^2 \kappa _{15}^A \lambda _2+2
   k\cdot p m_\eta^2 \kappa _{15}^A \lambda _2+m_\pi^2 p\cdot p_0 \kappa
   _{15}^A \lambda _2+2 k\cdot p p\cdot p_0 \kappa _{15}^A \lambda _2+2
   \left(p\cdot p_0\right){}^2 \kappa _{16}^A \lambda _2+m_\pi^2 m_{a_1}^2
   \kappa _{16}^A \lambda _2+2 k\cdot p m_{a_1}^2 \kappa _{16}^A
   \lambda _2\\
   +2 p\cdot p_0 m_{a_1}^2 \kappa _{16}^A \lambda
   _2+m_{a_1}^2 m_\eta^2 \kappa _{16}^A \lambda _2+p\cdot p_0 m_\eta^2 \kappa
   _{16}^A \lambda _2+m_\pi^2 p\cdot p_0 \kappa _{16}^A \lambda _2+2 k\cdot
   p p\cdot p_0 \kappa _{16}^A \lambda _2-2 \left(k\cdot p_0\right){}^2
   \left(\kappa _8^A-\kappa _{16}^A\right) \lambda _2\\
   +k\cdot p_0
   \left(\left(m_\pi^2-m_{a_1}^2+2 k\cdot p\right) \kappa _3^A-2 m_{a_1}^2
   \kappa _8^A-2 m_\eta^2 \kappa _8^A+m_\pi^2 \kappa _{15}^A-m_{a_1}^2 \kappa
   _{15}^A+2 k\cdot p \kappa _{15}^A+m_\pi^2 \kappa _{16}^A+2 m_{a_1}^2
   \kappa _{16}^A+m_\eta^2 \kappa _{16}^A\right.\\\left.
   +2 k\cdot p \kappa _{16}^A-4
   p\cdot p_0 \left(\kappa _8^A-\kappa _{16}^A\right)\right) \lambda
   _2-2 m_{\pi }^2 \left(2 \left(m_\pi^2-m_{a_1}^2+2 k\cdot p\right) \kappa
   _8^A \lambda _1-\left(2 \left(m_\pi^2-m_{a_1}^2+2 k\cdot p\right) \kappa
   _9^A-m_{a_1}^2 \kappa _{10}^A\right.\right.\\\left.\left.
   -p\cdot p_0 \kappa _{10}^A+2 m_\pi^2 \kappa
   _{11}^A-2 m_{a_1}^2 \kappa _{11}^A+4 k\cdot p \kappa
   _{11}^A-m_{a_1}^2 \kappa _{12}^A-p\cdot p_0 \kappa _{12}^A-k\cdot
   p_0 \left(\kappa _{10}^A+\kappa _{12}^A\right)\right) \lambda
   _2\right)+2 (k\cdot p)^2 \kappa _8^A \lambda _4\\\left.
   -k\cdot p m_{a_1}^2
   \kappa _8^A \lambda _4+m_\pi^2 k\cdot p \kappa _8^A \lambda _4+4 (k\cdot
   p)^2 \kappa _8^A \lambda _5-2 k\cdot p m_{a_1}^2 \kappa _8^A \lambda
   _5+2 m_\pi^2 k\cdot p \kappa _8^A \lambda _5\right)\\\left.
   -\sqrt{2} C_s \left(2
   m_K^2-m_{\pi }^2\right) \left(2 \left(m_\pi^2-m_{a_1}^2+2 k\cdot
   p\right) \kappa _9^A-\left(m_{a_1}^2+k\cdot p_0+p\cdot p_0\right)
   \kappa _{10}^A\right) \lambda _2\right)\,.
\end{multline}

We turn now to the vector factors, with the one-resonance exchange contributions (fig.~\ref{fig:1R-FV}) listed next:\\

\begin{multline}\label{v1-1R-RChL}
 v_1^{1R} =
 \frac{4\sqrt{2}C_q}{3F^2M_V^2m_\rho^2D_\rho[(p+k)^2]}\left(2 k\cdot p \left(4 \left(c_5+c_7\right) \left(m_{\rho }^2-2 m_{\pi }^2\right) 
 \left(\left(c_5+c_7\right) \left(k\cdot p_0+m_{\eta
   }^2\right)+4 c_3 \left(m_{\pi }^2-m_{\eta }^2\right)\right)\right.\right.\\\left.
   +p\cdot p_0 \left(8 \left(c_5+c_7\right) \left(2 c_3 m_{\eta }^2-\left(2
   c_3+c_5+c_7\right) m_{\pi }^2\right)+\left(4 \left(c_5+c_7\right){}^2+c_{1256}^2\right) m_{\rho }^2\right)\right)\\
   +4
   \left(c_5+c_7\right) m_{\pi }^2 \left(m_{\rho }^2-m_{\pi }^2\right) \left(\left(c_5+c_7\right) \left(k\cdot p_0+m_{\eta }^2\right)+4
   c_3 \left(m_{\pi }^2-m_{\eta }^2\right)\right)\\
   -16 \left(c_5+c_7\right) (k\cdot p)^2 \left(\left(c_5+c_7\right) \left(k\cdot p_0+m_{\eta
   }^2+p\cdot p_0\right)+4 c_3 \left(m_{\pi }^2-m_{\eta }^2\right)\right)\\\left.
   +p\cdot p_0 \left(-8 c_3 \left(m_{\pi }^2-m_{\eta }^2\right)
   \left(\left(c_{1256}-2 \left(c_5+c_7\right)\right) m_{\rho }^2+2 \left(c_5+c_7\right) m_{\pi }^2\right)+\left(4
   \left(c_5+c_7\right){}^2+c_{1256}^2\right) m_{\pi }^2 m_{\rho }^2
   -4 \left(c_5+c_7\right){}^2 m_{\pi }^4\right)\right)\\
  +\frac{4\sqrt{2}C_q}{3F^2M_V^2m_\omega^2D_\omega[(p_0+k)^2]}\left(16 c_3 m_{\pi }^2 k\cdot p_0 \left(\left(c_5+c_7\right) \left(2 k\cdot
   p_0+m_{\eta }^2\right)+4 c_3 \left(m_{\eta }^2-m_{\pi
   }^2\right)\right)\right.\\
   -m_{\omega }^2 \left(\left(c_{1256} \left(8 c_3+3
   c_{1256}\right) m_{\eta }^2+16 c_3 \left(c_5+c_7\right) m_{\pi
   }^2\right) k\cdot p_0+2 c_{1256}^2 \left(k\cdot p_0\right){}^2+64
   c_3^2 m_{\pi }^2 m_{\eta }^2\right.\\\left.\left.
   +c_{1256} \left(8 c_3+c_{1256}\right)
   m_{\eta }^4-64 c_3^2 m_{\pi }^4\right)\right)\\
 -\frac{8\sqrt{2}C_q\lambda_{15}^{S}}{3F^2D_{a_0}[(p+p_0)^2]}\left(c_mm_{\pi }^2 +c_d p\cdot p_0\right)\\
  -\frac{2C_qF_V}{3F^2m_\omega^2}\left(-\left(\lambda _{13}^V-\lambda _{14}^V-\lambda
   _{15}^V\right) \left(k\cdot p_0+m_\eta^2+p\cdot p_0\right)+4 m_{\pi }^2
   \lambda _6^V+2 p\cdot p_0 \left(\lambda _{11}^V+\lambda
   _{12}^V\right)\right)\,,
   \\
\end{multline}

\begin{multline}\label{v2-1R-RChL}
 v_2^{1R} =
 \frac{4\sqrt{2}C_q}{3F^2M_V^2m_\rho^2D_\rho[(p+k)^2]}\left(m_{\rho }^2 \left(-\left(\left(3 c_{1256}^2 m_{\pi }^2-8 c_3 \left(2 \left(c_5+c_7\right)+c_{1256}\right) 
 \left(m_{\pi }^2-m_{\eta
   }^2\right)\right) k\cdot p+2 c_{1256}^2 (k\cdot p)^2\right.\right.\right.\\\left.\left.\left.
   +c_{1256} m_{\pi }^2 \left(8 c_3 m_{\eta }^2+\left(c_{1256}-8 c_3\right) m_{\pi
   }^2\right)\right)\right)-16 c_3 \left(c_5+c_7\right) \left(m_{\pi }^2
   -m_{\eta }^2\right) k\cdot p \left(2 k\cdot p+m_{\pi }^2\right)\right)\\
  -\frac{16\sqrt{2}C_q}{3F^2M_V^2m_\omega^2D_\omega[(p_0+k)^2]}\left(-\frac{1}{2} k\cdot p_0 \left(p\cdot p_0 \left(8
   \left(c_5+c_7\right) \left(4 c_3 m_{\pi }^2-\left(2
   c_3+c_5+c_7\right) m_{\eta }^2\right)\right.\right.\right.\\\left.\left.
   +\left(4
   \left(c_5+c_7\right){}^2+c_{1256}^2\right) m_{\omega }^2\right)-4
   \left(4 c_3-c_5-c_7\right) m_{\pi }^2 \left(-2 \left(2
   c_3+c_5+c_7\right) m_{\eta }^2+\left(c_5+c_7\right) m_{\omega }^2+4
   c_3 m_{\pi }^2\right)\right)\\
   +\left(c_5+c_7\right) k\cdot p
   \left(\left(c_5+c_7\right) \left(2 k\cdot p_0+m_{\eta }^2\right)+4
   c_3 \left(m_{\eta }^2-m_{\pi }^2\right)\right) \left(2 k\cdot
   p_0+m_{\eta }^2-m_{\omega }^2\right)\\\left.+4 \left(c_5+c_7\right)
   \left(k\cdot p_0\right){}^2 \left(\left(c_5+c_7\right) \left(m_{\pi
   }^2+p\cdot p_0\right)-4 c_3 m_{\pi }^2\right)+\frac{1}{4} m_{\omega
   }^2 \left(16 c_3 m_{\pi }^2 \left(\left(c_5+c_7\right) \left(m_{\pi
   }^2+2 p\cdot p_0\right)-4 c_3 m_{\pi }^2\right)\right.\right.\\\left.
   -m_{\eta }^2 \left(4
   \left(\left(c_5+c_7\right){}^2-16 c_3^2\right) m_{\pi
   }^2+\left(c_{1256}^2+8 c_3 c_{1256}+4 \left(c_5+c_7\right) \left(4
   c_3+c_5+c_7\right)\right) p\cdot p_0\right)\right)\\\left.
   +\left(4 c_3 m_{\pi
   }^2-\left(4 c_3+c_5+c_7\right) m_{\eta }^2\right) \left(4 c_3 m_{\pi
   }^2 p\cdot p_0-m_{\eta }^2 \left(\left(c_5+c_7\right) \left(m_{\pi
   }^2+p\cdot p_0\right)-4 c_3 m_{\pi }^2\right)\right)\right)\\
+ \frac{8\sqrt{2}C_q\lambda_{15}^{S}}{3F^2D_{a_0}[(p+p_0)^2]}\left(c_mm_{\pi }^2 +c_d p\cdot p_0\right)\\
  -\frac{2C_qF_V}{3F^2m_\omega^2}\left(-\left(\lambda _{13}^V-\lambda _{14}^V-\lambda
   _{15}^V\right) \left(k\cdot p_0+m_\eta^2+p\cdot p_0\right)+4 m_{\pi }^2
   \lambda _6^V+2 p\cdot p_0 \left(\lambda _{11}^V+\lambda
   _{12}^V\right)\right)\,,
 \\
\end{multline}

\begin{multline}\label{v3-1R-RChL}
 v_3^{1R} = 
 -\frac{4\sqrt{2}C_q}{3F^2M_V^2m_\rho^2D_\rho[(p+k)^2]}\left(16 c_3 \left(c_5+c_7\right) \left(m_{\pi }^2-m_{\eta }^2\right)
   \left(2 k\cdot p+m_{\pi }^2\right)-m_{\rho }^2 \left(2 c_{1256}^2 k\cdot p\right.\right.\\\left.\left.-8 c_3 \left(2 \left(c_2+4 c_3+c_7\right)-c_{1256}\right)
   m_{\eta }^2+\left(c_{1256}^2+8 c_3 \left(2 \left(c_5+c_7\right)-c_{1256}\right)\right) m_{\pi }^2\right.\right.\\\left.\left.
   +2 \left(4 c_3-c_5\right) \left(2 c_5-2 c_6+c_{1256}\right) p\cdot p_0\right)\right)\\
  +\frac{16\sqrt{2}C_q}{3F^2M_V^2m_\omega^2D_\omega[(p_0+k)^2]}\left(c_5+c_7\right) \left(\left(c_5+c_7\right) \left(2 k\cdot p_0+m_{\eta
   }^2\right)+4 c_3 \left(m_{\eta }^2-m_{\pi }^2\right)\right) \left(2 k\cdot p_0+m_{\eta }^2-m_{\omega }^2\right)\\
  -\frac{2C_qF_V}{3F^2m_\omega^2} \left(\lambda _{13}^V-\lambda _{14}^V-\lambda_{15}^V\right)\,,
   \\
\end{multline}

\begin{multline}\label{v4-1R-RChL}
 v_4^{1R} = 
 -\frac{4\sqrt{2}C_q}{3F^2M_V^2m_\rho^2D_\rho[(p+k)^2]}4 \left(c_5+c_7\right){}^2 \left(2 k\cdot p+m_{\pi }^2\right) \left(2
   k\cdot p-m_{\rho }^2+m_{\pi }^2\right)\\
  -\frac{4\sqrt{2}C_q}{3F^2M_V^2m_\omega^2D_\omega[(p_0+k)^2]}\left(m_{\omega }^2 \left(c_{1256} \left(c_{1256} \left(2
   k\cdot p_0+m_{\eta }^2\right)+8 c_3 m_{\eta }^2\right)-16 c_3
   \left(c_5+c_7\right) m_{\pi }^2\right)\right.\\\left.
   +16 c_3 m_{\pi }^2
   \left(\left(c_5+c_7\right) \left(2 k\cdot p_0+m_{\eta }^2\right)+4
   c_3 \left(m_{\eta }^2-m_{\pi }^2\right)\right)\right)\\
  +\frac{2C_qF_V}{3F^2m_\omega^2} \left(\lambda _{13}^V-\lambda _{14}^V-\lambda_{15}^V\right)\,.
   \\
\end{multline}

Finally, we will give the two-resonance mediated contributions to the vector form factors (figure \ref{fig:2R-FV}):\\

\begin{multline}\label{v1-2R-RChL}
 v_1^{2R}=
  \frac{8F_VC_q}{3F^2M_Vm_\omega^2D_\rho[(p+p_0+k)^2]D_\omega[(p_0+k)^2]}\left(m_{\omega }^2 \left(k\cdot p_0 \left(2 \left(8
   c_3+3 c_{1256}\right) d_3 m_{\eta }^2\right.\right.\right.\\\left.
   +m_{\pi }^2 \left(-\left(16 c_3
   d_3+\left(2 \left(c_5+c_7\right)+c_{1256}\right)
   d_{12}\right)\right)+2 \left(2 \left(c_5+c_7\right)+c_{1256}\right)
   d_3 p\cdot p_0\right)+4 c_{1256} d_3 \left(k\cdot
   p_0\right){}^2\\\left.
   +\left(8 c_3 m_{\pi }^2-\left(8 c_3+c_{1256}\right)
   m_{\eta }^2\right) \left(d_{12} m_{\pi }^2-2 d_3 \left(m_{\eta
   }^2+p\cdot p_0\right)\right)\right)\\
   +2 d_3 k\cdot p \left(m_{\omega
   }^2 \left(\left(2 \left(c_5+c_7\right)+c_{1256}\right) k\cdot
   p_0+\left(8 c_3+c_{1256}\right) m_{\eta }^2-8 c_3 m_{\pi }^2\right)\right.\\\left.
   +2
   k\cdot p_0 \left(4 c_3 \left(m_{\pi }^2-m_{\eta
   }^2\right)-\left(c_5+c_7\right) \left(2 k\cdot p_0+m_{\eta
   }^2\right)\right)\right)\\\left.
   -2 k\cdot p_0 \left(2 d_3 p\cdot p_0-d_{12}
   m_{\pi }^2\right) \left(\left(c_5+c_7\right) \left(2 k\cdot
   p_0+m_{\eta }^2\right)+4 c_3 \left(m_{\eta }^2-m_{\pi
   }^2\right)\right)\right)\\
  -\frac{8F_VC_q}{3F^2M_Vm_\rho^2D_\rho[(p+p_0+k)^2]D_\rho[(p+k)^2]}\left(-16 c_5 d_2 m_{\pi }^6\right.\\
  +16 c_5
   d_2 m_{\rho }^2 m_{\pi }^4
   +16 c_7 d_2 m_{\rho }^2 m_{\pi }^4-16
   p\cdot p_0 c_5 d_2 m_{\pi }^4-16 p\cdot p_0 c_7 d_2 m_{\pi }^4-2
   p\cdot p_0 c_5 d_3 m_{\pi }^4-2 p\cdot p_0 c_7 d_3 m_{\pi }^4-2
   p\cdot p_0 c_5 d_4 m_{\pi }^4\\
   -2 p\cdot p_0 c_7 d_4 m_{\pi }^4+16
   p\cdot p_0 c_5 d_2 m_{\rho }^2 m_{\pi }^2+16 p\cdot p_0 c_7 d_2
   m_{\rho }^2 m_{\pi }^2-8 p\cdot p_0 c_{1256} d_2 m_{\rho }^2 m_{\pi
   }^2+2 p\cdot p_0 c_5 d_3 m_{\rho }^2 m_{\pi }^2\\
   +2 p\cdot p_0 c_7 d_3
   m_{\rho }^2 m_{\pi }^2-2 p\cdot p_0 c_{1256} d_3 m_{\rho }^2 m_{\pi
   }^2+2 p\cdot p_0 c_5 d_4 m_{\rho }^2 m_{\pi }^2+2 p\cdot p_0 c_7 d_4
   m_{\rho }^2 m_{\pi }^2-4 \left(p\cdot p_0\right){}^2 c_5 d_3 m_{\pi
   }^2-16 c_7 d_2 m_{\pi }^6\\
   -4 \left(p\cdot p_0\right){}^2 c_7 d_3 m_{\pi }^2+4 \left(p\cdot
   p_0\right){}^2 c_5 d_3 m_{\rho }^2+4 \left(p\cdot p_0\right){}^2 c_7
   d_3 m_{\rho }^2-2 \left(p\cdot p_0\right){}^2 c_{1256} d_3 m_{\rho
   }^2\\
   -8 (k\cdot p)^2 \left(c_5+c_7\right)
   \left(\left(-d_3+d_4+d_{12}\right) m_{\eta }^2+\left(k\cdot
   p_0+p\cdot p_0\right) \left(d_3+d_4\right)+8 d_2 \left(m_{\pi
   }^2-m_{\eta }^2\right)\right)\\
   +2 k\cdot p_0 \left(\left(c_5+c_7\right)
   \left(d_3+d_4\right) \left(m_{\rho }^2-m_{\pi }^2\right) m_{\pi
   }^2+p\cdot p_0 d_3 \left(\left(2 \left(c_5+c_7\right)-c_{1256}\right)
   m_{\rho }^2-2 \left(c_5+c_7\right) m_{\pi }^2\right)\right)\\
   -m_{\eta
   }^2 \left(2 \left(c_5+c_7\right) \left(8 d_2+d_3-d_4-d_{12}\right)
   m_{\pi }^2 \left(m_{\rho }^2-m_{\pi }^2\right)-p\cdot p_0 \left(8
   d_2-d_{12}\right) \left(2 \left(c_5+c_7\right) m_{\pi
   }^2+\left(c_{1256}-2 \left(c_5+c_7\right)\right) m_{\rho
   }^2\right)\right)\\
   +4 k\cdot p \left(-2 \left(c_5+c_7\right) d_3
   \left(p\cdot p_0\right){}^2+\left(\left(\left(c_5+c_7-c_{1256}\right)
   d_3\right.\right.\right.\\\left.\left.
   +\left(c_5+c_7\right) d_4\right) m_{\rho }^2+\left(c_5+c_7\right)
   \left(\left(8 d_2-d_{12}\right) m_{\eta }^2-2 \left(4
   d_2+d_3+d_4\right) m_{\pi }^2\right)\right) p\cdot
   p_0\\\left.\left.
   -\left(c_5+c_7\right) \left(8 d_2 m_{\pi }^2+\left(-8
   d_2-d_3+d_4+d_{12}\right) m_{\eta }^2\right) \left(2 m_{\pi
   }^2-m_{\rho }^2\right)-k\cdot p_0 \left(c_5+c_7\right) \left(2 p\cdot
   p_0 d_3+\left(d_3+d_4\right) \left(2 m_{\pi }^2-m_{\rho
   }^2\right)\right)\right)\right)\\
 +\frac{8F_VC_q\lambda^{SV}_3}{3F^2D_\rho[(p + p_0 + k)^2]D_{a_0}[(p+p_0)^2]}\left(c_d p\cdot p_0+c_m m_{\pi
   }^2\right)\\
  -\frac{16F_VC_q}{3F^2M_Vm_\rho^2 m_\omega^2D_\rho[(p+k)^2]}\left(k\cdot p \left(2 \left(\left(c_5+c_7\right)
   k\cdot p_0+\left(c_5+c_7\right) m_{\eta }^2+4 c_3 \left(m_{\pi
   }^2-m_{\eta }^2\right)\right)\right.\right.\\
   \left(m_{\pi }^2 (d_1+8
   d_2+d_3+2 d_4)-d_4 m_{\rho }^2\right)+p\cdot
   p_0 \left(2 c_5 \left(m_{\pi }^2 (d_1+8 d_2+d_3+2
   d_4)-d_4 m_{\rho }^2\right)+2 c_7 d_1 m_{\pi }^2+16
   c_7 d_2 m_{\pi }^2\right.\\\left.\left.
   +c_{1256} d_3 m_{\rho }^2+2 c_7
   d_3 m_{\pi }^2+8 c_3 d_4 \left(m_{\pi }^2-m_{\eta
   }^2\right)-2 c_7 d_4 m_{\rho }^2+4 c_7 d_4 m_{\pi
   }^2\right)\right)\\
   +m_{\pi }^2 \left(m_{\pi }^2-m_{\rho }^2\right)
   (d_1+8 d_2+d_3+d_4)
   \left(\left(c_5+c_7\right) k\cdot p_0+\left(c_5+c_7\right) m_{\eta
   }^2+4 c_3 \left(m_{\pi }^2-m_{\eta }^2\right)\right)\\
   +\frac{1}{2}
   p\cdot p_0 \left(8 c_3 \left(m_{\pi }^2-m_{\eta }^2\right)
   \left(m_{\pi }^2 (d_1+8
   d_2+d_3+d_4)-(d_3+d_4) m_{\rho
   }^2\right)+m_{\pi }^2 \left(2 c_5 \left(m_{\pi }^2-m_{\rho }^2\right)
   (d_1+8 d_2+d_3+d_4)\right.\right.\\\left.\left.
   +2 c_7 \left(m_{\pi
   }^2-m_{\rho }^2\right) (d_1+8
   d_2+d_3+d_4)+c_{1256} (-(d_1+8 d_2))
   m_{\rho }^2\right)\right)\\\left.
   +4 d_4 (k\cdot p)^2
   \left(\left(c_5+c_7\right) k\cdot p_0-4 c_3 m_{\eta }^2+c_5 m_{\eta
   }^2+c_7 m_{\eta }^2+4 c_3 m_{\pi }^2+\left(c_5+c_7\right) p\cdot
   p_0\right)\right)\\   
  +\frac{\sqrt{2}F_V^2C_q}{3F^2m_\rho^2m_\omega^2D_\rho[(p+p_0+k)^2]}\left(2 k\cdot p_0+m_{\eta }^2+m_{\rho
   }^2+2 p\cdot p_0\right) \\
   \left(\left(\lambda _3^{\text{VV}}+\lambda
   _4^{\text{VV}}+2 \lambda _5^{\text{VV}}\right) \left(k\cdot
   p_0+m_{\eta }^2+p\cdot p_0\right)+4 m_{\pi }^2 \lambda
   _6^{\text{VV}}+4 p\cdot p_0 \left(\lambda _1^{\text{VV}}+\lambda
   _2^{\text{VV}}\right)\right)\,, 
 \\
\end{multline}
where $d_{12}\equiv d_1+8d_2$ is fixed by the short-distance constraints (\ref{eq: Consistent set of relations}).\\

\begin{multline}\label{v2-2R-RChL}
 v_2^{2R}=
  \frac{8F_VC_q}{3F^2M_Vm_\omega^2D_\rho[(p+p_0+k)^2]D_\omega[(p_0+k)^2]}\left(8 c_3 d_3 m_{\omega }^2 m_{\pi }^4-8 c_3 d_4
   m_{\omega }^2 m_{\pi }^4+8 c_3 d_{12} m_{\omega }^2 m_{\pi }^4-8
   p\cdot p_0 c_3 d_{12} m_{\pi }^4\right.\\
   +64 p\cdot p_0 c_3 d_2 m_{\omega }^2
   m_{\pi }^2-16 p\cdot p_0 c_5 d_2 m_{\omega }^2 m_{\pi }^2+8 p\cdot
   p_0 c_3 d_3 m_{\omega }^2 m_{\pi }^2-8 p\cdot p_0 c_3 d_4 m_{\omega
   }^2 m_{\pi }^2-8 p\cdot p_0 c_3 d_{12} m_{\omega }^2 m_{\pi }^2\\
   -2
   p\cdot p_0 c_7 d_{12} m_{\omega }^2 m_{\pi }^2+p\cdot p_0 c_{1256}
   d_{12} m_{\omega }^2 m_{\pi }^2+8 p\cdot p_0 c_3 \left(d_{12}-8
   d_2\right) m_{\omega }^2 m_{\pi }^2-2 p\cdot p_0 c_5 \left(d_{12}-8
   d_2\right) m_{\omega }^2 m_{\pi }^2\\
   +16 \left(p\cdot p_0\right){}^2
   c_3 d_3 m_{\pi }^2-2 \left(4 c_3+c_5+c_7\right) \left(p\cdot p_0
   \left(d_3+d_4\right)-\left(d_3-d_4+d_{12}\right) m_{\pi }^2\right)
   m_{\eta }^4+4 \left(p\cdot p_0\right){}^2 c_5 d_3 m_{\omega }^2\\
   +4
   \left(p\cdot p_0\right){}^2 c_7 d_3 m_{\omega }^2-2 \left(p\cdot
   p_0\right){}^2 c_{1256} d_3 m_{\omega }^2+8 \left(k\cdot
   p_0\right){}^2 \left(c_5+c_7\right) \left(\left(d_3-d_4+d_{12}\right)
   m_{\pi }^2-p\cdot p_0 \left(d_3+d_4\right)\right)\\
   +2 m_{\eta }^2
   \left(-2 \left(4 c_3+c_5+c_7\right) d_3 \left(p\cdot
   p_0\right){}^2+\left(\left(\left(c_5+c_7\right) d_{12}+4 c_3
   \left(d_3+d_4+d_{12}\right)\right) m_{\pi }^2+\left(\left(-4
   c_3+c_5+c_7-c_{1256}\right) d_3\right.\right.\right.\\\left.\left.\left.
   +\left(4 c_3+c_5+c_7\right) d_4\right)
   m_{\omega }^2\right) p\cdot p_0-\left(d_3-d_4+d_{12}\right) m_{\pi
   }^2 \left(4 c_3 m_{\pi }^2+\left(4 c_3+c_5+c_7\right) m_{\omega
   }^2\right)\right)\\
   +4 k\cdot p_0 \left(-2 \left(c_5+c_7\right) d_3
   \left(p\cdot p_0\right){}^2+\left(\left(\left(c_5+c_7-c_{1256}\right)
   d_3+\left(c_5+c_7\right) d_4\right) m_{\omega }^2+4 c_3
   \left(d_3+d_4\right) \left(m_{\pi }^2-m_{\eta
   }^2\right)\right.\right.\\\left.\left.
   +\left(c_5+c_7\right) \left(d_{12} m_{\pi }^2-2
   \left(d_3+d_4\right) m_{\eta }^2\right)\right) p\cdot
   p_0+\left(d_3-d_4+d_{12}\right) m_{\pi }^2 \left(4 c_3 \left(m_{\eta
   }^2-m_{\pi }^2\right)+\left(c_5+c_7\right) \left(2 m_{\eta
   }^2-m_{\omega }^2\right)\right)\right)\\
   +2 k\cdot p \left(-4
   \left(c_5+c_7\right) \left(d_3+d_4\right) \left(k\cdot
   p_0\right){}^2+2 \left(\left(d_3+d_4\right) \left(4 c_3 m_{\pi }^2-2
   \left(2 c_3+c_5+c_7\right) m_{\eta }^2+\left(c_5+c_7\right) m_{\omega
   }^2\right)\right.\right.\\
   \left.-2 p\cdot p_0 \left(c_5+c_7\right) d_3\right) k\cdot
   p_0+\left(d_3+d_4\right) \left(4 c_3 m_{\pi }^2-\left(4
   c_3+c_5+c_7\right) m_{\eta }^2\right) \left(m_{\eta }^2-m_{\omega
   }^2\right)\\
   \left.\left.
   +p\cdot p_0 d_3 \left(8 c_3 m_{\pi }^2-2 \left(4
   c_3+c_5+c_7\right) m_{\eta }^2+\left(2
   \left(c_5+c_7\right)-c_{1256}\right) m_{\omega
   }^2\right)\right)\right)\\
   -\frac{8F_VC_q}{3F^2M_Vm_\rho^2D_\rho[(p+p_0+k)^2]D_\rho[(p+k)^2]}\\
   \left(4 (k\cdot p)^2 \left(c_{1256} d_3 m_{\rho
    }^2-\left(c_5+c_7\right) \left(2 d_3 \left(k\cdot p_0+p\cdot
    p_0\right)+\left(d_{12}-8 d_2\right) m_{\eta }^2+8 d_2 m_{\pi
    }^2\right)\right)\right.\\
    +k\cdot p \left(m_{\rho }^2 \left(2 \left(2
    \left(c_5+c_7\right)+c_{1256}\right) d_3 \left(k\cdot p_0+p\cdot
    p_0\right)-\left(2 \left(c_5+c_7\right)+c_{1256}\right) \left(8
    d_2-d_{12}\right) m_{\eta }^2\right.\right.\\
     \left.\left.
    +2 m_{\pi }^2 \left(4 \left(2
    \left(c_5+c_7\right)+c_{1256}\right) d_2+3 c_{1256}
    d_3\right)\right)-2 \left(c_5+c_7\right) m_{\pi }^2 \left(2 d_3
    \left(k\cdot p_0+p\cdot p_0\right)+\left(d_{12}-8 d_2\right) m_{\eta
    }^2+8 d_2 m_{\pi }^2\right)\right)\\\left.
    +c_{1256} m_{\pi }^2 m_{\rho }^2
    \left(2 d_3 \left(k\cdot p_0+m_{\pi }^2+p\cdot
    p_0\right)+\left(d_{12}-8 d_2\right) m_{\eta }^2+8 d_2 m_{\pi
    }^2\right)\right)\\   
  +\frac{8F_VC_q\lambda^{SV}_3}{3F^2D_\rho[(p + p_0 + k)^2]D_{a_0}[(p+p_0)^2]}\left(c_d p\cdot p_0+c_m m_{\pi
    }^2\right)\\
   -\frac{8F_VC_q}{3F^2M_V m_\rho^2 m_\omega^2D_\rho[(p+k)^2]}\left(k\cdot p \left(8 c_3 \left(m_{\pi
    }^2-m_{\eta }^2\right) \left(m_{\pi }^2 (d_1+8
    d_2+d_3+d_4)+(d_3-d_4) m_{\rho
    }^2\right)\right.\right.\\\left.
    +c_{1256} m_{\pi }^2 m_{\rho }^2 (d_1+8 d_2-2
    d_3)\right)+m_{\pi }^2 (d_1+8 d_2) m_{\rho }^2
    \left(c_{1256} m_{\pi }^2-8 c_3 \left(m_{\pi }^2-m_{\eta
    }^2\right)\right)\\\left.
    +2 (k\cdot p)^2 \left(8 c_3 d_4 \left(m_{\pi
    }^2-m_{\eta }^2\right)-c_{1256} d_3 m_{\rho }^2\right)\right)\\
   +\frac{\sqrt{2}F_V^2C_q}{3F^2m_\rho^2m_\omega^2D_\rho[(p+p_0+k)^2]}\left(2 k\cdot p_0+m_{\eta }^2+m_{\rho
    }^2+2 p\cdot p_0\right)\\
    \left(\left(\lambda _3^{\text{VV}}+\lambda
    _4^{\text{VV}}+2 \lambda _5^{\text{VV}}\right) \left(k\cdot p+m_{\pi
    }^2+p\cdot p_0\right)+4 m_{\pi }^2 \lambda _6^{\text{VV}}+4 p\cdot
    p_0 \left(\lambda _1^{\text{VV}}+\lambda _2^{\text{VV}}\right)\right)\,,
    \\
\end{multline}

\begin{multline}\label{v3-2R-RChL}
 v_3^{2R}=
  -\frac{16F_VC_q(d_3-d_4)}{3F^2M_Vm_\omega^2D_\rho[(p+p_0+k)^2]D_\omega[(p_0+k)^2]}\left(4c_3(m_\pi^2-m_\eta^2)-(c_5+c_5)(2k\cdot p_0+m_\eta^2)\right)
  (m_\omega^2-m_\eta^2-2k\cdot p_0)\\
  +\frac{16F_VC_q}{3F^2M_Vm_\rho^2D_\rho[(p+p_0+k)^2]D_\rho[(p+k)^2]}\left(\frac{m_\rho^2}{2}\left(2 d_3 \left(c_{1256}
   \left(2 k\cdot p+k\cdot p_0+p\cdot p_0\right)\right.\right.\right.\\\left.\left.
   -2 \left(c_5+c_7\right)
   \left(k\cdot p_0+p\cdot p_0\right)\right)+\left(2
   \left(c_5+c_7\right)-c_{1256}\right) \left(8 d_2-d_{12}\right)
   m_{\eta }^2+2 m_{\pi }^2 \left(4 \left(c_{1256}-2
   \left(c_5+c_7\right)\right) d_2+c_{1256}
   d_3\right)\right)\\\left.
   +\left(c_5+c_7\right) \left(2 k\cdot p+m_{\pi
   }^2\right) \left(2 d_3 \left(k\cdot p_0+p\cdot
   p_0\right)+\left(d_{12}-8 d_2\right) m_{\eta }^2+8 d_2 m_{\pi
   }^2\right)\frac{}{}\right)\\
-\frac{8F_VC_q}{3F^2M_V m_\rho^2 m_\omega^2D_\rho[(p+k)^2]} \left(8 c_3 \left(m_{\pi }^2-m_{\eta }^2\right)
   \left(m_{\pi }^2 (d_1+8 d_2+d_3+d_4)-(d_3+d_4) m_{\rho }^2\right)\right.\\
   \left.+c_{1256} m_{\pi }^2
   (-(d_1+8 d_2)) m_{\rho }^2+2 k\cdot p \left(c_{1256} d_3 m_{\rho }^2+8 c_3 d_4 \left(m_{\pi }^2-m_{\eta
   }^2\right)\right)\right) \\
  -\frac{\sqrt{2}F_V^2C_q}{3F^2m_\rho^2m_\omega^2D_\rho[(p+p_0+k)^2]}\left(\lambda _3^{\text{VV}}+\lambda _4^{\text{VV}}+2 \lambda _5^{\text{VV}}\right) 
   \left(2 k\cdot p_0+m_{\eta }^2+m_{\rho }^2+2 p\cdot p_0\right)\,,
 \\
\end{multline}

\begin{multline}\label{v4-2R-RChL}
 v_4^{2R}=
  \frac{16F_VC_q(d_3-d_4)}{3F^2M_Vm_\omega^2D_\rho[(p+p_0+k)^2]D_\omega[(p_0+k)^2]}
  \left(\frac{1}{2} m_{\omega }^2 \left(4 c_{1256} d_3 k\cdot
   p_0+2 \left(c_{1256}-2 \left(c_5+c_7\right)\right) d_3 k\cdot p\right.\right.\\\left.
   +2
   \left(8 c_3+c_{1256}\right) d_3 m_{\eta }^2-16 c_3 d_3 m_{\pi }^2+2
   c_5 d_{12} m_{\pi }^2+2 c_7 d_{12} m_{\pi }^2-c_{1256} d_{12} m_{\pi
   }^2+2 \left(c_{1256}-2 \left(c_5+c_7\right)\right) d_3 p\cdot
   p_0\right)\\\left.
   +\left(2 d_3 \left(k\cdot p+p\cdot p_0\right)-d_{12} m_{\pi
   }^2\right) \left(\left(c_5+c_7\right) \left(2 k\cdot p_0+m_{\eta
   }^2\right)+4 c_3 \left(m_{\eta }^2-m_{\pi }^2\right)\right)\frac{}{}\right)\\
  -\frac{16F_VC_q\left(c_5+c_7\right) \left(d_3-d_4\right)}{3F^2M_Vm_\rho^2D_\rho[(p+p_0+k)^2]D_\rho[(p+k)^2]} 
   \left(2 k\cdot p+m_{\pi }^2\right) \left(2 k\cdot p-m_{\rho }^2+m_{\pi }^2\right)\\
  +\frac{16F_VC_q\left(c_5+c_7\right)}{3F^2M_V m_\rho^2 m_\omega^2D_\rho[(p+k)^2]}\left(-2 k\cdot p+m_{\rho
   }^2-m_{\pi }^2\right) \left(m_{\pi }^2 (d_1+8
   d_2+d_3+d_4)+2 d_4 k\cdot p\right)\\
  +\frac{\sqrt{2}F_V^2C_q}{3F^2m_\rho^2m_\omega^2D_\rho[(p+p_0+k)^2]}\left(\lambda _3^{\text{VV}}+\lambda _4^{\text{VV}}+2 \lambda _5^{\text{VV}}\right) 
   \left(2 k\cdot p_0+m_{\eta }^2+m_{\rho }^2+2 p\cdot p_0\right)\,.
 \\
\end{multline}
}

\section*{Appendix C: Off-shell width of meson resonances}
For completeness we explain in this appendix the expressions that we have used for the off-shell width of meson resonances relevant to our study. The $\rho(770)$ width is basically driven by Chiral Perturbation Theory results
\begin{equation}
 \Gamma_\rho(s)\,=\,\frac{s M_\rho}{96\pi F^2}\left[\sigma_\pi^{3/2}(s)\theta(s-4m_\pi^2)+\frac{1}{2}\sigma_K^{3/2}(s)\theta(s-4m_K^2)\right]\,,
\end{equation}
where $\sigma_P(s)=\sqrt{1-4\frac{m_P^2}{s}}$ and we note that the definition of the vector meson width is independent of the realization of the spin-one fields \cite{GomezDumm:2000fz}. Given the narrow character of the $\omega(782)$ resonance the 
off-shellness of its width can be neglected. A similar comment would apply to the $\phi(1020)$ meson, although it does not contribute to the considered processes in the ideal-mixing scheme for the $\omega-\phi$ mesons that we are following.\\

The $a_1(1260)$ meson energy-dependent width was derived in Ref.~\cite{Dumm:2009va} applying the Cutkosky rules to the analytical results for the form factors into $3\pi$~\cite{Dumm:2009va} and $KK\pi$ channels~\cite{Dumm:2009kj} that are the main 
contributions to this width. Since its computation requires the time-consuming numerical calculation of the corresponding correlator over phase-space, we computed $\Gamma_{a_1}(s)$ at 800 values of $s$ and use linear interpolation to obtain the 
width function at intermediate values.\\

Finally, the $a_0(980)$ meson is also needed as an input in the analyses. We have used the functional dependence advocated in eqs.~(19) and (20) of Ref.~\cite{Escribano:2016ntp} which take into account the main absorptive parts given by the 
$\pi\eta$, $K\bar{K}$ and $\pi\eta^\prime$ cuts. The very low-energy (G-parity violating) $\pi\pi$ cut has been neglected.\\

We point out that we are considering only the imaginary parts of the meson-meson loop functions giving rise to the resonance widths. On the contrary, we are disregarding the corresponding real parts. Although this procedure violates analyticity 
at NNLO in the chiral expansion, the numerical impact of this violation is negligible (see e.g. Ref.~\cite{Boito:2008fq}) and, for simplicity, we take this simplified approach in our study.\\


\begin{thebibliography}{99}

\bibitem{Leroy:1977pq}
  C.~Leroy and J.~Pestieau,
  Phys.\ Lett.\  B {\bf 72}, 398 (1978).

\bibitem{weinberg}
S.~Weinberg,
  Phys.\ Rev.\  {\bf 112}, 1375 (1958).

\bibitem{pi0eta}  
R.~F. Dashen, Phys.\ Rev.\ {\bf 183}, 1245 (1969);
  R.~Urech,
  Nucl.\ Phys.\ B {\bf 433} (1995) 234;
  B.~Moussallam,
  Nucl.\ Phys.\ B {\bf 504} (1997) 381;
  G.~Ecker, G.~Muller, H.~Neufeld and A.~Pich,
  Phys.\ Lett.\ B {\bf 477} (2000) 88;
  G.~Amor\'os, J.~Bijnens and P.~Talavera,
  Nucl.\ Phys.\ B {\bf 602} (2001) 87.
  
 \bibitem{PDG}
C. Patrignani \textit{et al.} (Particle Data Group), Chin. Phys. C, 40, 100001 (2016).

\bibitem{Aubert:2009an}
  B.~Aubert {\it et al.} [BaBar Collaboration],
  Phys.\ Rev.\ Lett.\  {\bf 103} (2009) 041802.

  \bibitem{Paver:2012tq}
  N.~Paver and Riazuddin,
  Phys.\ Rev.\ D {\bf 86} (2012) 037302.
  
\bibitem{Grenacs:1985da}
  L.~Grenacs,
  Ann.\ Rev.\ Nucl.\ Part.\ Sci.\  {\bf 35}, 455 (1985).

\bibitem{derrick1987}
M. Derrick {\it et al}, Phys. Lett. B{\bf 189}, 260 (1987).

\bibitem{estimates-scc}
C.~A.~Dom\'inguez,
  Phys.\ Rev.\ D {\bf 20}, 802 (1979);
  S.~Tisserant and T.~N.~Truong,
  Phys.\ Lett.\ B {\bf 115} (1982) 264;
A.~Pich,
  Phys.\ Lett.\ B {\bf 196}, 561 (1987);
A.~Bramon, S.~Narison and A.~Pich,
  Phys.\ Lett.\ B {\bf 196}, 543 (1987);
E. Berger and H. Lipkin, Phys. Lett. B{\bf 189}, 226 (1987);
  Y.~Meurice,
  Mod.\ Phys.\ Lett.\ A {\bf 2}, 699 (1987); Phys.\ Rev.\ D {\bf 36}, 2780 (1987); 
C.~K.~Zachos and Y.~Meurice,
  Mod.\ Phys.\ Lett.\ A {\bf 2}, 247 (1987); 
 S.~Fajfer and R.~J.~Oakes,
  Phys.\ Lett.\ B {\bf 213}, 376 (1988);
 H.~Pietschmann and H.~Rupertsberger,
  Phys.\ Rev.\ D {\bf 40}, 3115 (1989); 
J.~O.~Eeg and O.~Lie-Svendsen,
  Phys.\ Lett.\ B {\bf 200}, 182 (1988);
M. Suzuki, Phys. Rev. D{\bf 36}, 950 (1987); E.~Braaten, R.~J.~Oakes and S.~M.~Tse,
  Int.\ J.\ Mod.\ Phys.\ A {\bf 5}, 2737 (1990);
  J.~L.~D\'iaz- Cruz and G.~L\'opez Castro,
  Mod.\ Phys.\ Lett.\  A {\bf 6}, 1605 (1991);
  H.~Neufeld and H.~Rupertsberger,
  Z.\ Phys.\ C {\bf 68} (1995) 91.

\bibitem{babar2011}
P. del amo S\'anchez et al [BaBar collaboration], Phys. Rev. D{\bf 83}, 032002 (2011).
 
 \bibitem{Aubert:2008nj}
  B.~Aubert {\it et al.} [BaBar Collaboration],
  Phys.\ Rev.\ D {\bf 77} (2008) 112002.

\bibitem{recent}
  S.~Nussinov and A.~Soffer,
  Phys.\ Rev.\  D {\bf 78}, 033006 (2008);
{\bf 80} (2009) 033010;
  N.~Paver and Riazuddin,
  Phys.\ Rev.\ D {\bf 82}, 057301 (2010);
{\bf 84}, 017302 (2011);
  M.~K.~Volkov and D.~G.~Kostunin,
  Phys.\ Rev.\ D {\bf 86}, 013005 (2012).

\bibitem{Hayasaka:2009zz} 
  K.~Hayasaka [Belle Collaboration],
  PoS EPS {\bf -HEP2009}, 374 (2009).

\bibitem{B2TIPReport}
Belle-II Physics Book, Belle-II Collaboration and B2TIP-Community, to be published in Progress of Theoretical and Experimental Physics.
  
  \bibitem{Descotes-Genon:2014tla}
  S.~Descotes-Genon and B.~Moussallam,
  Eur.\ Phys.\ J.\ C {\bf 74} (2014) 2946.
  
  \bibitem{Escribano:2016ntp}
 R.~Escribano, S.~Gonz\'alez-Sol\'is and P.~Roig,
 Phys.\ Rev.\ D {\bf 94} (2016) 034008.
 
\bibitem{Kroll:2005sd} 
  P.~Kroll,
  Mod.\ Phys.\ Lett.\ A {\bf 20}, 2667 (2005).
 
 \bibitem{Guo:2012yt}
  Z.~H.~Guo, J.~A.~Oller and J.~Ruiz de Elvira,
  Phys.\ Rev.\ D {\bf 86} (2012) 054006.

  \bibitem{Simon}
  Talk given by Simon Eidelman at the MIAPP: FLAVOUR PHYSICS WITH HIGH-LUMINOSITY EXPERIMENTS, 24 Oct. to 18 Nov. 2016, Munich, Germany.

  \bibitem{Bevan:2014iga}
  A.~J.~Bevan {\it et al.} [BaBar and Belle Collaborations],
  Eur.\ Phys.\ J.\ C {\bf 74} (2014) 3026.
  
 \bibitem{Inami:2008ar}
   K.~Inami {\it et al.} [Belle Collaboration],
  Phys.\ Lett.\ B {\bf 672} (2009) 209.

  \bibitem{Hayashii}
 H.~Hayashii {\it et al.} [Belle Collaboration], 
talk given at the TAU'14 Conference, Aachen, Germany, 14-19 September, 2014.
 
 \bibitem{Denis}
 D.~Epifanov {\it et al.} [Belle Collaboration]. Work in progress. See, for instance, the poster presented at TAU'16 by Yifan Jin. To appear in the proceedings of the conference.
 
 \bibitem{Fujikawa:2008ma}
  M.~Fujikawa {\it et al.} [Belle Collaboration],
  Phys.\ Rev.\ D {\bf 78} (2008) 072006.
 
 \bibitem{Escribano:2013bca}
R.~Escribano, S.~Gonz\'alez-Sol\'is and P.~Roig,
  JHEP {\bf 1310}, 039 (2013).
  
\bibitem{Escribano:2014joa} 
  R.~Escribano, S.~Gonz\'alez-Sol\'is, M.~Jamin and P.~Roig,
  JHEP {\bf 1409}, 042 (2014).
 
 \bibitem{Dumm:2012vb}
  D.~G.~Dumm and P.~Roig,
  Phys.\ Rev.\ D {\bf 86} (2012) 076009.
 
 \bibitem{Guo}
 Z.~H.~Guo and P.~Roig,
 Phys.\ Rev.\ D \textbf{82} (2010) 113016.
 
 \bibitem{Dumm:2009va}
  D.~G.~Dumm, P.~Roig, A.~Pich and J.~Portol\'es,
  Phys.\ Lett.\ B {\bf 685} (2010) 158.

 \bibitem{Dumm:2013zh}
  D.~G\'omez Dumm and P.~Roig,
  Eur.\ Phys.\ J.\ C {\bf 73} (2013) no.8,  2528.
 
 \bibitem{Shekhovtsova:2012ra}
  O.~Shekhovtsova, T.~Przedzinski, P.~Roig and Z.~Was,
  Phys.\ Rev.\ D {\bf 86} (2012) 113008.
 
 \bibitem{Nugent:2013hxa}
  I.~M.~Nugent, T.~Przedzinski, P.~Roig, O.~Shekhovtsova and Z.~Was,
  Phys.\ Rev.\ D {\bf 88} (2013) 093012.

  \bibitem{Was:2015laa}
  Z.~Was and J.~Zaremba,
  Eur.\ Phys.\ J.\ C {\bf 75} (2015) no.11,  566
   Erratum: [Eur.\ Phys.\ J.\ C {\bf 76} (2016) no.8,  465].
  
 \bibitem{Lees:2012ks}
  J.~P.~Lees {\it et al.} [BaBar Collaboration],
  Phys.\ Rev.\ D {\bf 86} (2012) 092010.

  \bibitem{Low:1958sn}
  F.~E.~Low,
  Phys.\ Rev.\  {\bf 110} (1958) 974.
  
  \bibitem{BEG}
  J.~Bijnens, G.~Ecker and J.~Gasser, Nucl.\ Phys.\ B\textbf{396}, 81 (1993).
 
 \bibitem{Cirigliano:2002pv}
 V.~Cirigliano, G.~Ecker and H.~Neufeld, JHEP 0208, 002 (2002).
 
 \bibitem{Cirigliano:2001er}
  V.~Cirigliano, G.~Ecker and H.~Neufeld,
  Phys.\ Lett.\ B {\bf 513} (2001) 361.
  
 \bibitem{Asner:1999kj}
  D.~M.~Asner {\it et al.} [CLEO Collaboration],
  Phys.\ Rev.\ D {\bf 61} (2000) 012002.

  \bibitem{Uehara:2009cf}
  S.~Uehara {\it et al.} [Belle Collaboration],
  Phys.\ Rev.\ D {\bf 80} (2009) 032001.
 
 \bibitem{FloresTlalpa:2008zz}
  A.~Flores-Tlalpa,
  Ph. D. Thesis Cinvestav, Mexico City, 2008.
 
  \bibitem{ChPT}
  S.~Weinberg,
  Physica A {\bf 96} (1979) 327;
  J.~Gasser and H.~Leutwyler,
  Annals Phys.\  {\bf 158} (1984) 142;
  Nucl.\ Phys.\ B {\bf 250} (1985) 465;
  J.~Bijnens, G.~Colangelo and G.~Ecker,
  JHEP {\bf 9902} (1999) 020;
  Annals Phys.\  {\bf 280} (2000) 100;
  J.~Bijnens, L.~Girlanda and P.~Talavera,
  Eur.\ Phys.\ J.\ C {\bf 23} (2002) 539;
  B.~Ananthanarayan and B.~Moussallam,
  JHEP {\bf 0205} (2002) 052.
  
  \bibitem{Colangelo:1996hs}
  G.~Colangelo, M.~Finkemeier and R.~Urech,
  Phys.\ Rev.\ D {\bf 54} (1996) 4403.
  
  \bibitem{RChL}
  J.~F.~Donoghue, C.~Ram\'irez and G.~Valencia,
  Phys.\ Rev.\ D {\bf 39} (1989) 1947;
  G.~Ecker, J.~Gasser, A.~Pich and E.~de Rafael,
  Nucl.\ Phys.\ B {\bf 321} (1989) 311;
  G.~Ecker, J.~Gasser, H.~Leutwyler, A.~Pich and E.~de Rafael,
  Phys.\ Lett.\ B {\bf 223} (1989) 425.
  
  \bibitem{HLS}
  M.~Bando, T.~Kugo, S.~Uehara, K.~Yamawaki and T.~Yanagida,
  Phys.\ Rev.\ Lett.\  {\bf 54} (1985) 1215;
  M.~Bando, T.~Kugo and K.~Yamawaki,
  Phys.\ Rept.\  {\bf 164} (1988) 217.
  
  \bibitem{BL}
  S.~J.~Brodsky and G.~R.~Farrar,
  Phys.\ Rev.\ Lett.\  {\bf 31} (1973) 1153;
  G.~P.~Lepage and S.~J.~Brodsky,
  Phys.\ Rev.\ D {\bf 22} (1980) 2157.

  \bibitem{LargeNc}
  G.~'t Hooft,
  Nucl.\ Phys.\ B {\bf 72} (1974) 461;
 {\bf 75} (1974) 461;
  E.~Witten,
  Nucl.\ Phys.\ B {\bf 160} (1979) 57.

  \bibitem{Pich:2002xy}
  A.~Pich,
  hep-ph/0205030. Phenomenology of large N(c) QCD. Proceedings, Tempe, USA, January 9-11, 2002. (River Edge, USA: World Scientific (2002), 239-258).

  \bibitem{NLO_1/N_C}
  J.~J.~Sanz-Cillero and A.~Pich,
  Eur.\ Phys.\ J.\ C {\bf 27} (2003) 587;
  I.~Rosell, J.~J.~Sanz-Cillero and A.~Pich,
  JHEP {\bf 0408} (2004) 042;
{\bf 0701} (2007) 039;
  A.~Pich, I.~Rosell and J.~J.~Sanz-Cillero,
  JHEP {\bf 0807} (2008) 014; 
{\bf 1102} (2011) 109;
  I.~Rosell, P.~Ruiz-Femen\'ia and J.~J.~Sanz-Cillero,
  Phys.\ Rev.\ D {\bf 79} (2009) 076009;
  J.~J.~Sanz-Cillero,
  Phys.\ Lett.\ B {\bf 681} (2009) 100;
  J.~J.~Sanz-Cillero and J.~Trnka,
  Phys.\ Rev.\ D {\bf 81} (2010) 056005.
  
  \bibitem{CCWZ}
  S.~R.~Coleman, J.~Wess and B.~Zumino,
  Phys.\ Rev.\  {\bf 177} (1969) 2239;
  C.~G.~Callan, Jr., S.~R.~Coleman, J.~Wess and B.~Zumino,
  Phys.\ Rev.\  {\bf 177} (1969) 2247.
  
  \bibitem{KL}
  R.~Kaiser and H.~Leutwyler,
  In *Adelaide 1998, Nonperturbative methods in quantum field theory* 15-29; 
  Eur.\ Phys.\ J.\ C {\bf 17} (2000) 623.

  \bibitem{qfbasis}
  J.~Schechter, A.~Subbaraman and H.~Weigel,
  Phys.\ Rev.\ D {\bf 48} (1993) 339;
  T.~Feldmann, P.~Kroll and B.~Stech,
  Phys.\ Rev.\ D {\bf 58} (1998) 114006;
  Phys.\ Lett.\ B {\bf 449} (1999) 339;
  T.~Feldmann,
  Int.\ J.\ Mod.\ Phys.\ A {\bf 15} (2000) 159.

    \bibitem{ChiralAnomaly}
  J.~Wess and B.~Zumino,
  Phys.\ Lett.\ B {\bf 37} (1971) 95;
  E.~Witten,
  Nucl.\ Phys.\ B {\bf 223} (1983) 422.

  \bibitem{Sakurai:1960ju}
  J.~J.~Sakurai,
  Annals Phys.\  {\bf 11} (1960) 1.
  
  \bibitem{Cirigliano:2006hb}
  V.~Cirigliano, G.~Ecker, M.~Eidem\"uller, R.~Kaiser, A.~Pich and J.~Portol\'es,
  Nucl.\ Phys.\ B {\bf 753} (2006) 139.

  \bibitem{Kampf:2011ty}
  K.~Kampf and J.~Novotny,
  Phys.\ Rev.\ D {\bf 84} (2011) 014036.

  \bibitem{Ecker:2007us}
  G.~Ecker and C.~Zauner,
  Eur.\ Phys.\ J.\ C {\bf 52} (2007) 315.
  
  \bibitem{Shekhovtsova:2016luj}
  O.~Shekhovtsova, J.~J.~Sanz-Cillero and T.~Przedzinski,
  arXiv:1610.00153 [hep-ph].
  
  \bibitem{GomezDumm:2003ku}
  D.~G\'omez Dumm, A.~Pich and J.~Portol\'es,
  Phys.\ Rev.\ D {\bf 69} (2004) 073002.
  
  \bibitem{RuizFemenia:2003hm}
  P.~D.~Ruiz-Femen\'ia, A.~Pich and J.~Portol\'es,
  JHEP {\bf 0307} (2003) 003.
  
  \bibitem{Roig:2013baa}
  P.~Roig and J.~J.~Sanz Cillero,
  Phys.\ Lett.\ B {\bf 733} (2014) 158.

  \bibitem{Weinberg:1967kj}
  S.~Weinberg,
  Phys.\ Rev.\ Lett.\  {\bf 18} (1967) 507.

  \bibitem{Wilson:1969zs}
  K.~G.~Wilson,
  Phys.\ Rev.\  {\bf 179} (1969) 1499.

  \bibitem{Guo:2008sh}
  Z.~H.~Guo,
  Phys.\ Rev.\ D {\bf 78} (2008) 033004.
  
  \bibitem{Guo:2010dv}
  Z.~H.~Guo and P.~Roig,
  Phys.\ Rev.\ D {\bf 82} (2010) 113016.
  
  \bibitem{Dumm:2009kj}
  D.~G.~Dumm, P.~Roig, A.~Pich and J.~Portol\'es,
  Phys.\ Rev.\ D {\bf 81} (2010) 034031.
  
  \bibitem{Guevara:2013wwa}
  A.~Guevara, G.~L\'opez Castro and P.~Roig,
  Phys.\ Rev.\ D {\bf 88} (2013) no.3,  033007.

  \bibitem{Ananthanarayan:2004qk}
  B.~Ananthanarayan and B.~Moussallam,
  JHEP {\bf 0406} (2004) 047.
  
  \bibitem{Peris}
  S.~Peris, M.~Perrottet and E.~de Rafael,
  JHEP {\bf 9805} (1998) 011;
  M.~Knecht, S.~Peris, M.~Perrottet and E.~de Rafael,
  Phys.\ Rev.\ Lett.\  {\bf 83} (1999) 5230;
  S.~Peris, B.~Phily and E.~de Rafael,
  Phys.\ Rev.\ Lett.\  {\bf 86} (2001) 14.
  
 \bibitem{JOP}
  M.~Jamin, J.~A.~Oller and A.~Pich,
  Nucl.\ Phys.\ B {\bf 587} (2000) 331;
{\bf 622} (2002) 279.
  
  \bibitem{Cirigliano:2004ue}
  V.~Cirigliano, G.~Ecker, M.~Eidem\"uller, A.~Pich and J.~Portol\'es,
  Phys.\ Lett.\ B {\bf 596} (2004) 96.

  \bibitem{Cirigliano:2005xn}
  V.~Cirigliano, G.~Ecker, M.~Eidem\"uller, R.~Kaiser, A.~Pich and J.~Portol\'es,
  JHEP {\bf 0504} (2005) 006.

  \bibitem{Bijnens:2011tb}
  J.~Bijnens and I.~Jemos,
  Nucl.\ Phys.\ B {\bf 854} (2012) 631.

  \bibitem{Jiang:2009uf}
  S.~Z.~Jiang, Y.~Zhang, C.~Li and Q.~Wang,
  Phys.\ Rev.\ D {\bf 81} (2010) 014001.
 
 \bibitem{Chen:2012vw}
  Y.~H.~Chen, Z.~H.~Guo and H.~Q.~Zheng,
  Phys.\ Rev.\ D {\bf 85} (2012) 054018.
 
 \bibitem{Chen:2013nna}
  Y.~H.~Chen, Z.~H.~Guo and H.~Q.~Zheng,
  Phys.\ Rev.\ D {\bf 90} (2014) no.3,  034013.

  \bibitem{Chen:2014yta}
  Y.~H.~Chen, Z.~H.~Guo and B.~S.~Zou,
  Phys.\ Rev.\ D {\bf 91} (2015) 014010.
 
 
 \bibitem{v1}
  A.~Guevara, G.~L\'opez-Castro and P.~Roig,
  arXiv:1612.03291v1 [hep-ph].
 Version one of this paper (including the $MDM/R\chi L$ simulations sampling the parameter space uniformly within a one-sigma range) can be found in http://arxiv.org/pdf/1612.03291v1.pdf.

  \bibitem{Pich:2013lsa}
  A.~Pich,
  Prog.\ Part.\ Nucl.\ Phys.\  {\bf 75} (2014) 41.
  
  \bibitem{GomezDumm:2000fz}
  D.~G\'omez Dumm, A.~Pich and J.~Portol\'es,
  Phys.\ Rev.\ D {\bf 62} (2000) 054014.
  
  \bibitem{Boito:2008fq}
  D.~R.~Boito, R.~Escribano and M.~Jamin,
  Eur.\ Phys.\ J.\ C {\bf 59} (2009) 821.
  
  \bibitem{Denner:1991kt}
  A.~Denner,
  Fortsch.\ Phys.\  {\bf 41} (1993) 307.
  \end{thebibliography}
\end{document}